\documentclass[twocolumn]{aastex61}
\usepackage{url}
\usepackage{graphicx}
\usepackage{natbib}
\usepackage{subfigure}
\usepackage{gensymb}

\newcommand{\Msun}{$M_{\odot}$}
\newcommand{\lascarI}{ACT$-$CL\,J0102$-$4915}
\newcommand{\lascarII}{ACT$-$CL\,J0215$-$5212}
\newcommand{\lascarIII}{ACT$-$CL\,J0232$-$5257}
\newcommand{\lascarIV}{ACT$-$CL\,J0235$-$5121}
\newcommand{\lascarV}{ACT$-$CL\,J0245$-$5302}
\newcommand{\lascarVI}{ACT$-$CL\,J0330$-$5227}
\newcommand{\lascarVII}{ACT$-$CL\,J0438$-$5419}
\newcommand{\lascarVIII}{ACT$-$CL\,J0546$-$5345}
\newcommand{\lascarIX}{ACT$-$CL\,J0559$-$5249}
\newcommand{\lascarX}{ACT$-$CL\,J0616$-$5227}

\newcommand{\macsB}{MACS\,J1115.8+0129}

\shortauthors{Aguirre et al.}
\begin{document}

\title{ The LABOCA/ACT Survey of Clusters at All Redshifts: multi-wavelength analysis of background submillimeter galaxies.}

%\correspondingauthor{Paula Aguirre}
%\email{paaguirr@ing.puc.cl}

\author{Paula Aguirre}
%\email{paaguirr@ing.puc.cl}
\affiliation{Escuela de Ingenier\'ia, Pontificia Universidad Cat\'olica de Chile, Av. Vicu\~na Mackenna 4860, Macul, Santiago, Chile}

\author{Robert R. Lindner}
\affiliation{Department of Physics and Astronomy, Rutgers, The State University of New Jersey, 136 Frelinghuysen Road, Piscataway, NJ 08854-8019, USA}
\affiliation{Department of Astronomy, The University of Wisconsin-Madison, 475 N. Charter Street, Madison, WI 53706-1582, USA}

\author{Andrew J. Baker}
\affiliation{Department of Physics and Astronomy, Rutgers, The State University of New Jersey, 136 Frelinghuysen Road, Piscataway, NJ 08854-8019, USA}

\author{J. Richard Bond}
\affiliation{Canadian Institute for Theoretical Astrophysics, 60 St. George Street,University of Toronto, Toronto, ON, M5S 3H8, Canada}

\author{Rolando D\"unner}
\affiliation{Instituto de Astrof\'isica, Facultad de F\'isica, Pontificia Universidad Cat\'olica de Chile, Av.Vicu\~na Mackenna 4860, Macul, Santiago, Chile}

\author{Gaspar Galaz}
\affiliation{Instituto de Astrof\'isica, Facultad de F\'isica, Pontificia Universidad Cat\'olica de Chile, Av.Vicu\~na Mackenna 4860, Macul, Santiago, Chile}

\author{Patricio Gallardo}
\affiliation{Department of Physics, Cornell University, Ithaca, NY, 14853, USA}

\author{Matt Hilton}
\affiliation{Astrophysics and Cosmology Research Unit, School of Mathematics, Statistics \& Computer Science, University of KwaZulu-Natal, Durban 4041, South Africa}
\affiliation{Centre for Astronomy and Particle Theory, School of Physics and Astronomy, University of Nottingham, Nottingham NG7 2RD, UK}

\author{John P. Hughes}
\affiliation{Department of Physics and Astronomy, Rutgers, The State University of New Jersey, 136 Frelinghuysen Road, Piscataway, NJ 08854-8019, USA}
\affiliation{Center for Computational Astrophysics, Flatiron Institute, 162 Fifth Avenue, New York, NY 10010, USA}

\author{Leopoldo Infante}
\affiliation{Instituto de Astrof\'isica and Centro de Astroingenier\'ia, Facultad de F\'isica, Pontificia Universidad Cat\'olica de Chile, Av.Vicu\~na Mackenna 4860, Macul, Santiago, Chile}

\author{Marcos Lima}
\affiliation{Departamento de F\'isica Matem\'atica, Instituto de F\'isica, Universidade de S\~ao Paulo, S\~ao Paulo SP, Brazil}

\author{Karl M. Menten}
\affiliation{Max-Planck-Institut f{\"u}r Radioastronomie, Auf dem H{\"u}gel 69, 53121, Bonn, Germany}

\author{Jonathan Sievers}
\affiliation{School of Chemistry and Physics, University of KwaZulu-Natal, Private Bag X54001, Durban 4000, South Africa}

\author{Axel Wei{\ss}}
\affiliation{Max-Planck-Institut f{\"u}r Radioastronomie, Auf dem H{\"u}gel 69, 53121, Bonn, Germany}

\author{Edward J. Wollack}
\affiliation{NASA Goddard Space Flight Center, 8800 Greenbelt Road, Greenbelt, MD 20771, USA}

\begin{abstract}

We present a multi-wavelength analysis of 48 submillimeter galaxies (SMGs) detected in the LABOCA/ACT Survey of Clusters at All Redshifts, LASCAR, which acquired new 870 $\mu$m and ATCA 2.1 GHz observations of ten galaxy clusters detected through their Sunyaev-Zel'dovich effect (SZE) signal by the Atacama Cosmology Telescope. Far-infrared observations were also conducted with the  PACS (100/160 \micron)  and SPIRE (250/350/500 \micron) instruments on {\it Herschel} for sample subsets of five and six clusters. LASCAR 870 \micron\, maps were reduced using a multi-scale iterative pipeline that removes the SZE increment signal, yielding point-source sensitivities of $\sigma\sim2\rm{\,mJy\,beam}^{-1}$. We detect in total 49 sources at the $4\sigma$ level, and conduct a detailed multi-wavelength analysis considering our new radio and far-IR observations plus existing near-IR and optical data. One source is identified as a foreground galaxy, 28 SMGs are matched to single radio sources, 4 have double radio counterparts, and 16 are undetected at 2.1 GHz but tentatively associated in some cases to near-IR/optical sources. We estimate photometric redshifts for 34 sources with secure (25) and tentative (9) matches at different wavelengths, obtaining a median $z=2.8^{+2.1}_{-1.7}$. Compared to previous results for single-dish surveys, our redshift distribution has a comparatively larger fraction of sources at $z>3$ and the high-redshift tail is more extended. This is consistent with millimeter spectroscopic confirmation of a growing number of high-$z$ SMGs and relevant for testing of cosmological models. Analytical lens modeling is applied to estimate magnification factors for 42 SMGs at cluster-centric radii $>1.2$\arcmin; with the demagnified flux densities and source-plane areas, we obtain integral number counts that agree with previous submillimeter surveys.
\end{abstract}


\section{Introduction} \label{s-intro}

Since the initial measurements of the cosmic infrared background (CIB; for a review see \citealt{hauser2001}) revealed that the amount of energy radiated in the far-infrared (IR) and submillimeter spectral windows is comparable to that measured at ultraviolet (UV) and optical wavelengths, it has been widely recognized that one of the keys to a comprehensive understanding of the star formation history of the Universe is the study of the multi-wavelength properties of dusty star-forming galaxies (DSFGs) whose integrated radiation produce the CIB. These systems host intense star-forming activity obscured by large columns of dust, which re-emit the UV radiation of young hot stars at longer wavelengths, so that the peak of their rest-frame spectral energy distribution (SED) falls in the far-IR. In the local Universe, DSFGs are typically identified as luminous or ultra-luminous infrared galaxies (LIRGs/ULIRGs; \citealt{sanders96}), whereas more distant DSFGs' emission can be redshifted into the submillimeter domain, allowing many to manifest as submillimeter galaxies (SMGs; \citealt{blain02,casey2014}). SMGs were first detected with the Submillimeter Common-User Bolometer Array (SCUBA; \citealt{holland99}) on the James Clerk Maxwell Telescope (JCMT) both in blank-field surveys (e.g., \citealt{hughes98,barger99a,scott2002,serjeant03,webb03,coppin06}) and behind galaxy clusters (e.g., \citealt{smail1997,chapman2002b,cowie2002,knudsen08}). With the subsequent advent of comparable single-dish telescopes and larger format instruments covering the $870~\mu$m atmospheric window like the Large APEX Bolometer Camera (LABOCA; \citealt{siringo09}) on the 12-meter Atacama Pathfinder Experiment telescope (APEX; \citealt{gusten2006}), and more recently SCUBA-2 \citep{holland2013}, the number of known SMGs is now of the order of a few thousand (e.g., \citealt{weiss09b,johansson11, chen2013,hsu2016, geach2017}), and intensive observational efforts have been devoted to understanding their physical properties. In this quest, one of the main challenges has been the coarse (15-20\arcsec) resolution of single-dish observations, which hinders identification of counterparts at different wavelengths and can result in the blending of multiple, fainter SMGs into a single brighter object. However, persistence of the local radio-FIR correlation to higher redshifts (e.g., \citealt{condon92}) allows determination of accurate SMG positions from deep radio imaging obtained with the Very Large Array (VLA) at 1.4 GHz and the Australia Telescope Compact Array (ATCA) at 2.1 GHz (e.g., \citealt{ivison1998,ivison2000,ivison2002, smail2000,chapman2002b}), thus enabling identification of optical and near-IR counterparts, determination of photometric and spectroscopic redshifts (e.g., \citealt{chapman2005}), modeling of SEDs, analysis of individual morphologies, and characterization of dust and stellar components (for full reviews of these results, see \citealt{blain02} and \citealt{casey2014}). The general picture derived from such studies is that SMGs are massive, gas-rich galaxies with high IR luminosities ($L_{\rm{IR}}\geq10^{12}\,L_{\sun}$) and complex optical/near-IR morphologies, in which respects they resemble the local ULIRG population. However, SMGs have a median redshift $z\sim2.5$ \citep{chapman2005} and significantly higher number density than ULIRGs. Complementary observations with centimeter and (sub)millimeter telescopes have been used as well to study the cool, molecular gas of SMGs (e.g., \citealt{carilli2013}), an effort that has been transformed in recent years thanks to the exceptional spatial resolution and sensitivity provided by the Jansky Very Large Array (VLA) and the Atacama Large Millimeter/submillimeter Array (ALMA). Moreover, high-resolution continuum imaging at 870 \micron\, with ALMA has made it possible to resolve the structure of the dust emission from SMGs and identify their counterparts in an unbiased way, revealing that a large fraction of bright single-dish detections actually ``break up'' into multiple, fainter ($S_{870}\lesssim9$ mJy) SMGs blended together at the coarse resolution of the maps in which they were detected (\citealt{hodge13,karim13,simpson2015a,simpson2015b}).

In addition to the classical SMG population selected at $\sim$850 \micron, wide-field (sub)millimeter experiments like the Atacama Cosmology Telescope (ACT; \citealt{swetz11}) and the South Pole Telescope (SPT; \citealt{carlstrom11}) have detected significant numbers of strongly lensed, high-$z$ DSFGs (\citealt{vieira2010,marriage11a,mocanu2013,marsden2014,su2017}), whose apparent luminosities and sizes are highly magnified by foreground galaxies or galaxy groups, thus enabling examination of their internal structures. This new subpopulation of DSFGs opens a new window for the study of high-$z$ star-formation. 

We note, however, that ACT and SPT surveys were primarily designed to measure cosmic microwave background (CMB) anisotropies and to detect the Sunyaev-Zel'dovich Effect (SZE; \citealt{sunyaev72}) signals of galaxy clusters, which manifest as  decrements relative to the CMB at observed frequencies $\nu<220$ GHz and increments at $\nu>220$ GHz. ACT and SPT have produced sizable catalogs of new, massive clusters over a $z\sim0.1-1.4$ redshift range (e.g., \citealt{marriage11b, reichardt2013, hasselfield2013,bleem2015}) that can be used to study the formation and evolution of the largest virialized structures in the universe, and to set constraints on cosmological parameters. Multiband follow-up of SZE-selected galaxy clusters is required to confirm the detections and investigate their internal physical properties; in the case of the ACT galaxy cluster sample, optical imaging and spectroscopy have been used in combination with X-ray data for assessment of sample purity, determination of redshifts, measurement of dynamical masses, and characterization of  scaling relations  (\citealt{menan10b,menanteau12,menan13,sifon2013,sifon2016}). Near-IR  photometry (\citealt{hilton13}) from the \emph{Spitzer Space Telescope} (\citealt{werner04}) has been obtained as well to measure the stellar mass components of ACT clusters.

In \citet{lindner14} we introduced the LABOCA/ ACT Survey of Clusters at All Redshifts, denoted LASCAR in honor of the homonymous active volcano in the north of Chile. This project comprises new observations at 870 $\mu$m obtained with LABOCA (19.6\arcsec\, resolution) and at 2.1 GHz with ATCA ($\sim5.0$\arcsec\, resolution) of a set of ten massive clusters from the southern ACT sample. At submillimeter wavelengths, follow-up observation of SZE-selected clusters at higher spatial resolution than the $\sim1$\arcmin\, FWHM of the original detection maps has twofold appeal. First, at 870 \micron\, one can measure the SZE increment and constrain the shape of the thermal SZE spectrum; this information is used in turn to estimate the clusters' peculiar velocities and evaluate the scatter they introduce to scaling relations between mass and kinetic Sunyaev-Zel'dovich (kSZ) signal. Additionally, galaxy clusters serve as gravitational lenses for background point sources, thus allowing detection of magnified SMGs. Since point source emission is a significant contributor to the surface
brightness in clusters at ACT frequencies, measurement of background SMG flux densities is essential to determine the degree of contamination of 148 GHz decrements, and thus avoid biases in the SZE-mass relation and in the estimated number counts of clusters as function of mass and $z$ (\citealt{sehgal07,sehgal11}). This contamination directly affects the cross-calibration and interpretation of SZE surveys, and therefore those surveys' ability to derive robust conclusions about cosmology. 

Our initial paper (\citealt{lindner14}) focused on the measurement of the clusters' spatially resolved SZE increments, on the analysis of background and foreground contamination of this signal, and on estimation  of the cluster's peculiar velocities. It was predicted in this work that  the combined signals from 2.1 GHz-selected radio sources and 870 $\mu$m-selected SMGs contaminate the 148GHz SZE decrement signal by $\sim5\%$, and the 345 GHz SZE increment by $\sim5\%$.

In this work, we present a multi-wavelength study of the submillimeter point source population detected in LASCAR, and use our original 870 $\mu$m and 2.1 GHz observations in combination with new far-IR imaging obtained with the Spectral and Photometric Receiver (SPIRE; \citealt{griffin10}) and the Photodetector Array Camera and Spectrometer (PACS; \citealt{pogli10}) on board the \emph{Herschel Space Observatory}\footnote{{\it Herschel} is an ESA space observatory with science instruments provided by European-led Principal Investigator consortia and with important participation from NASA.} (\citealt{pilbratt10}), plus existing optical and near-IR $Spitzer$ data, to identify their plausible counterparts and investigate the general properties of background SMGs. 

This paper is  organized as follows: in Section 2 we give a detailed description of the clusters targeted in our submillimeter survey, and in Section 3 we present new submillimeter, radio, and far-IR observations of the LASCAR sample, and report on the data reduction techniques applied in each case. In Section 4 we describe additional datasets used in the analysis of SMGs, and Section 5 focuses on source extraction and photometry algorithms. In Section 6 we analyze sources (individually and collectively)  detected in our data, including counterpart identification, estimation of redshifts and gravitational magnifications, and calculation of the redshift distribution and number counts for the SMG sample. Finally, Section 7 summarizes the conclusions derived from our survey. Throughout this work, we assume a flat $\Lambda$CDM cosmology with $H_0=70$ km s$^{-1}$ Mpc$^{-1}$, $\Omega_M=0.27$, and $\Omega_{\Lambda}=0.73$ \citep{komatsu11}.

\section{Cluster Sample}\label{sec:sample}

\indent The ACT southern survey obtained 148 GHz observations of a 455 deg$^2$ strip centered at Dec.=$-52.5^{\circ}$, which resulted in the identification of 23 cluster decrements (\citealt{marriage11b}), nine of them with signal to noise ratios S/N$>6$, and 14 with S/N between 3 and 6. Optical follow-up (\citealt{menan09a,menan10a}, see \S\,4.2 ) confirmed that all decrements corresponded to rich galaxy clusters, ten of them newly discovered, and it was shown that the sample is $80\%$ complete for clusters with masses larger that $6\times10^{14}$ \Msun\ (\citealt{menan10b}).

\indent From the set of 15 ACT clusters with highest S/N, we targeted 10 clusters that had not been previously mapped at submillimeter wavelengths, 9 of which were unknown before ACT or SPT. These clusters span a redshift range $z=0.3-1.1$ and have dynamical masses $M_{500}=5.2-11.3\times10^{14}$ \Msun\ (\citealt{sifon2016}), where $M_{500}=500(4\pi/3)\rho_cr^3_{500}$, and $r_{500}$ is the radius enclosing a mass density equal to 500 times the critical density ($\rho_c$) of the Universe at the cluster's redshift.  Their physical properties are detailed in Table \ref{tab:clusters}. The most studied cluster in this sample is \lascarI, identified by \citet{menanteau12} as the most massive known cluster at high redshift ($z=0.87$), and also known as ``El Gordo''.

%TABLE 1
\begin{deluxetable*}{lcccccc}
\tablecolumns{7}
\tablewidth{0pt}
\tabletypesize{\scriptsize}
\tablecaption{Galaxy clusters mapped with LABOCA at 870 \micron.  \label{tab:clusters}}
\tablehead{	
\colhead{Target} & \colhead{R.A. (J2000)} & \colhead{Dec. (J2000) }   &  \colhead{z}  &  \colhead{Dynamical mass (10$^{14}$\Msun )} &\colhead{ACT S/N} } 
\startdata
\hline \\
\lascarI   & 01:02:52.5  & -49:14:58.0   & 0.87008$\pm$0.00010 & $11.3\pm2.9$ &9.0 \\
\lascarII &  02:15:12.3  & -52:12:25.3   & 0.48009$\pm$0.00012 & \, $7.6\pm2.3$&4.9 \\
\lascarIII & 02:32:46.2  & -52:57:50.0   & 0.55595$\pm$0.00009 & \,  $5.2\pm1.4$&4.7\\
\lascarIV & 02:35:45.3  & -51:21:05.2  &  0.27768$\pm$0.00006 & \, $8.0\pm2.0$&6.2 \\
\lascarV  & 02:45:35.8  & -53:02:16.8  & 0.30280$\pm$0.00001  & \nodata & 9.1\\
\lascarVI & 03:30:56.8  & -52:28:13.7  & 0.44173$\pm$0.00009 & $11.6\pm2.7$&6.1 \\
\lascarVII & 04:38:17.7  & -54:19:20.7  & 0.42141$\pm$0.00011 & $12.9\pm3.2$&8.0 \\
\lascarVIII & 05:46:37.7 & -53:45:31.1  & 1.06628$\pm$0.00020 & \, $5.5\pm2.3$&6.5\\
\lascarIX& 05:59:43.2 & -52:49:27.1  &0.60910$\pm$0.00026  & \, $8.3\pm3.0$&5.1 \\
\lascarX  & 06:16:34.2  & -52:27:13.3 &0.68380$\pm$0.00044 &\, $9.5\pm4.5$ &5.9\\
\enddata
\tablecomments{Column 1 indicates the cluster name; Columns 2 to 5 correspond to the coordinates, redshift, and mass estimate for each cluster. Column 6 indicates the signal to noise ratio of the ACT 148 GHz detections (from \citealt{marriage11b}). Coordinates for all LASCAR clusters correspond to the positions of the brightest cluster galaxies (BCGs) identified by \citet{menan10b}. For all clusters the redshifts and masses are based on the spectroscopic measurements of \citet{sifon2016}, except for \lascarV, for which we use the spectroscopic measurement of \citet{ruel2014}.}
\end{deluxetable*}

\section{Observations and Data Reduction} \label{sec:obs}

\subsection{ LABOCA 870 \micron}

The LASCAR dataset consists of new 870 \micron\, mapping of the ten galaxy clusters listed in Table \ref{tab:clusters} obtained with the LABOCA  instrument at the APEX telescope. In the following subsections we give details on the observations and the reduction algorithm applied to these data.
 
\subsubsection{ LABOCA 870 \micron\, observations}

LABOCA observations of our cluster sample took place over three semesters (2010B, 2011A, 2011B) and were split among several  proposals that amounted to a total of 140 hours of on-source time. All mapping was carried out in the standard spiral raster mode to concentrate on the central region ($<8$\arcmin) in which the lensing magnification is expected to be highest. We aimed for a 1$\sigma$ point source sensitivity of $\sim$1.5 mJy beam$^{-1}$, so as to allow $3\sigma$ detection of the $S(850\mu$m$)\geq 5$ mJy population. This is similar to the depths achieved  by previous  LABOCA and SCUBA surveys (\citealt{smai97,knudsen08,weiss09b,johansson11}).

Several types of calibration data were taken every $\sim1-2$ hours to correct for instrumental and atmospheric effects. First, we obtained single spiral scans of planets and bright QSOs to measure and correct for deviations from the telescope pointing model, and also observed bright planets (Mars, Venus, Saturn, and Jupiter) regularly to verify and adjust the telescope focus. Second, we used regular sky dip observations to determine atmosphere opacity, which attenuates the astronomical signal detected in the LABOCA passband. Finally, planets and secondary calibrators with well known fluxes (\citealt{siringo09}) were observed  in a raster-spiral mode to refine the canonical factor used to convert the raw data to physical units and improve the accuracy of the flux calibration. In general, the observing conditions encountered during our observations were good, with $\sim80\%$ of our science scans obtained with wind speeds under 10 ms$^{-1}$ and precipitable water vapor (PWV) under 1.0 mm.

%TABLE 2
\begin{center}
\begin{deluxetable*}{llclccc}
\tablecolumns{7}
%\tablewidth{0pt}
\tabletypesize{\scriptsize}
\tablecaption{ LABOCA 870 \micron\, observations.\label{tab:laboca_obs}} 
\tablehead{ \colhead{Target} &  \colhead{Period}  & \colhead{P.I.}& \colhead{Observation dates} &  \colhead{On-source time (hr)} &  \colhead{$\sigma$ (mJy beam$^{-1}$)} &  \colhead{SZE S/N}}
\startdata
\hline \\
\lascarI  & 2011B  \tablenotemark{a}&  A. Baker & 2011 Aug 20-28 &  11.3 & 2.4 & 9.7\\
\lascarII & 2011B &  L. Infante & 2011 Oct 1-4 &  17.6 & 2.4 & 8.2\\
\lascarIII  & 2010B  & A. Baker & 2010 Aug 6-10   &  17.0 & 2.0 & 4.3\\
\lascarIV & 2011A &   A. Baker & 2011 Jul 14-26  & 12.2 & 1.7 & 2.2\\
\lascarV & 2011B &   A. Wei\ss & 2011 Oct 21-26 &  11.6 & 2.1 &2.9 \\
\lascarVI  &  2010B &  A. Baker & 2010 Oct 16-21  &   8.1& 1.9 &2.1  \\
\lascarVII  & 2010B   &  A. Baker & 2010 Aug 23-24 & 28.3 & 1.6& 8.8\\
	       &	     &		  & 2010 Oct 13-21  &       &   &   \\
\lascarVIII & 2010B   & L. Infante & 2010 Aug 26-27  & 16.3& 1.6 &8.2  \\
	       &	     	  &  & 2010 Oct 26-29       &    &  &\\
\lascarIX & 2011A  & L. Infante & 2011 Jul 26-30  & 13.3 & 2.8 & 1.8\\
\lascarX & 2011B  & A. Baker & 2011 Sep 7-12 &  14.8 & 2.1 & 7.7 \\
	       &	     &		    & 2011 Nov 4-5  &         & & \\
\enddata
\tablenotetext{a}{Director's Discretionary Time (DDT) proposal.}
\end{deluxetable*}
\end{center}

\subsubsection{ LABOCA 870 \micron\, data reduction}

Our LABOCA observing campaign was designed to study simultaneously the 870 \micron\, emission of SMGs and SZE clusters, and our data reduction process was therefore guided by the need to detect significant signals at different spatial scales. With this aim, we developed an iterative multi-scale algorithm that maximizes our sensitivity to low-level extended emission by applying a series of matched filters to search for signal at spatial extent of various scalers.. This pipeline is based on several standard data reduction routines implemented in the Bolometer Array Analysis Software (BoA\footnote{\url{http://www.apex-telescope.org/bolometer/laboca/boa}}), and is described in full detail in \citet{lindner14}.

\begin{figure*}
\caption{LABOCA 870 \micron\, signal-to-noise maps for the sample of ten clusters targeted by LASCAR. Circles indicate the locations of all detected point sources, as listed in Table \ref{tab:laboca_dets}.}

	\subfigure{\includegraphics[width=6cm]{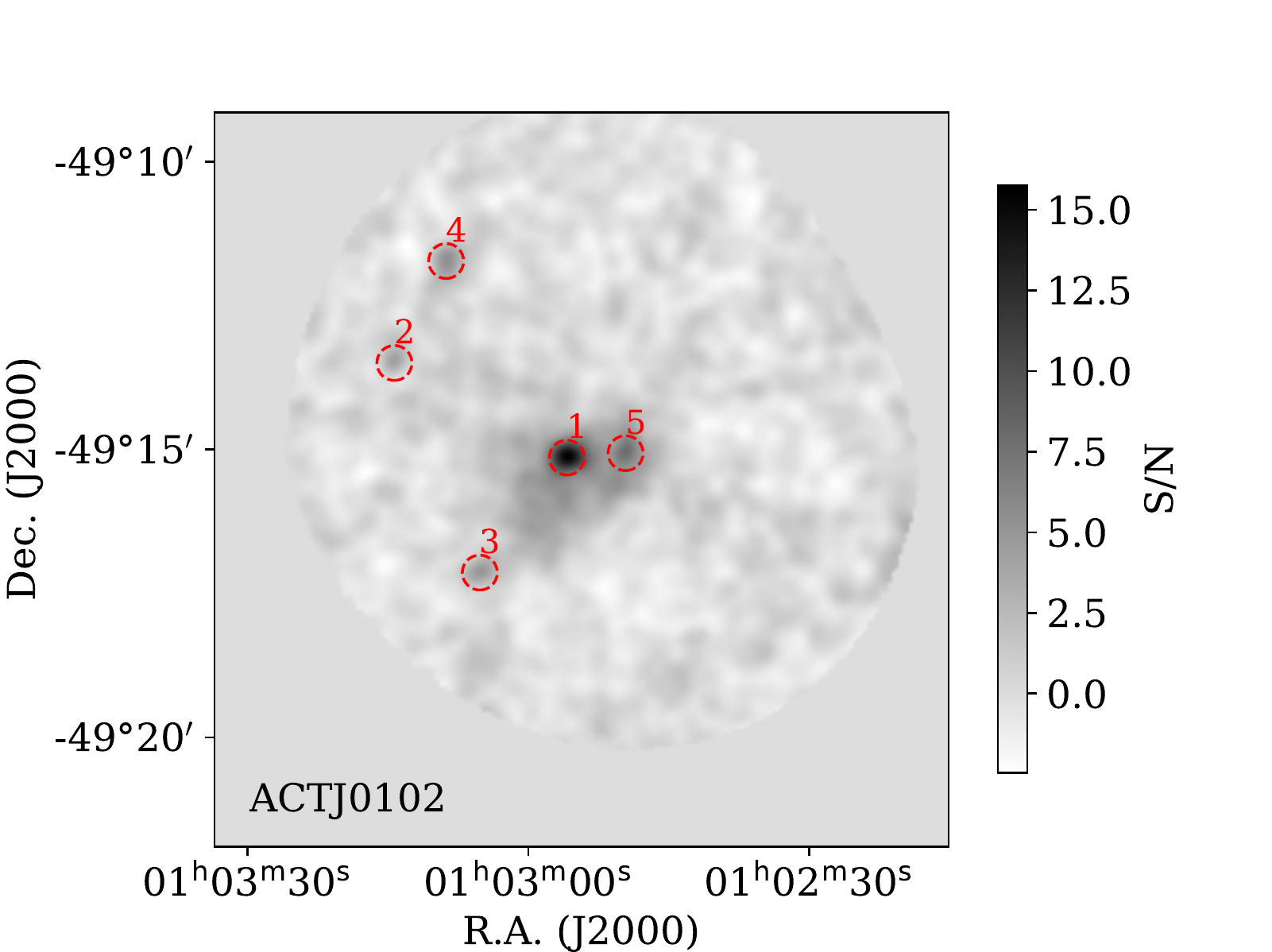}}
	\subfigure{\includegraphics[width=6cm]{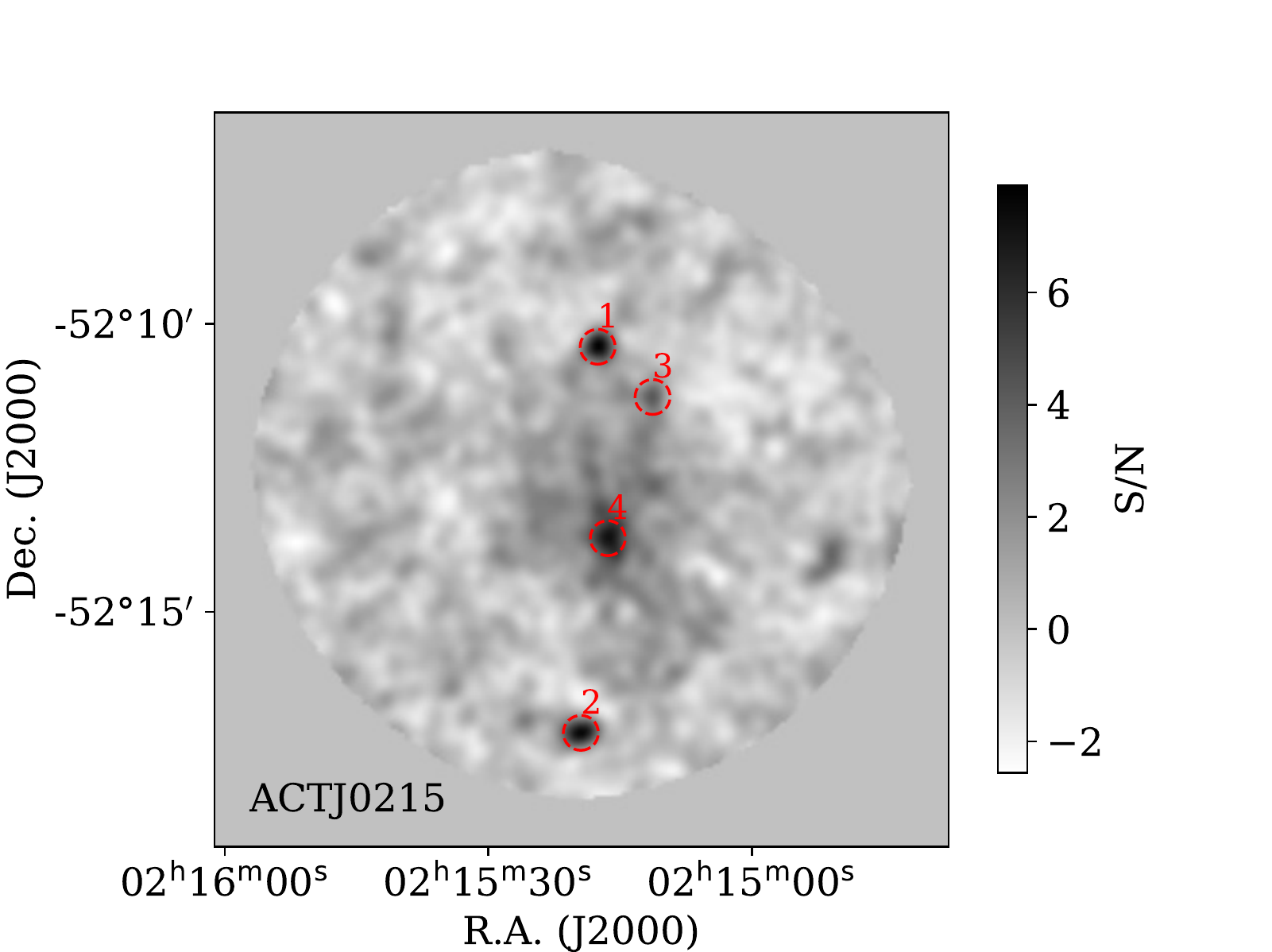}}
	\subfigure{\includegraphics[width=6cm]{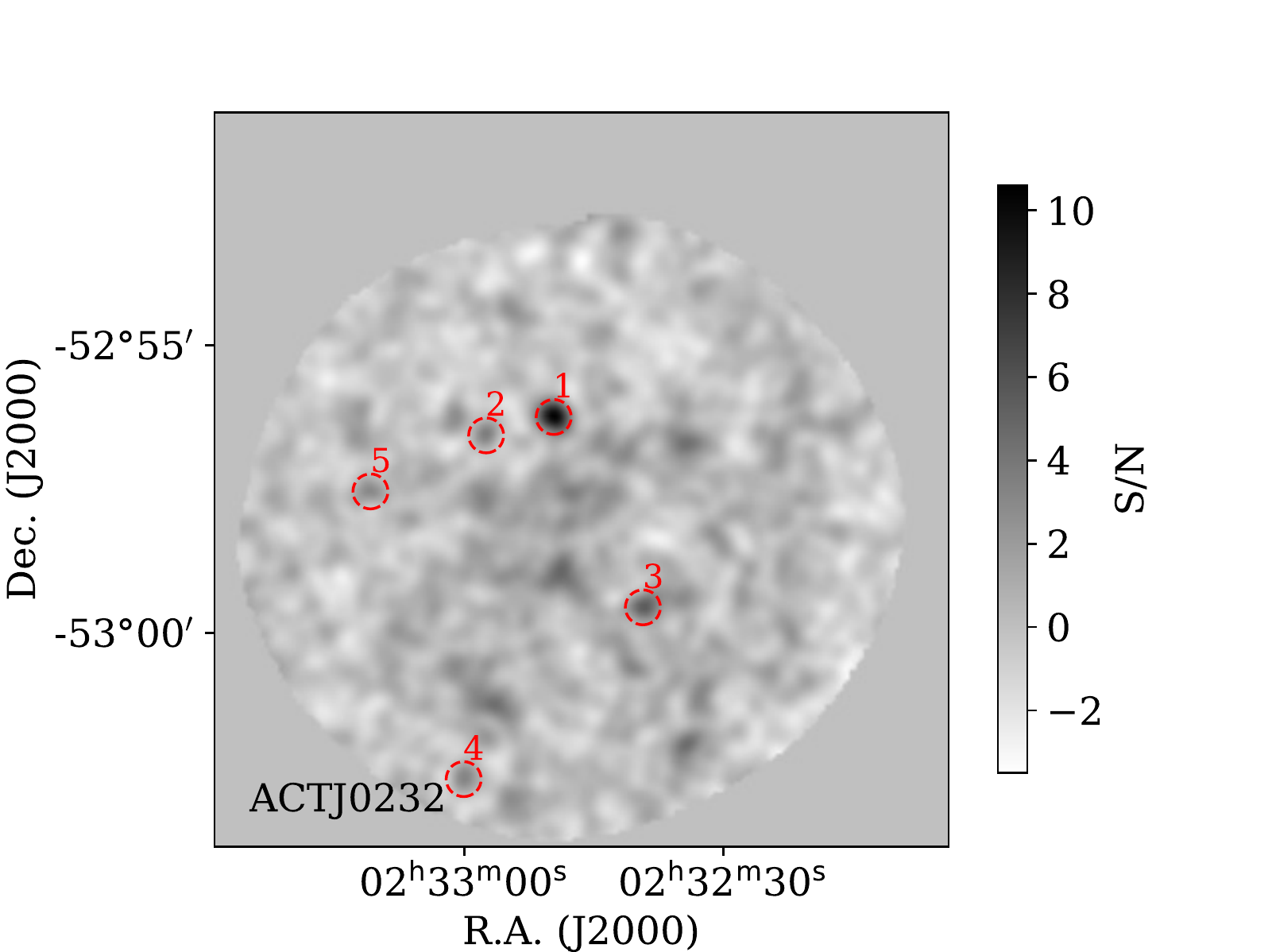}}
	\subfigure{\includegraphics[width=6cm]{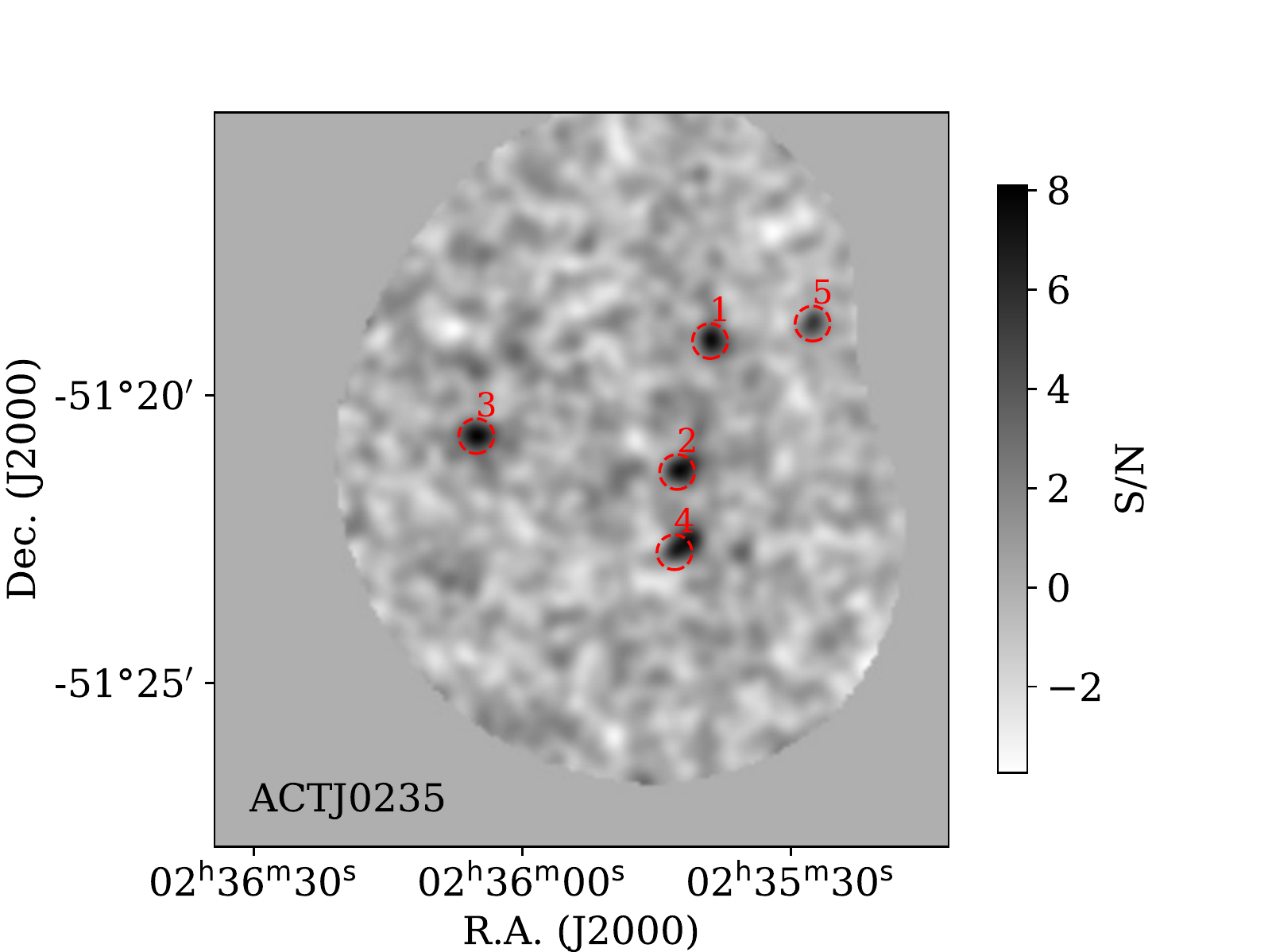}}
	\subfigure{\includegraphics[width=6cm]{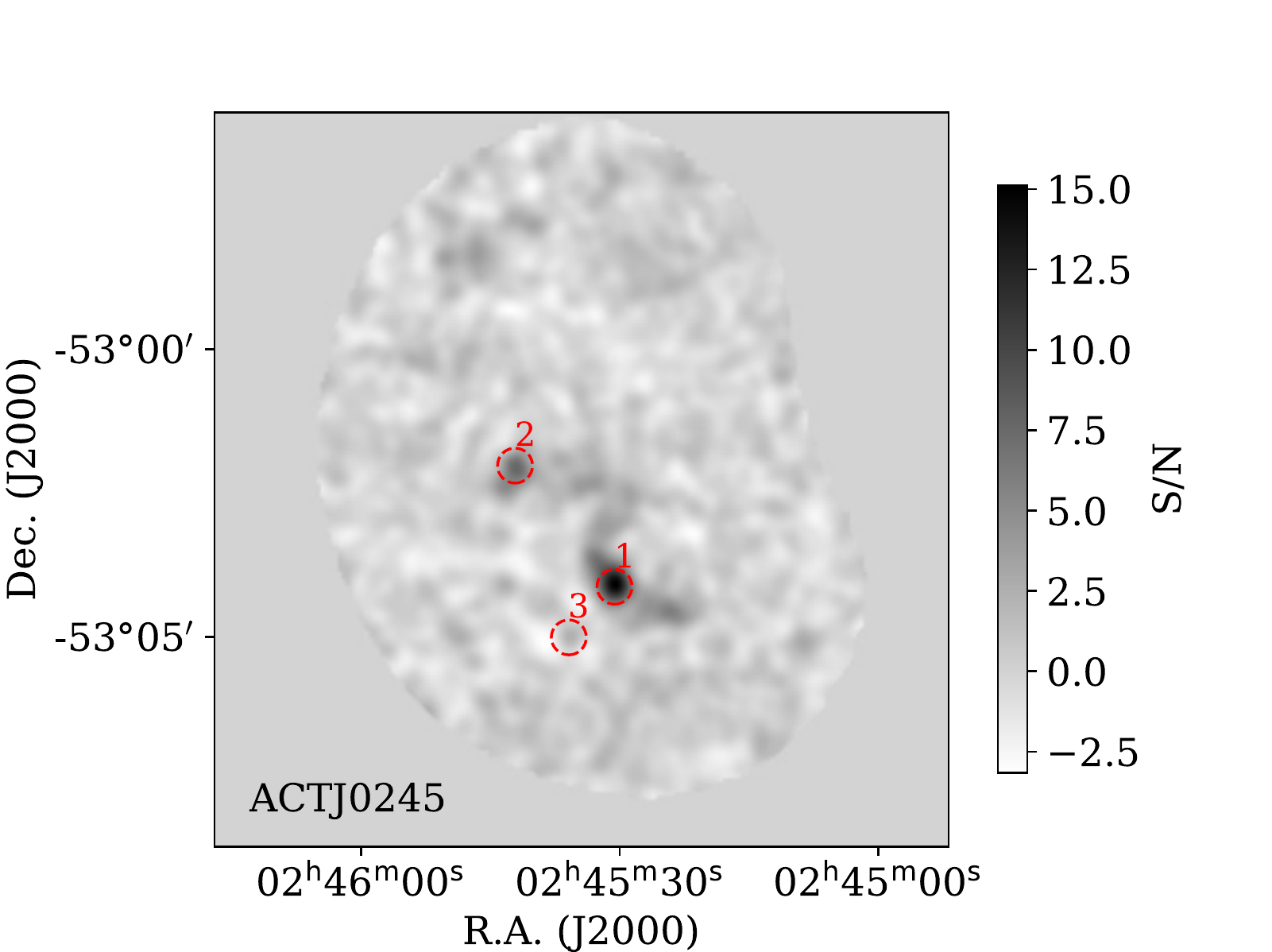}}
	\subfigure{\includegraphics[width=6cm]{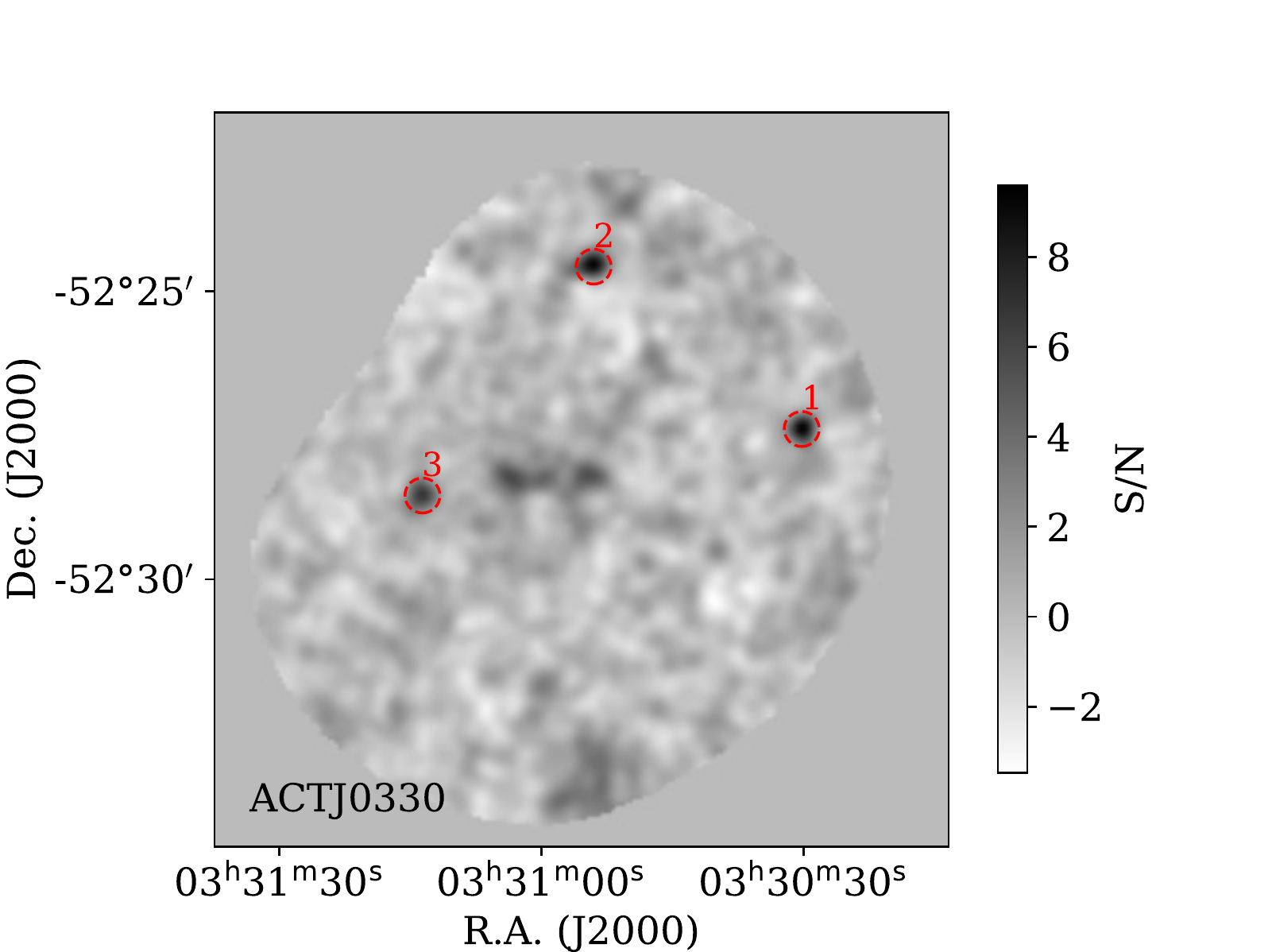}}
	\subfigure{\includegraphics[width=6cm]{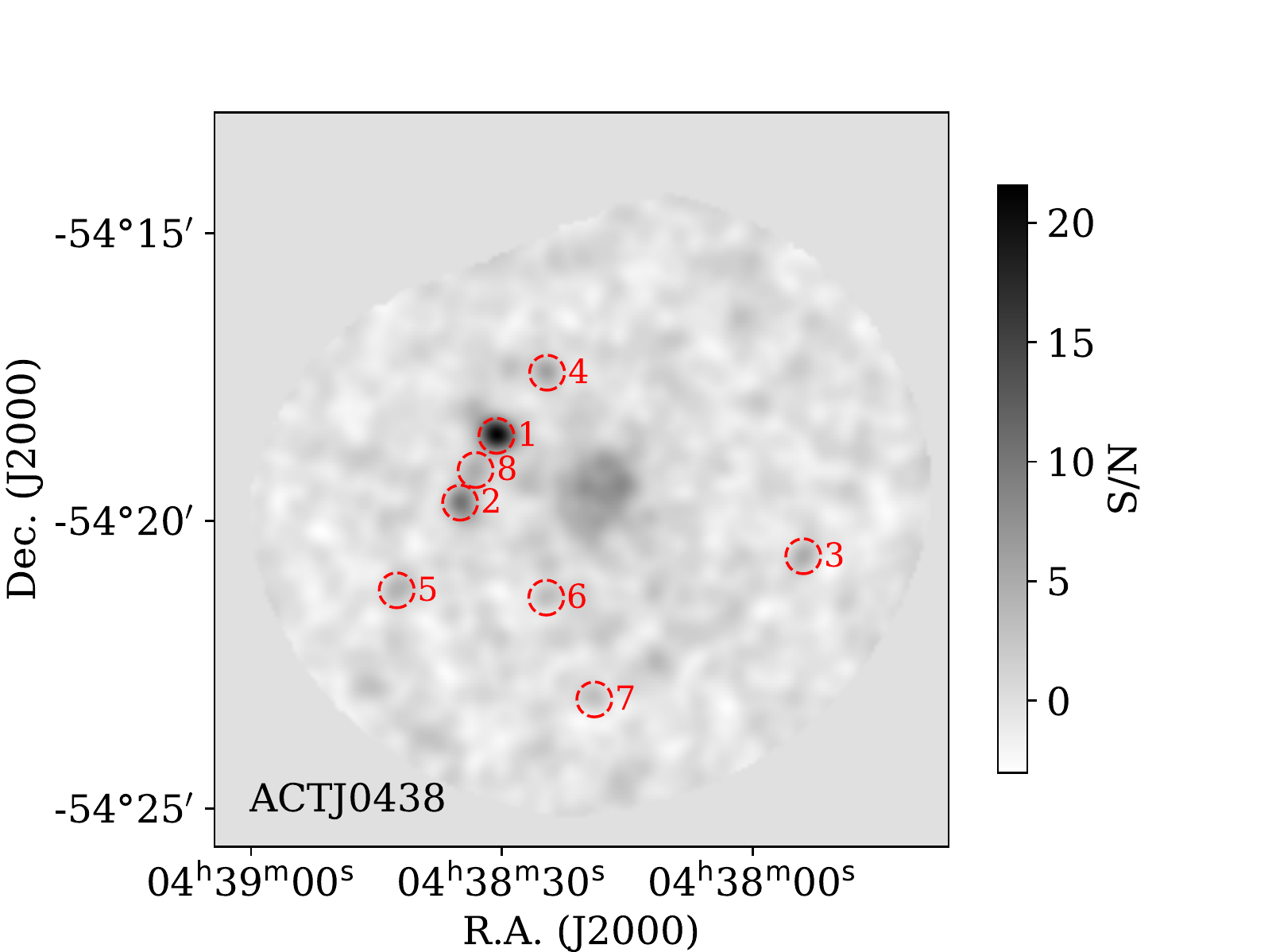}}
	\subfigure{\includegraphics[width=6cm]{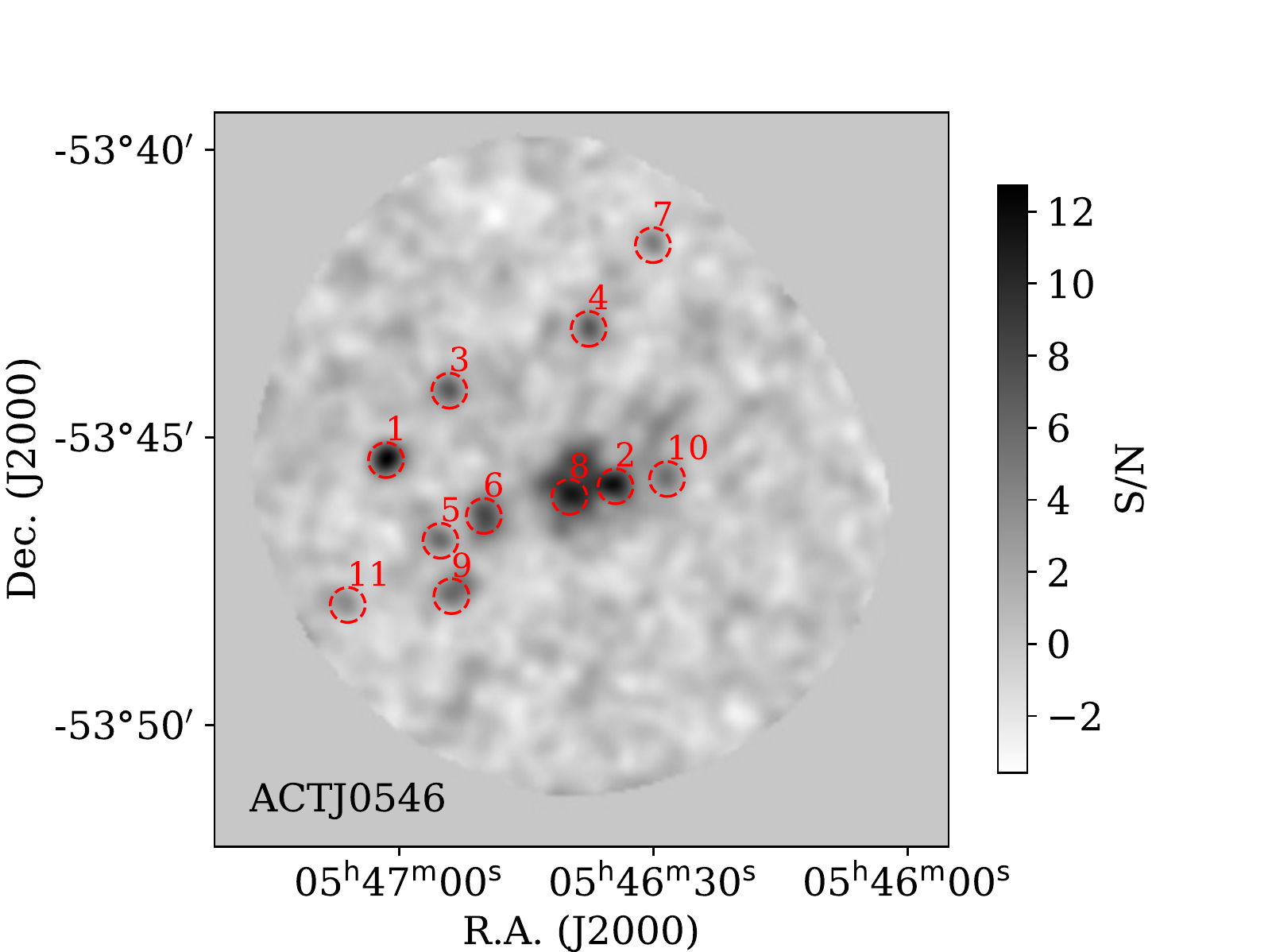}}
	\subfigure{\includegraphics[width=6cm]{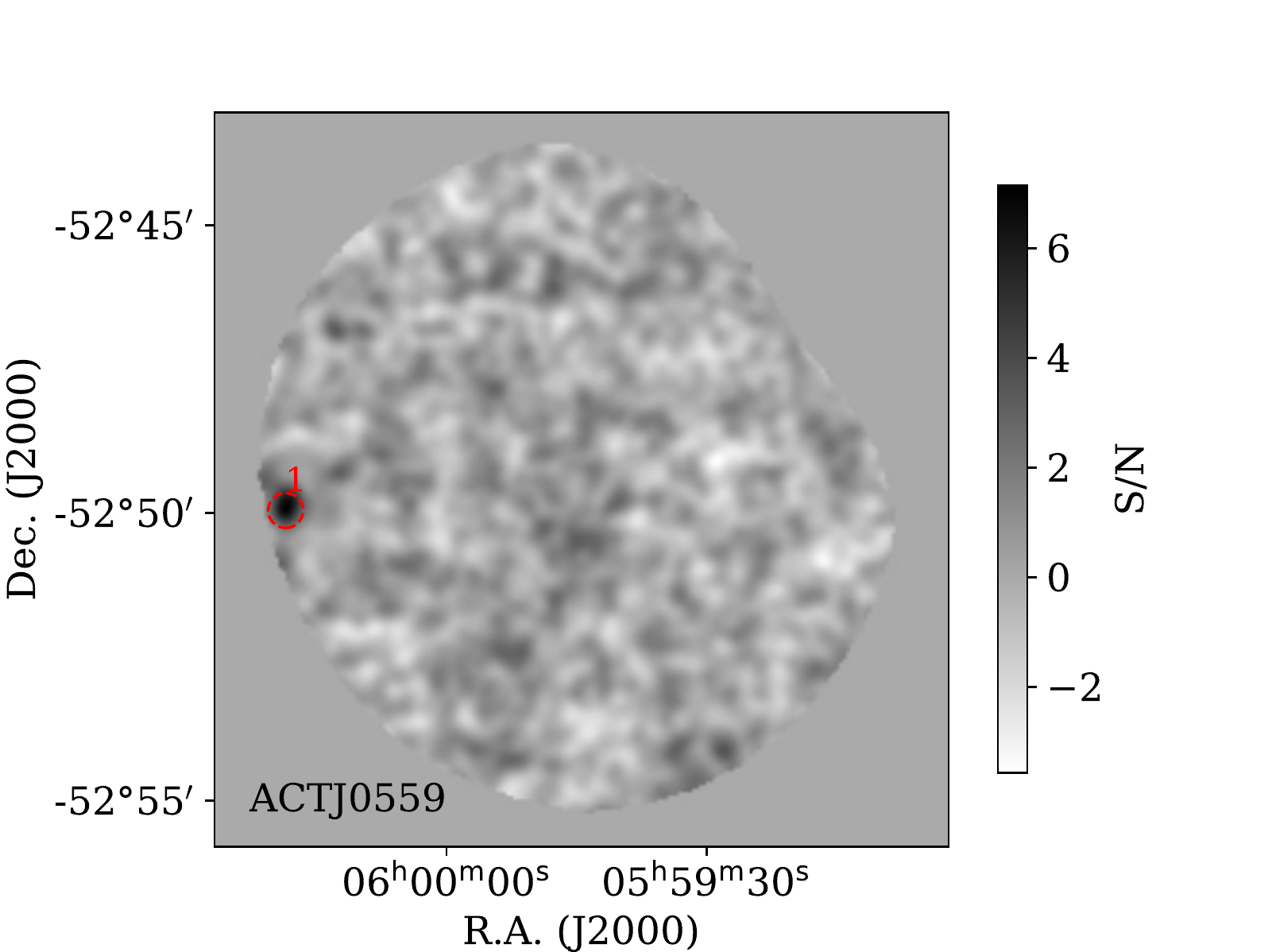}}
	\subfigure{\includegraphics[width=6cm]{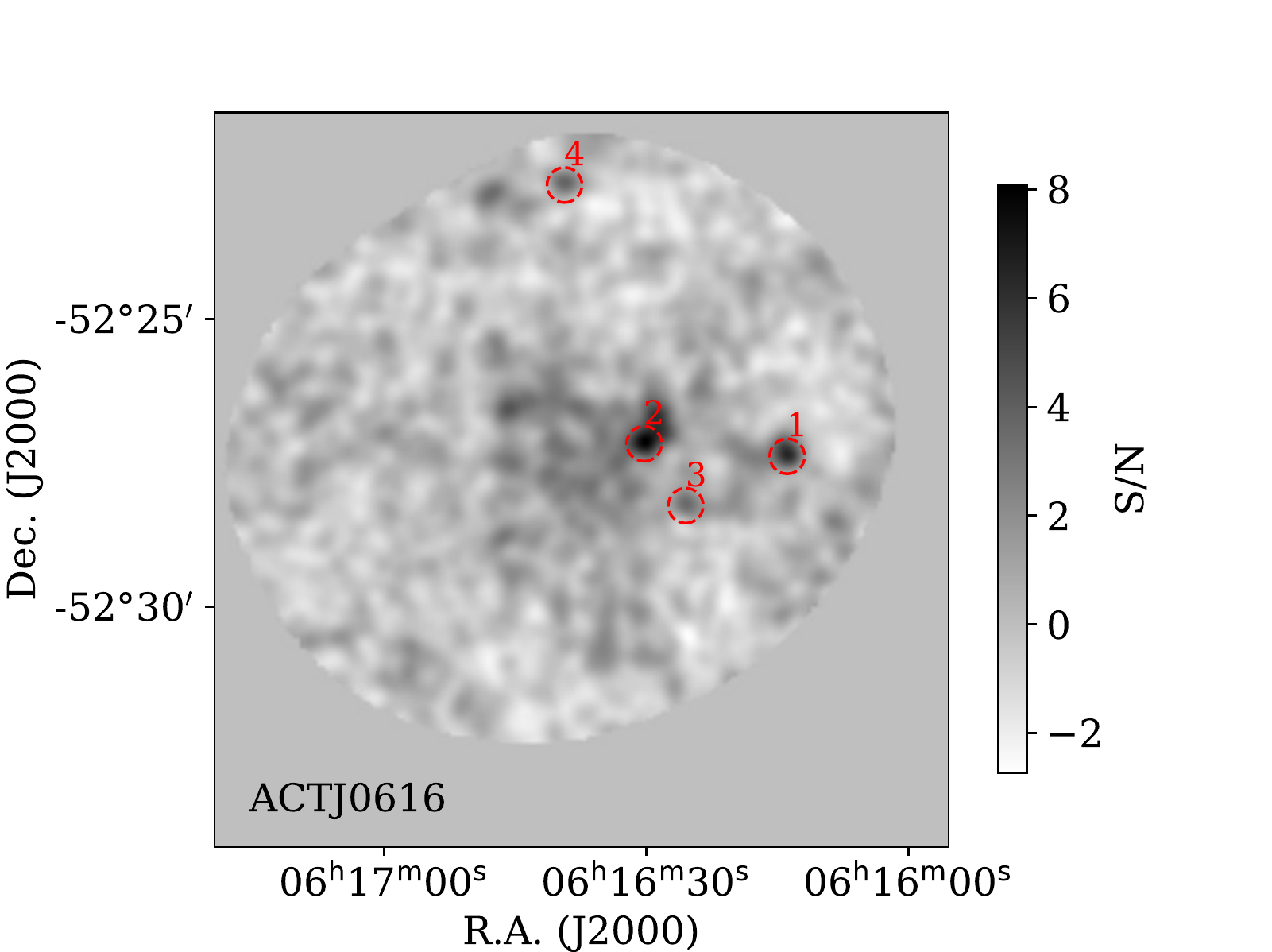}}
\label{fig:labocamaps}
\end{figure*}

Final flux density maps have a pixel scale of 3.6\arcsec\, pix$^{-1}$, and the beam has a FWHM of 19.2\arcsec. The iterative reduction applied to our LABOCA data resulted in an average root-mean-square (RMS) noise ($\sigma$) of $\sim$2 mJy beam$^{-1}$, as measured from the final beam-smoothed maps; individual $\sigma$ values for each field are given in Table \ref{tab:laboca_obs} and final signal to noise maps are presented in Figure \ref{fig:labocamaps}. The effectiveness and reliability of our pipeline was tested on ESO archival LABOCA data for the "Bullet" cluster (1E0657-56; \citealt{marke02}), which was previously reduced by \citet{johansson10} using the CRUSH software.\footnote{\url{http://www.submm.caltech.edu/sharc/crush/}} In their processing, they intentionally filter out the extended emission and therefore do not recover the SZE increment signal, but detect 17 point sources with S/N$>$4. Comparison with results of our detection and photometry algorithm for this cluster is presented in Section \ref{sec:extraction}.

For the LASCAR sample, we detect point sources in all fields, and strong SZE increment signals (S/N$>$3.5) in six clusters (see Table \ref{tab:laboca_obs}). \citet{lindner14}  present a thorough multi-wavelength analysis of the SZE increments, and in the following sections we focus on the extraction, photometry, and analysis of unresolved sources.

%FIGURE 1

\subsection{ ATCA 2.1 GHz observations}

We have conducted deep, high-resolution continuum mapping at 2.1 GHz of all ten clusters with ATCA. These data are crucial to facilitate identification of counterparts for the submillimeter emission detected with LABOCA. The observations were carried out in January 2011, December 2011, and April 2012 using the 16 cm-band receiver with the CFB 1M-0.5k correlator mode, which gives a bandwidth of 2 GHz with 2048 $\times$ 1-MHz channels.\footnote{\url{http://www.narrabri.atnf.csiro.au/observing/CABB.html}}All clusters have been observed with the 6A antenna configuration; ACTJ0102-4915 was also observed in the 1.5B configuration.  The integration time per cluster is indicated in Table \ref{tab:radioMaps}. For flux and bandpass calibration we used PKS 1934-638 \citep{reynolds94}; for phase calibration, we selected an appropriate bright, nearby and compact source for each cluster.

For ATCA data reduction we used the MIRIAD software package \citep{sault95}; details are given in \citet{lindner14}. Final maps have RMS sensitivities from 6.9 to 12.0 $\mu$Jy beam$^{-1}$, and synthesized beams have major axes $\sim4-6$\arcsec (see Table \ref{tab:radioMaps}). We used the Common Astronomy Software Applications (CASA; \citealt{mcmullin07}) to extract point sources above a $4\sigma$ threshold and measured their flux densities by fitting 2-D Gaussian profiles with the beam shapes and position angles given in Table \ref{tab:radioMaps}. The combined number counts over all fields follow a power law $dN/dS\propto S^{-\delta}$, with index $\delta = 1.7$ \citep{lindner14}, where $S$ is the flux density.

%TABLE 3
\begin{center}
\begin{deluxetable*}{lcccc}
\tablecolumns{5}
%\tablewidth{0pt}
\tabletypesize{\scriptsize}
\tablecaption{Properties of 2.1 GHz ATCA Radio Maps for LASCAR Clusters. \label{tab:radioMaps}}
\tablehead{	
\colhead{Target} & \colhead{Integration time } & \colhead{$\sigma$ } & \colhead{Beam}  & \colhead{P.A.  } \\
\colhead{ } & \colhead{  (hr)} & \colhead{ ($\mu$Jy beam$^{-1}$)} & \colhead{}  & \colhead{($^{\circ}$) } 
}
\startdata
\hline 
ACT$-$CL\,J0102$-$4915   & 12.1 &\, 7.5 & 6.1\arcsec$\times$3.1\arcsec & \, -1.9 \\
ACT$-$CL\,J0215$-$5212 & \, 8.6 & 11.0& 4.7\arcsec$\times$2.9\arcsec & -21.0\\
ACT$-$CL\,J0232$-$5257 & 19.8 &\, 8.1 & 4.8\arcsec$\times$3.1\arcsec & \,\,\, 6.0\\
ACT$-$CL\,J0235$-$5121& \, 8.5 & 10.9 & 5.3\arcsec$\times$2.7\arcsec & -10.0\\
ACT$-$CL\,J0245$-$5302  & 10.3 & 10.5 & 4.4\arcsec$\times$3.0\arcsec & \,\,\, 4.3\\
ACT$-$CL\,J0330$-$5227 & \, 8.8& 11.7 & 5.1\arcsec$\times$2.7\arcsec & \, 17.7\\
ACT$-$CL\,J0438$-$5419 & \, 8.1 & 11.9 & 5.3\arcsec$\times$2.8\arcsec & -19.6\\
ACT$-$CL\,J0546$-$5345 & 21.0 &\, 6.9 & 4.7\arcsec$\times$3.2\arcsec & \, -3.1\\
ACT$-$CL\,J0559$-$5249& \, 8.9 &\, 9.6 & 5.4\arcsec$\times$2.9\arcsec & -13.3\\
ACT$-$CL\,J0616$-$5227  & \, 7.8 & 12.0 & 5.0\arcsec$\times$3.0\arcsec &\,  26.6\\
\enddata
\end{deluxetable*}
\end{center}

\subsection{$Herschel$ observations}

In our efforts to characterize the far-IR through submillimeter SEDs of  point sources detected by LABOCA and constrain at the same time the clusters' SZE spectra, we obtained new \emph{Herschel Space Observatory} observations for the subset of six clusters from our full LASCAR sample listed in Table \ref{tab:pacs}. These targets were selected based on their high masses and strong SZE decrement signals, and all six of them were mapped at 250, 350, and 500 \micron\, with SPIRE; five of the six were also observed at 100 and 160 \micron\, using PACS.

\subsubsection{ $Herschel$/PACS observations and data reduction}

New PACS observations of five LASCAR clusters followed the observational strategy of the $Herschel$ Lens Survey (HLS, \citealt{egami10}), using the  scan-map mode with medium speed and 13 scan legs of 4\arcmin\ each with a 20\arcsec\ cross-scan step that result in 8\arcmin$\times$8\arcmin\,maps. To control optimally for systematics, each cluster was observed twice by orthogonal scan maps with orientation angles of 45\degree\ and 315\degree, with 18 repetitions each. The total on-source time for each cluster was 1.6 hours. 

All observations were reduced interactively using version 11.1.0 of the Herschel Interactive Processing Environment (HIPE, \citealt{ott10}) using PACS calibration version 41 and scripts supplied with HIPE.  The standard processing of raw data includes pixel flagging, flux density conversion, coordinate registration, high-pass filtering to remove background structure, and spatial deglitching. Individual maps were mosaiced to remove $1/f$ noise from scanning. Final maps have an image scale of 1\arcsec\ pix$^{-1}$ in both bands, RMS noise of $\sim$3.8 (3.3) mJy beam$^{-1}$, and angular resolution of 6\arcsec (13\arcsec) at 100 (160) \micron. For detection and photometry of point sources, we applied the same extraction algorithm used for construction of the 870 \micron\, catalog, which is described in detail in Section 5.

%TABLE 4
\begin{center}
\begin{deluxetable*}{lccccccc}
\tablecolumns{8}
\tablewidth{0pt}
\tabletypesize{\scriptsize}
\tablecaption{$Herschel$ PACS and SPIRE observations of LASCAR clusters.  \label{tab:pacs}}
\tablehead{	
\colhead{Target} & \colhead{Observation ID} &\colhead{Observation date} & \colhead{$\sigma_{100}$ } & \colhead{$\sigma_{160}$}  & \colhead{$\sigma_{250}$} & \colhead{$\sigma_{350}$ } & \colhead{$\sigma_{500}$ } 
}
\startdata
\hline 
ACT$-$CL\,J0102$-$4915 & 1342256977 & 2012-12-11 	&  3.79	& 3.03  	&7.4 &7.2 & 7.4\\
ACT$-$CL\,J0235$-$5121  & 1342262209 & 2013-01-27 	&  3.83 	& 3.26 	&7.4 &7.2 & 7.4\\
ACT$-$CL\,J0245$-$5302   & 1342262466  & 2013-01-28 & 3.83 	& 3.39 	&7.4 &7.2 & 7.4\\
ACT$-$CL\,J0330$-$5227   &1342259282 & 2013-01-17 	&  \nodata & \nodata &9.4 &8.2 & 8.4 \\
ACT$-$CL\,J0438$-$5419   &1342259282 & 2013-01-17 	&  3.71 	& 3.42 	&7.4 &7.2 & 7.4 \\
ACT$-$CL\,J0546$-$5345   & 1342261752 & 2013-01-21 	&  3.75 	&  3.25 	&7.4 &7.2 & 7.4 \\
\enddata
\tablecomments{For each target we give dataset specifications, and the RMS noise ($\sigma_{\lambda}$) of final maps for observation wavelengths $\lambda=$100, 160, 250, 350, and 500 \micron, in mJy beam$^{-1}$.}
\end{deluxetable*}
\end{center}

\subsubsection{ $Herschel$/SPIRE observations and data reduction}

SPIRE observations of the six clusters in Table \ref{tab:pacs} used the Large Map mode, and in all cases except for \lascarVI\, consisted of four repetitions of 8$ ^{\prime}\times$8\arcmin\ maps at four dithered positions, which account for a total of 2224 s on-source time. For \lascarVI, we performed only four repetitions of a 6\arcmin$\times$6\arcmin\ map at a single position, with a total on-source time of 492 s. Data reduction and map-making follow standard procedures implemented within HIPE, and resulted in angular resolutions of 17.6, 23.9, and 35.0$ ^{\prime\prime}$,  pixel scales of 6, 10 and 14\arcsec\, pix$^{-1}$, and RMS map sensitivities of 7.4, 7.2, and 7.2 mJy beam$^{-1}$ at 250, 350, and 500 \micron, respectively (excluding \lascarVI; see Table \ref{tab:pacs}). To remove the SZE signal and extract point sources, we used the 250 \micron\, map to derive a model for the confused SMG background at 350 and 500 \micron, as detailed in Section 3.7 of \citet{lindner14}. The final confusion noise-free maps have mean RMS sensitivities $\sim3$ mJy beam$^{-1}$, and we apply a $4.5\sigma$ threshold for detection of point sources.

\section{Complementary data}\label{sec:complementary}

As part of the ACT collaboration's efforts to achieve complete multi-wavelength follow-up of SZE-detected clusters, targets in the LASCAR sample had been previously observed in the near-IR with the \emph{Spitzer Space Telescope} and in the optical with ground based telescopes. In the following subsections we summarize relevant information regarding these observations; details of their reduction and analysis have been published in previous work.

\subsection{$Spitzer$/IRAC  data}

All clusters in the LASCAR sample were imaged with the InfraRed Array Camera (IRAC; \citealt{fazio04}) on \emph{Spitzer}, at wavelengths of 3.6 and 4.5 \micron, through a proposal that targeted 14 confirmed ACT clusters with $0.27<z<1.07$ (PI: Menanteau, PID: 70149). Observations took place in August 2010 - July 2011;  they were designed to provide coverage out to the clusters' virial radii, using  $2\times2$ grids of IRAC pointings centered on the cluster positions. Details on the observations, data reduction, and photometry algorithms have been published in \cite{menanteau12} and \cite{hilton13}. The resulting catalog is 80$\%$ complete for point sources at $m_{\rm{AB}}\sim 22.6$ mag in both channels.

\subsection{Optical data}\label{sec:optical}

The original catalog of ACT SZE-detected cluster candidates  was followed up in the optical to confirm their nature and determine the purity of the ACT sample. The optical campaign started in semester 2009B and consisted of 7 nights of observations at 4m class telescopes in Chile (3.6 m NTT at La Silla and 4.1 m SOAR Telescope at Cerro Pach\'on). All targets were observed using either the Gunn or SDSS $griz$ filter sets, and the final images have pixel scales of 0.24\arcsec\ pix$^{-1}$ (0.15\arcsec\ pix$^{-1}$) for NTT (SOAR). The full details of observations, data reduction, and analysis are given in \citet{menan10b}. The final catalogs are estimated to be $80\%$ complete down to magnitudes $i=23.5$ (NTT) and $i=24.3$ (SOAR), and the multi-band photometry was used to estimate photometric redshifts, although the associated errors are relatively large due to the limited number of filters. In this work, we use $griz$ photometric catalogs to investigate the existence of optical counterparts for our SMGs at least down to the quoted limiting magnitudes. 

In addition to $griz$ imaging, \citet{sifon2013} carried out deep multi-object spectroscopic observations for a sample of 16 massive ACT clusters and obtained intermediate-resolution ($R\sim700-800$) spectra and redshifts for $\sim60$ member galaxies per cluster. These data were used to measure dynamical masses ($M_{200}$) and radii ($r_{200}$) for their clusters, which include all LASCAR targets except \lascarV, a previously known cluster (Abell S0295). As will be described in Section \ref{sec:lensing}, these results are of great utility for modeling the lensing properties of our clusters and estimating the magnifications of the detected background sources. The full tables containing the magnitudes and spectroscopic redshifts of the galaxies targeted by \citet{sifon2013} are available online, and are also used in this work to check if any of our submillimeter sources coincide with known cluster members. 

\section{Source extraction and photometry}\label{sec:extraction}

The signal in each of our LABOCA and $Herschel$ maps is a combination of point source emission from DSFGs located in front of, in, and behind each galaxy cluster, possibly magnified by its gravitational potential in the latter case, and the cluster's SZE increment signal. In \citet{lindner14}, we report the extraction and measurement of the integrated SZE flux densities from LABOCA and SPIRE maps, and here we focus on the detection and photometry of submillimeter point sources and measurement of their multi-wavelength properties.

To optimize extraction and photometry of point sources in LABOCA and $Herschel$ data, we use a hybrid algorithm that combines median filtering with the matched-filter technique of \citet{serjeant03} that is typically applied to maps with non-uniform noise. First, we preprocess the final pipeline-produced flux density image by subtracting the median filtered version with a kernel size equal to three times the beam's FWHM, so as to optimize detection of point sources above even bright diffuse background signals, like those we expect in the clusters' central region.  Next, we generate a minimal-$\chi^2$ signal-to-noise map according to the formula 

\begin{equation}
\text{S/N}=\frac{(S W)\otimes P}{\sqrt{W \otimes P^2}}
\end{equation}

\noindent where $S$ is the pre-processed image signal, $W$ is the pixel weight map generated in the data reduction pipeline, $P$ is the image Gaussian point-spread function and $\otimes$ denotes a convolution (\citealt{serjeant03}). We use this optimal map to locate the positions of unresolved sources with S/N$>4$, and then measure their flux densities in the original image $without$ median filtering.\footnote{Median filtering is beneficial for suppression of diffuse background noise, but may remove flux from sources in a complex morphology-dependent way.} At each point-source position, we fit a 2-D Gaussian with a fixed center and a FWHM equal to that of the map's beam, and allow for a varying, non-negative peak amplitude plus a constant offset to account for the background signal.

With the described algorithm we have detected a total of 49 new submillimeter point sources over ten cluster fields, with 870 $\mu$m flux densities ranging from 6.6 to 33.9 mJy. All LABOCA source positions, flux densities, and S/N ratios are listed in Table \ref{tab:laboca_dets}; sources are circled in the LABOCA S/N maps presented in Figure \ref{fig:labocamaps}. To test our data reduction and source extraction pipeline, we have applied the same methodology to archival data for the extensively studied ``Bullet" cluster, which has been observed with LABOCA (\citealt{johansson10}), PACS, and SPIRE (\citealt{egami10}), among many other instruments. We have generated an optimal-S/N 870 \micron\,  point source catalog in the exact manner as for the LASCAR clusters, and compare our results with those of \citet{johansson10}. We detect in total seven point sources, five of which are also contained in their catalog and have flux density measurements that are in good agreement, with a mean relative difference of $16\%$. The two remaining sources that are not selected by \citet{johansson10} are relatively fainter ($S_{870}\sim9.5$ mJy) and located within the 148 GHz decrements measured by ACT \citep{marriage11b}, so the discrepancy may be explained by the  aggressive filtering applied by those authors to remove the extended SZE signal. Inspection of archival PACS data (\citealt{egami10}) reveals that one of them also has a 100/160 $\mu$m counterpart. 

On the other hand, \citet{johansson10} report in total 13 point sources within the central 10\arcmin\, of which we do not recover one $8.2$ mJy source that is identified with a foreground galaxy, and seven fainter sources with $S_{870}\leq6.2$ mJy. From their multiwavelength analysis, four of these sources lack IRAC or 24$\mu$m counterparts, and we verify that they are also undetected in PACS imaging. Hence, there is no further confirmation of their authenticity. The three remaining sources do have 24$\mu$m or PACS counterpart candidates and would be extracted with our hybrid algorithm using a lower threshold of $3\sigma$, which increases however the number of spurious detections. Overall, we find that our data reduction and source extraction methods are in good agreement with the results of \citet{johansson10} in terms of detection and photometry for sources with flux densities above $\sim6.5$ mJy, but in both cases there is a tradeoff between detection of faint SMGs and possible contamination of spurious sources below this limit. Hence, we favor the more conservative results obtained with our pipeline with a $4\sigma$ threshold.  Additionally, the careful treatment of the extended 870 \micron\, emission and removal of the cluster's SZE increment implemented in our pipeline improves extraction of point sources that overlap with the SZE signal.  

\begin{figure}
\includegraphics[width=8.5cm]{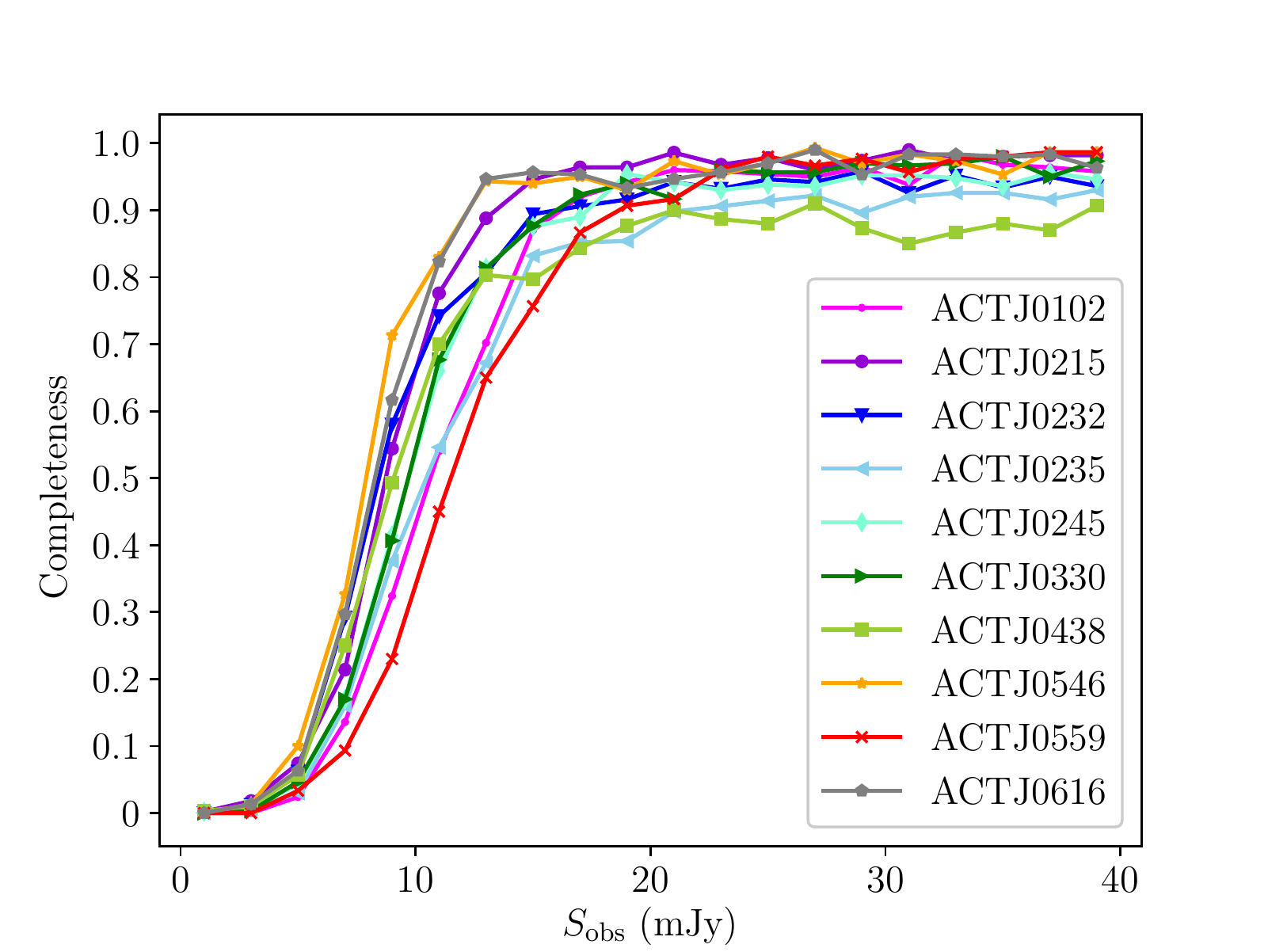}
\caption{Completeness of LABOCA point source detections for each cluster, estimated from simulations, and represented in terms of the fraction of detected over total sources versus observed flux density $S_{\rm obs}$.}
\label{fig:completeness}
\end{figure}

To estimate the completeness of our detections, we follow a procedure similar to that used by \citet{knudsen08}, \citet{weiss09b}, and \citet{johansson11}: we add point sources of varying flux densities at random positions to the flux density maps shown in Figure \ref{fig:labocamaps}, run the detection pipeline with the same settings as for the actual cluster maps, and compare the resulting detections with the input source catalog to determine the fraction of recovered sources. The artificial point sources are modeled by a Gaussian profile with a FWHM equal to the LABOCA beam (19.2\arcsec), peak amplitudes from 0.5 to 40 mJy in steps of 0.5 mJy, and random ($x,y$) positions following a uniform distribution. For each field, at each flux density step, we simulate 100 sets of 10 sources each, apply our hybrid extractor to produce a S/N map and detect point sources at the 4$\sigma$ level, and calculate the mean detection rate. In comparing the simulated and detected catalogs, we consider as matching detections those within one beam width of the input positions. The results of our completeness analysis for each cluster are shown in Figure \ref{fig:completeness}; on average we reach a 90$\%$ detection rate at a flux density $\sim18$ mJy, and as expected, the completeness at a given flux limit is higher in those maps with lower average noise levels.

%FIGURE 2

%TABLE 5
\startlongtable
\begin{deluxetable*}{lcccccccr}
\tablecolumns{9}
%\tablewidth{0pt}
\tabletypesize{\scriptsize}
\tablecaption{ LABOCA 870\micron\, point sources detected in the fields of the clusters in our sample. \label{tab:laboca_dets}} 
\tablehead{ 
 \colhead{Source ID} &  \colhead{R.A.} & \colhead{Dec.} &  \colhead{$S(870\,\mu$m$)$  }   &  \colhead{$\Delta S(870\,\mu$m$)$  }  &  \colhead{S/N} & \colhead{$\theta$} &\colhead{$z_{\rm{phot}}$} & \colhead{$\mu$} \\
\colhead{}                    &  \colhead{(J2000)}       &  \colhead{(J2000)}        & \colhead{(mJy) }         & \colhead{(mJy) }             &  \colhead{}  & \colhead{(arcmin) } & \colhead{ } &\colhead{} } 
\startdata
ACTJ0102$-$1 & 01:02:55.9 & -49:15:09.1 & 33.9 &1.44 & 13.8 & 0.6 & 4.5$\pm$0.6           &5.6\tablenotemark{c} \\
ACTJ0102$-$2 & 01:03:14.2 & -49:13:30.3 & 20.2 &1.60 & 6.6 & 3.8   & \, 4.5$\pm$0.8\tablenotemark{a}&1.0\tablenotemark{c}\\
ACTJ0102$-$3 & 01:03:05.1 & -49:17:08.9 & 13.0 &1.28 & 6.0 & 3.0   & 4.2$\pm$0.5 		   &1.9\tablenotemark{c}\\
ACTJ0102$-$4 & 01:03:08.7 & -49:11:44.1 & 14.6& 1.56 & 5.9 & 4.2 	& 2.0$\pm$0.6		   &1.0\tablenotemark{c}\\
ACTJ0102$-$5 & 01:02:49.6 & -49:15:04.4 & 18.9& 1.66 & 5.1 & 0.5 	& 1.1$\pm$0.7		   &1.6\tablenotemark{c}\\ \hline
ACTJ0215$-$1 & 02:15:17.6 & -52:10:24.5 & 18.5 &1.19 & 9.0 & 2.2 	& \nodata &1.1$\pm$0.1 \\
ACTJ0215$-$2 & 02:15:19.6 & -52:17:07.0 & 15.4 &1.09 & 9.2 & 4.8 	& \nodata &1.0$\pm$0.1  \\
ACTJ0215$-$3 & 02:15:11.3 & -52:11:16.8 & 10.2 &1.11 & 4.4 & 1.2	& \nodata &\nodata \\
ACTJ0215$-$4 & 02:15:16.4 & -52:13:44.1 & \, 7.5 &1.04 & 4.1 & 1.5  & \nodata &1.3$\pm$0.1 \\ \hline
ACTJ0232$-$1 & 02:32:49.6 & -52:56:15.7 & 16.9 &1.0 & 10.3 & 1.7   & 4.5$\pm$0.2 &1.1$\pm$0.1 \\
ACTJ0232$-$2 & 02:32:57.4 & -52:56:34.9 & 10.4&1.0 & 5.0 & 2.1      & 2.8$\pm$0.6  &1.1$\pm$0.1 \\
ACTJ0232$-$3 & 02:32:39.3 & -52:59:34.2 &\, 8.2&0.9 & 5.1 & 2.0	&  \nodata 	&1.1$\pm$0.1 \\
ACTJ0232$-$4 & 02:33:00.1 & -53:02:33.2 &\, 8.9&1.1 & 4.7 & 5.2 	& 1.5$\pm$0.7  &1.0$\pm$0.0\\
ACTJ0232$-$5 & 02:33:10.8 & -52:57:33.1 &\, 7.1&1.2 & 4.1 & 3.7	& 1.7$\pm$0.4 &1.0$\pm$0.1\\ \hline
ACTJ0235$-$1 & 02:35:39.1 & -51:19:04.6 & 15.3&1.4 & 6.0 & 2.2	& \nodata 			  &1.2$\pm$0.1 \\ 
ACTJ0235$-$2 & 02:35:42.7 & -51:21:21.0 & 16.2&1.5 & 6.2 & 0.5	& \, 2.9$\pm$0.7\tablenotemark{a}&\nodata\\
ACTJ0235$-$3 & 02:36:05.1 & -51:20:43.6 & 11.3&1.2 & 5.6 & 3.1      & 0.018\tablenotemark{b} 		  &\nodata \\
ACTJ0235$-$4 & 02:35:43.0 & -51:22:44.7 & 12.1&1.1 & 5.7 & 1.7      & \, 4.0 $\pm$0.7\tablenotemark{a} &1.3$\pm$0.1\\
ACTJ0235$-$5 & 02:35:27.7 & -51:18:46.2 & 16.9&1.6 & 5.7 & 3.6	& \, 3.7$\pm$0.6\tablenotemark{a}  &1.1$\pm$0.1 \\ \hline
ACTJ0245$-$1 & 02:45:30.6 & -53:04:08.7 & 22.4&1.3 & 11.1 & 2.0    & \, 4.0$\pm$0.6\tablenotemark{a}  & 1.2$\pm$0.1\\
ACTJ0245$-$2 & 02:45:42.1 & -53:02:02.4 & 11.0&1.2 & 4.8 & 1.0      & 2.1$\pm$0.6  		    & \nodata\\
ACTJ0245$-$3 & 02:45:35.9 & -53:05:01.5 & \, 9.6&1.2 & 3.9 & 2.7	& 3.8$\pm$0.5 		    & 1.1$\pm$0.1\\ \hline
ACTJ0330$-$1 & 03:30:30.3 & -52:27:24.1 & 21.9&1.2 & 11.4 & 4.1    & \, 3.0$\pm$0.5\tablenotemark{a}  & 1.1$\pm$0.1 \\
ACTJ0330$-$2 & 03:30:54.0 & -52:24:35.0 & 20.9&1.3 & 9.0 & 3.7	& 3.2$\pm$1.0 		    & 1.1$\pm$0.1\\
ACTJ0330$-$3 & 03:31:13.6 & -52:28:33.5 & 13.1&1.2 & 6.5 & 2.6	& 1.7$\pm$0.7 		    & 1.1$\pm$0.1 \\ \hline
ACTJ0438$-$1 & 04:38:30.6 & -54:18:32.4 & 33.3 &1.0 & 21.4 & 2.0  & \, 3.4$\pm$0.7\tablenotemark{a}   & 1.3$\pm$0.1\\
ACTJ0438$-$2 & 04:38:34.9 & -54:19:42.0 & 16.8 & 1.0 & 10.3 & 2.5 & \, 3.2$\pm$0.6\tablenotemark{a}   &1.2$\pm$0.1\\
ACTJ0438$-$3 & 04:37:54.0 & -54:20:37.8 & \, 8.1 & 0.8 & 6.6 & 3.7  & \nodata 			    & 1.1$\pm$0.1\\
ACTJ0438$-$4 & 04:38:24.6 & -54:17:26.6 & 12.6 &1.0 & 6.1 & 2.1    & \nodata 			    &1.3$\pm$0.1\\
ACTJ0438$-$5 & 04:38:42.5 & -54:21:13.3 & \,9.1 &1.1 & 5.2 & 4.1    & 3.5$\pm$0.8 		    &1.1$\pm$0.0\\
ACTJ0438$-$6 & 04:38:24.7 & -54:21:21.0 & \,7.6 & 0.9 & 4.4 & 2.2   & \nodata 			    & 1.2$\pm$0.1\\
ACTJ0438$-$7 & 04:38:18.9 & -54:23:07.2 & \,6.6 & 0.9 & 4.1 & 3.8   & \nodata 			    & 1.1$\pm$0.0\\
ACTJ0438$-$8 & 04:38:33.1 & -54:19:07.9 & \,7.4 &1.0 & 4.0 & 2.3    & \nodata 			    & 1.2$\pm$0.1\\ \hline
ACTJ0546$-$1 & 05:47:01.4 & -53:45:24.5 & 21.4 & 1.0 & 14.3 & 3.5 & 4.9$\pm$0.9  &1.0$\pm$0.0\\
ACTJ0546$-$2 & 05:46:34.5 & -53:45:51.9 & 12.9 & 1.0 & 7.5 & 0.6	& 1.9$\pm$0.7 & \nodata\\
ACTJ0546$-$3 & 05:46:53.9 & -53:44:12.3 & 12.2 & 1.0 & 7.5 & 2.7	& 2.2$\pm$0.8 & 1.0$\pm$0.0\\
ACTJ0546$-$4 & 05:46:37.6 & -53:43:07.9 & 12.7 & 0.9 & 7.4 & 2.4	& 4.5$\pm$0.9 & 1.0$\pm$0.1\\
ACTJ0546$-$5 & 05:46:55.0 & -53:46:48.7 & 10.9 & 0.9 & 6.2 & 2.9	& 2.7$\pm$1.0 & 1.0$\pm$0.1\\
ACTJ0546$-$6 & 05:46:49.9 & -53:46:22.9 & \, 7.2 & 0.9 & 6.1 & 2.0	&  \nodata  	& 1.0$\pm$0.1\\
ACTJ0546$-$7 & 05:46:30.1 & -53:41:40.6 & 13.9 &1.4 & 5.5 & 4.0	& 3.2$\pm$0.1 & 1.0$\pm$0.0\\
ACTJ0546$-$8 & 05:46:39.9 & -53:46:02.9 &\, 8.3 &1.0 & 4.9 & 0.6	& \, 2.3$\pm$0.5\tablenotemark{a} &\nodata\\
ACTJ0546$-$9 & 05:46:53.8 & -53:47:46.5 & \, 7.0 & 0.9 & 4.8 & 3.3	& 2.1$\pm$0.4 & 1.0$\pm$0.0\\
ACTJ0546$-$10 & 05:46:28.4 & -53:45:44.4 &\, 7.1& 0.9 & 4.3 & 1.4	& 3.2$\pm$1.0 & 1.1$\pm$0.2\\
ACTJ0546$-$11 & 05:47:05.9 & -53:47:55.5 & \, 7.6 & 1.0 & 4.2 & 4.8	& 1.9$\pm$0.7 & 1.0$\pm$0.0\\ \hline
ACTJ0559$-$1 & 06:00:18.5 & -52:49:57.8& 18.4&1.9 & 5.4 & 5.3	& 2.9$\pm$1.0 & 1.0$\pm$0.0 \\ \hline
ACTJ0616$-$1 & 06:16:13.9 & -52:27:23.6 & 12.4&1.1 & 6.6 & 3.1	& 1.1$\pm$0.4 & 1.0$\pm$0.1\\ 
ACTJ0616$-$2 & 06:16:30.3 & -52:27:10.6 & 13.7&1.1 & 6.4 & 0.6 	& \nodata &\nodata\\
ACTJ0616$-$3 & 06:16:25.5 & -52:28:15.1 & \, 7.7&1.0 & 4.3 & 1.7	&  \nodata & 1.2$\pm$0.1\\
ACTJ0616$-$4 & 06:16:39.3 & -52:22:41.0 & 12.5&1.3 & 4.4 & 4.6      & 3.9$\pm$0.7 & 1.0$\pm$0.0\\
\enddata
\tablecomments{}{The first column indicates the source identifier, formed by the name of the cluster in whose field it was detected, plus a correlative ID number that ranks sources in a field according to their S/N ratio. The equatorial coordinates indicate the location of the source's centroid in the 870 \micron\, S/N map; the integrated flux density S(870 \micron) is measured  from the reduced, non-smoothed flux density map, and the signal to noise ratio S/N is obtained from the Gaussian-filtered map. The last columns indicate the angular distance to the cluster's center ($\theta$), the photometric redshift estimated from SED modeling when possible, and the magnification factor ($\mu$) estimated from the models described in Section \ref{sec:lensing}.}
\tablenotetext{a}{\hspace{0.3cm}Tentative counterpart identifications.}
\tablenotetext{b}{\hspace{0.3cm}Redshift of foreground galaxy ESO 198$-$G021, see Appendix A. No magnification is considered for this source.}
\tablenotetext{c}{\hspace{0.2cm} From \citet{zitrin2013} magnification maps.}
\end{deluxetable*}

\section{Analysis of 870 \micron\, point sources}\label{sec:sources}

In this section we study the multi-wavelength properties of the set of SMGs detected by LASCAR at 870 \micron. First we identify their counterparts in the radio, far-IR, near-IR, and optical regimes, model their SEDs and estimate photometric redshifts when possible, estimate their magnification due to lensing by the clusters, and analyze the resulting redshift distribution and number counts.

\subsection{Counterpart identification}

To identify the multi-wavelength counterparts of LASCAR SMGs, we searched  the radio, PACS, SPIRE, IRAC and optical catalogs described above and identified as preliminary counterpart candidates all sources located within a circle of radius equal to half the LABOCA beam's FWHM (19.2\arcsec) centered on a position in Table \ref{tab:laboca_dets}. We then applied the following criteria:\\

\noindent $\bullet$ Among the available data,  the best resource for accurate SMG localization is the deep 2.1 GHz imaging, so we start by identifying a matching radio source when possible, and then use the radio centroid as a reference for  counterpart identification at shorter wavelengths. To evaluate the reliability of the association of the SMG with a radio source of  flux density  $S_{2.1}$ located within  the LABOCA beam, we calculate the corrected Poisson probability of a chance detection within a beam's area around the submillimeter centroid ($P_C$) as:

\begin{equation}
P_C=1-\exp \left\{- P^*\left[1+\ln\left(\frac{P}{P^*} \right) \right] \right\}
\end{equation}

\noindent where $P^*$ is the raw Poisson probability of finding a source brighter than $S_{2.1}$ inside a search radius $r=9.6$\arcsec, and $P$ is the raw probability of finding a source brighter than the critical radio detection flux density in the same region (\citealt{downes86}). Both are calculated as $1-\exp(-\pi n r^2)$, where $n$ is the integral number density of radio sources above a given flux. For the radio source number counts, we scale the 2.1 GHz flux densities to 1.4 GHz assuming a radio spectral index $\alpha=-0.75$ (\citealt{ibar10}), and adopt the 1.4 GHz differential number counts function obtained by \citet{bondi08} for the VLA-COSMOS survey, which yielded a catalog of $\sim3600$ radio sources over a 2 deg$^{2}$ region, down to a 1$\sigma$ sensitivity limit of about 11 $\mu$Jy. This catalog has a resolution of 1.5\arcsec, but only $\sim7\%$ of all sources form pairs with angular separation smaller than the resolution of our radio imaging.
modeling
The traditional approach is to require $P_C< 0.05$ for a reliable radio-submillimeter association. In our analysis, we consider a single radio counterpart candidate with $P_C<0.05$ as a secure identification, and if there is more than one radio source within the LABOCA beam meeting this requirement, we consider them to define a multiple-component source. We then use the radio positions to look for PACS, IRAC, and optical matches within a search radius of 2\arcsec, which is the mean circularized radius of the synthesized beam for our ATCA imaging. Finally, if there are no reliable radio detections within the LABOCA beam, we set an upper limit for the 2.1 GHz flux density and proceed to the analysis of FIR candidate counterparts.\\

\noindent $\bullet$ In the absence of a significant radio detection, we move on to inspection of PACS images, which have been shown to have SMG detection rates of  $\sim40-50\%$  at 160 \micron, for a $3\sigma$ detection limit $\sim5.7$ mJy and sample median redshifts $\langle z \rangle \sim 2$ (\citealt{danner10,magnelli10}). Again, we apply the $P_C<0.05$ criterion for a reliable association and calculate the probability of chance association between the LABOCA and PACS source based on the PACS differential number counts obtained by \citet{berta10} for the GOODS-N and COSMOS fields at 100 and 160 \micron. The PACS coordinates are used as the reference to search for matching sources at shorter wavelengths. However, given that 100 and 160 \micron\, maps have beam FWHMs of 7.2 and 12$ ^{\prime\prime}$, respectively, in some cases two or more IRAC/optical matches may be blended together by the far-IR beam, hindering precise counterpart identification.\\

\noindent $\bullet$ SPIRE 250/350/500 \micron\, maps have spatial resolutions that are comparable to or poorer than that of our LABOCA imaging. Therefore, SPIRE detections within the LABOCA beam are considered counterparts to the 870 \micron\, emission and add data points to the SED modeling, but do not provide improved source positioning or aid in the identification of counterparts in other bands.\\

\noindent $\bullet$ For clusters with no $Herschel$/PACS observations, we are only able to pinpoint the near-IR/optical counterpart if there is a radio detection. Otherwise, we are only able to select a candidate counterpart among IRAC sources that fall within the LABOCA beam based on their  $S_{4.5}/S_{3.6}$ color, which we expect to be comparable to the median $1.27\pm0.24$  observed for an SMG sample by \citet{hainline09}. These are recorded as ``tentative'' identifications.\\

In Appendix A, we give a brief description of the counterpart identification process for each of the 49 submillimeter sources in our catalog, and in Figures \ref{fig:count0102} to \ref{fig:count0616} we show multi-wavelength postage stamps for the SMGs in each cluster. All results are summarized in Tables \ref{tab:counterparts} and \ref{tab:counterparts_coords}, where we respectively list the flux densities and coordinates of counterparts identified in radio, SPIRE, PACS, IRAC, and optical data.

In total, we find that out of 49 submillimeter sources, one coincides with a foreground galaxy (ACTJ0235$-$3). Of the remaining 48 SMGs, 4 have no identifiable counterparts in the available bands, 26 have single radio counterparts, 4 have double radio counterparts, and 14 are not detected in radio mapping, but have tentative counterpart identifications in the far or near-IR. The last column of Table \ref{tab:counterparts} indicates in which category each source falls.

\subsection{SED Modeling and Photometric Redshifts}\label{sec:photoz}

\citet{micha10} used a set of 76 SMGs ($S_{850}\gtrsim3$ mJy) with spectroscopic redshifts $z=0.080-3.623$ and dust temperatures $T_D=11.4-113.3$ K from the sample of \citet{chapman2005}  to model the entire UV-to-radio spectral SEDs of  a statistically significant sample in a self-consistent way. Their modeling was based on a library of 35,000 models from \citet{iglesias07}, which were developed in GRASIL (\citealt{silva98}), cover a broad range of galaxy properties from quiescent to starburst, and include a set of templates based on nearby ULIRGs (\citealt{silva98}) and gamma-ray burst host galaxies (\citealt{micha08}). \citet{micha10} matched these templates to all available UV to radio photometry simultaneously for each source, obtaining a library of  best-fitting SMG SEDs that exhibits a wide range of stellar population properties. \citet{smolcic12} tested these templates on eight SMGs with spectroscopic redshifts and found that the implied photometric redshifts were in better agreement than those obtained with other models drawn from the \citet{bruz03} library or provided by the public code Hyper-z \citep{bolzonella2000}.

We used this library and our radio, submillimeter, far-IR, and near-IR measurements to estimate photometric redshifts for all LASCAR SMGs with secure radio, PACS, and/or SPIRE counterparts, and also for LASCAR sources with no radio counterparts but with tentative near-IR matches, as indicated  in Table \ref{tab:counterparts}.  For each SMG, we redshifted all templates from $z=0$ to $z=8$ in even steps of $\Delta z=0.05$, scaled the spectra to match the observed  870 \micron\, flux density, and calculated the resulting $\chi^2$ statistic. We find that with the limited number of data points currently available, is it not possible to discriminate with confidence between different SEDs that give comparably good fits at different redshifts, and determination of a ``best-fitting'' redshift based solely on $\chi^2$ minimization can be misleading. Therefore, our preferred approach is to find the best-fitting redshift for each model in the \citet{micha10} library, select those templates for which the resulting $\chi^2$ is among the 10$\%$ lowest values, and average the corresponding photometric redshifts to determine a mean optimal redshift, with an associated dispersion. Although this procedure results in photometric estimates with rather large errors, of order $\Delta z\sim1$, it gives a realistic representation of the uncertainties in the determination of SMG redshifts based on a limited number of observations. If we relax the $10\%$ limit to include a larger set of best-fitting redshifts, the mean results are similar but dispersion increases; for example, if the $25\%$ lowest values are selected instead, photometric redshifts vary on average by $\sim9\%$, but the associated errors are $\sim12\%$ larger.

\begin{figure}[tb]
\includegraphics[width=8.5cm]{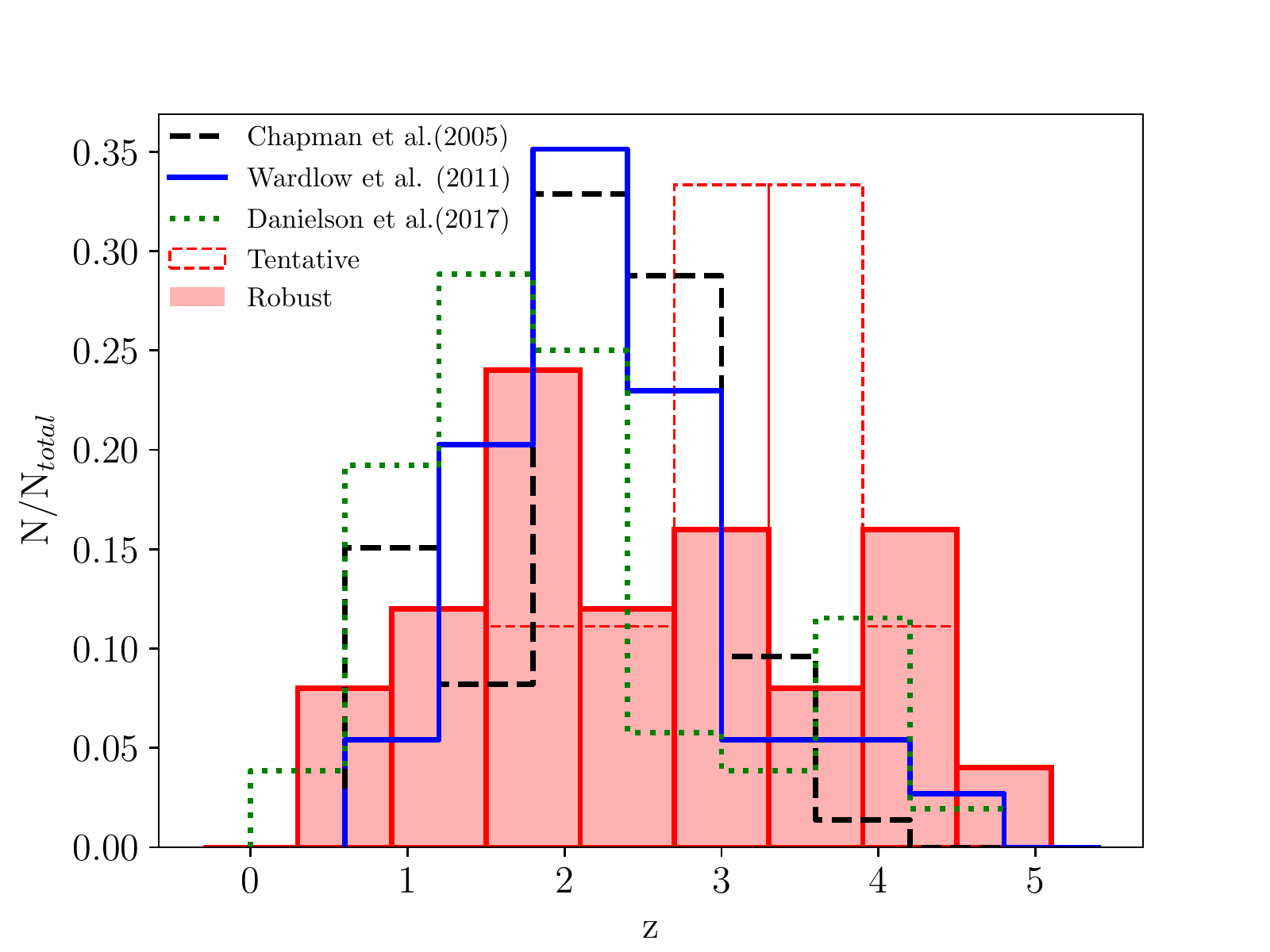}
\caption{Redshift distribution for SMGs detected in LASCAR. Results of this work are plotted in red. The filled histogram shows the redshift distribution for 25 sources with robust radio/near-IR counterparts, for which the median value is $z=2.8\pm0.7$. The dashed histogram corresponds to the redshift distribution of sources with tentative counterparts, with a median $z=3.4^{+1.1}_{-1.1}$. For comparison we include the normalized results from \citet{chapman2005} (black, dashed), \citet{wardlow11} (blue, solid) and \citet{danielson2017} (green, dotted). }

\label{fig:zdist}
\end{figure}

In total, we model the SEDs  of 34 sources, 25 of which correspond to secure counterpart identifications and 9 of which are only tentative estimates. In Appendix A, Figure \ref{fig:seds1}, we show our SED fits. For each source we plot observed flux densities, flux density upper limits for non-detections when appropriate, and the range of best SED fits that we consider in our calculations of the mean photometric redshift. In Figure \ref{fig:zdist} we plot the resulting $N(z)$ distribution for sources with robust and tentative counterparts, together with previous literature results for comparison.

For the sample with robust  and tentative counterpart identifications, the median redshifts are $z=2.8^{+2.1}_{-1.7}$ and $z=3.4^{+1.1}_{-1.1}$, respectively. These results are in reasonable agreement with those from \citet{chapman2005} (median $z=2.2\pm0.1$ for a radio pre-selected sample); from the LABOCA Extended Chandra Deep Field South (ECDFS) submillimeter survey (LESS; \citealt{wardlow11}; median $z = 2.2\pm0.1$); and also with the more recent ALMA surveys of \citet{danielson2017} (median $z = 2.4\pm0.1$) and \citet{brisbin2017} (median $z = 2.48\pm0.05$). In our photometric redshift distribution there is a comparatively larger fraction of sources at $z >$ 3, and the high-redshift tail extends further out than that of the LESS distribution including one SMG with $z_{\rm{phot}}\sim4.9$. The existence of this tail is consistent with millimeter spectroscopic confirmation of a growing number of SMGs at $z\gtrsim4$ (e.g., \citealt{capak08, weiss13, smolcic12, smolcic15, strandet2016}), and with the spectroscopic and photometric redshift distributions obtained by \citet{danielson2017} and \citet{brisbin2017} for samples of SMGs observed at the spatial resolution of ALMA. \citet{su2017} also estimate photometric redshifts for nine strongly lensed DSFGs selected by ACT at 218 GHz (1.4 mm) and obtain a higher median redshift $z=4.1^{+1.1}_{-1.0}$, which is expected since the highest-redshift sources remain bright at this lower frequency due to the negative $K$-correction. 

Our photometric redshift estimates support the existence of a $z\gtrsim4$ SMG population, but it is unclear whether these sources' number counts are consistent with the predictions of cosmological models. For example, semianalytic models that assume a top-heavy IMF and star formation driven by merger-triggered starbursts \citep{baug05} predict a redshift distribution centered at $z\sim2$, with very few sources at much higher redshifts. As a reference, the expected redshift distribution from \citet{baug05} predicts that only $\sim5\%$ of SMGs with $S_{850} = 8$ mJy ($S_{870}\sim7.5$ mJy) lie at $z > 4$, so for our subset of 25 SMGs with robust counterpart identifications we would expect only $1-2$ objects in this redshift range, in contrast to the $\sim5$ that we estimate (Table \ref{tab:laboca_dets}). Additional data are critical to confirm the existence and abundance of very high-$z$ SMGs in our sample and confirm or refute our preliminary results; in particular, high-resolution submillimeter imaging is key to confirm counterpart identification for SMGs that are undetected at radio wavelengths.

%FIGURE 3

\subsection{Gravitational magnification}\label{sec:lensing}

The detection of SMGs in LASCAR may be facilitated by the foreground galaxy clusters acting as gravitational lenses, but the resulting flux magnifications also affect the analysis of the intrinsic counts distribution and enhance the scatter in the contamination of the SZ signal. The amplification of each source depends exclusively on the mass distribution and geometrical configuration of the intervening structures relative to the source position, with no need for additional assumptions regarding their dynamical state  (\citealt{limousin07}). The effect is strongest close to a cluster's core, where the mass surface density is high enough to produce strong lensing features like arcs and multiple images, which can be used to reconstruct the gravitational potential and then calculate the resulting magnification as a function of position. However, such analysis requires very deep and high-resolution optical imaging  combined with extensive spectroscopy to detect lensed images and measure the redshifts of a lens and of a collection of background  objects. Datasets typically used to build successful models include multi-band $HST$ imaging and ground-based spectroscopy using 8-10 m telescopes (e.g.  \citealt{jullo07,limousin07}, and references therein). Among LASCAR clusters, such data are partially available only for \lascarI, for which \citet{zitrin2013} present a strong-lensing analysis. For the remaining clusters, we are unable to construct detailed lensing models at present, but we can use an analytic approach to approximate the magnifications affecting our detected SMGs, and estimate how strongly lensing affects our number counts and other results. From inspection of our optical and near-IR imaging, we find no evidence of superposition with individual lens galaxies that might boost local magnification, so in our calculations we account only for the effects of the clusters' gravitational potentials. 

To estimate magnifications for the LASCAR clusters, we follow the rationale and equations presented in \citet{lima10,lima10a}: we adopt a density profile to model the cluster's dark matter halo, derive the analytical form of lensing observables like the shear and convergence, and finally use the basic lensing equations to calculate the magnification at each projected cluster-centric radius. We assume that the cluster dark matter halo mass density is well represented by a Navarro-Frenk-White (NFW; \citealt{nfw97}) spherical profile with a characteristic virial mass  $M_{200}$, corresponding to the mass within a sphere of a radius $r_{200}$ whose mean interior is density $200\rho_{\rm{crit}}$, for $\rho_{\rm{crit}}$ the critical density for a flat universe. Our calculations are based on the results of \citet{sifon2016}, who used deep optical spectroscopy to determine $M_{200}$ and $r_{200}$ for all clusters in our sample except \lascarV, which was known before the ACT detection (e.g., \citealt{abell1989, voges1999, edge1994, wu97}) and therefore not included in their observations. To evaluate $M_{200}$ for \lascarV,  we apply the scaling relation in \citet{sifon2016}:

\begin{equation}
\sigma_{200}=A_{\rm 1D} \left[ \frac{h E(z) M_{200}}{10^{15}M_{\odot}} \right]^{\alpha}
\end{equation}

\noindent where $\sigma_{200}$ is the line-of-sight velocity dispersion in a spherical cluster of galaxies within $r_{200}$, $E(z)=[ \Omega_{\Lambda}+(1+z)^3\Omega_m ]^{1/2}$, $A_{\rm 1D}=1177\pm 4.2$ km s$^{-1}$,  and $\alpha=0.0364\pm0.002$ \citep{sifon2016}. Using the velocity dispersion measurement of \citet{ruel2014} for \lascarV, $\sigma_{200}=1245\pm210$ km s$^{-1}$, we estimate a virial mass $M_{200}=(14.48\pm0.82) \times10^{14} M_{\odot}$ for this cluster.

This analytical procedure is expected to provide a reasonable estimate of the magnification produced by clusters with approximately spherical dark matter mass density profiles. However, it is not appropriate for the case of \lascarI, which is undergoing a major merger between two clumps with mass ratio 2:1 \citep{menanteau12} that form a very elongated lens with axis ratio $\sim5.5$ \citep{zitrin2013}. Hence, for sources in the field of \lascarI, we obtain magnification factors directly from lensing maps generated by \citet{zitrin2013} at the corresponding source redshifts (private communication).

\begin{figure}[tb]
\centering 
\includegraphics[width=8.5cm]{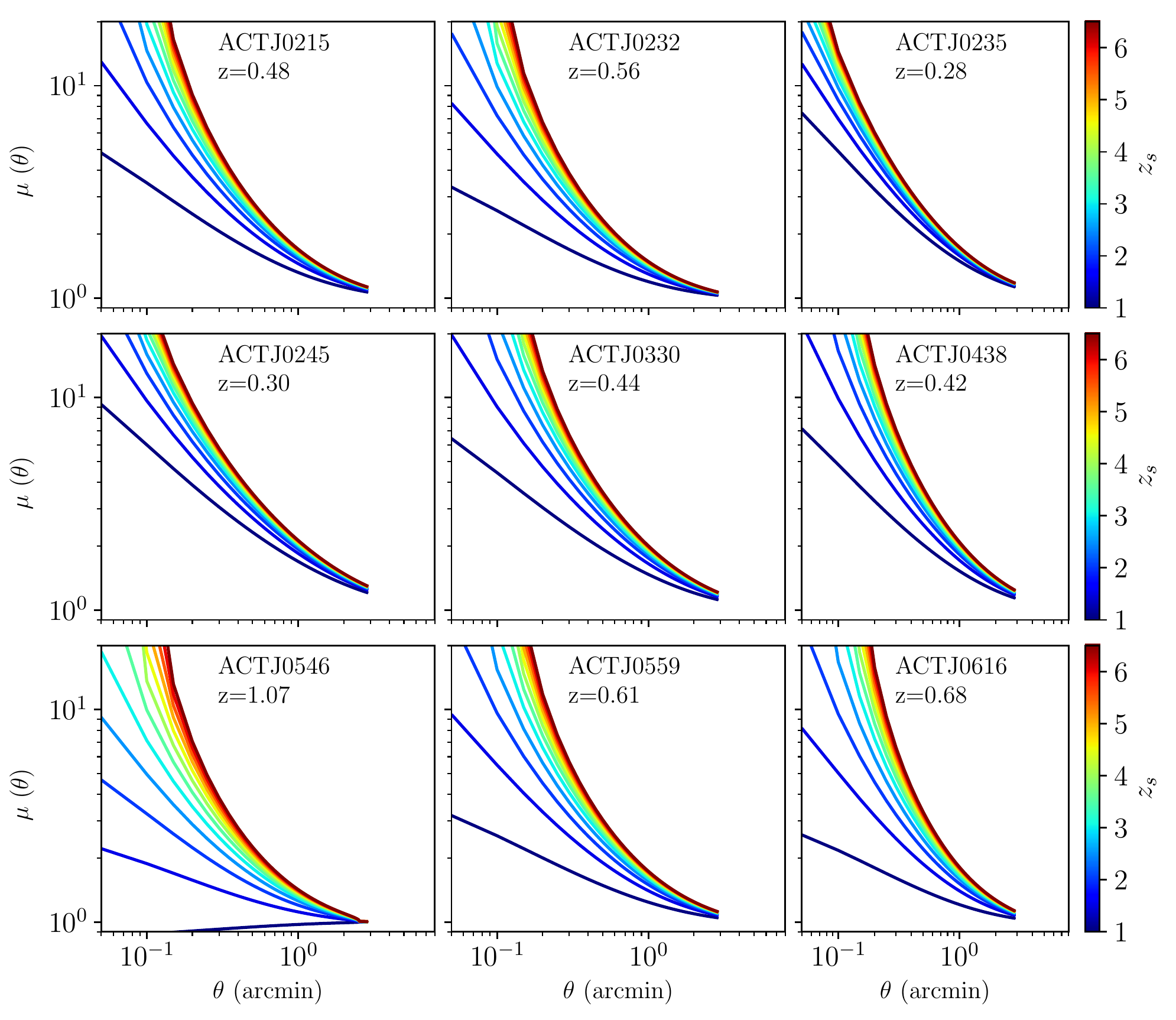}
\caption{Gravitational magnification vs. radius for clusters with virial mass and radius measured by \citet{sifon2013}. In each case, we calculate the expected magnification as a function of angular distance to the cluster's center for a range of source redshifts from $z_s$=1.0 to $z_s$=7.0, in steps of $\Delta z_s=0.5$.}
 \label{fig:lensing}
\end{figure}

 The results of our lensing models for all LASCAR clusters except \lascarI\, are shown in Figure \ref{fig:lensing}, where we plot the resulting $\mu(\theta)$ curves for a set of source redshifts ($z_s$) from $z_s=1.0$ to $z_s=7.0$, so as to cover the full range expected for the SMG population. We find that in general the maximum magnification factor can reach up to $\sim10-11$ close to the cluster's core, but at radii larger than $\sim4$ arcmin the magnification becomes negligible. We also see that for a given radius, sources at higher redshifts are more strongly lensed, but this dependence becomes less significant beyond $z_s\sim3$. For very distant sources, the magnification factor is not strongly influenced by redshift, i.e., sources at $z_s=4.0$ and $z_s=7.0$ will experience similar lensing. 

%FIGURE 4

To estimate the magnification factor for each source, we use the models described above, the cluster-centric angular distances $\theta$ given in Table \ref{tab:laboca_dets}, and the photometric redshifts calculated from our SED modeling. For sources with no photometric redshift estimates, we assume $z_s = 2.8\pm0.7$, the median value for our sample with robust counterpart identifications. The propagated errors due to uncertainties in the virial mass and radius and in the source and cluster ($z_c$) redshifts were calculated separately and added in quadrature; for each variable we performed 1000 Monte Carlo simulations using random values generated from a Gaussian probability distribution with mean and standard deviation equal to the experimental measurement and error. Mathematically, from the random simulations we obtain the individual errors $\sigma_{M_{\rm{200}}}$, $\sigma_{r_{\rm{200}}}$,   $\sigma_{z_c}$ and $\sigma_{z_s}$, and sum them in quadrature to calculate the total error $\sigma_{\rm{total}}$

%$\sigma_{\rm{total}}=\sqrt{\sigma_{M_{\rm{vir}}}^2+\sigma_{r_{\rm{vir}}}^2+\sigma_{z_c}^2+\sigma_{z_s}^2}.$

\begin{figure}[tb]
\centering 
{\includegraphics[width=8.5cm]{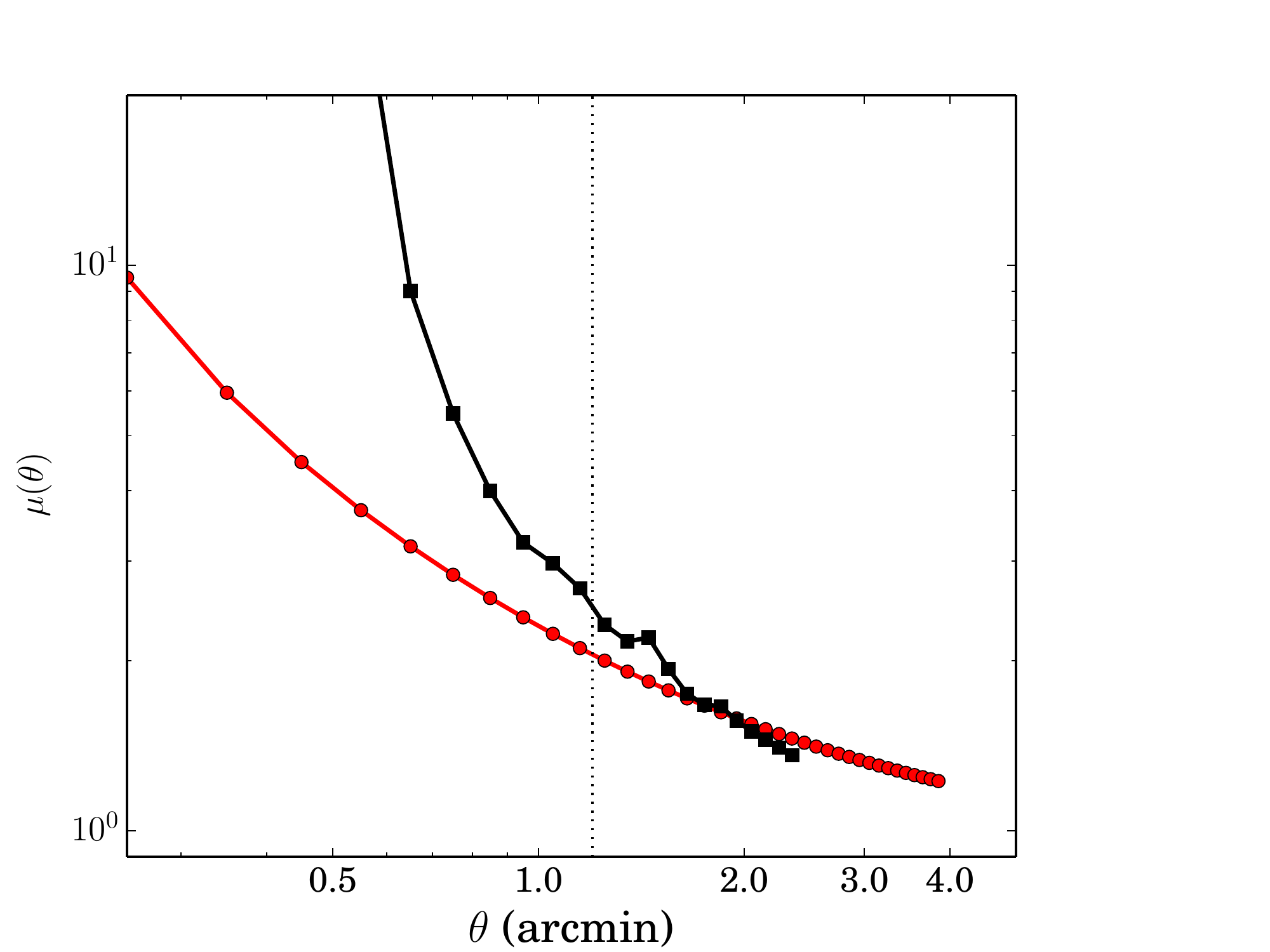}}
\caption{Comparison of magnification radial profiles predicted by the CLASH lensing model for cluster \macsB\ (black, squares) and our calculations (red, circles), for a background source at $z=2.3$, the median redshift of a purely submillimeter flux-limited SMG sample as predicted by \citet{chapman2005}. The vertical dotted line at $\theta=1.2$\arcmin\, marks the radius where discrepancies between the two models are reduced to $\sim10\%$. }
 \label{fig:magmap}
\end{figure}

We tested the reliability of our magnification estimates by performing a similar analysis for cluster \macsB, which has been observed in 16 bands with $HST$ by the Cluster Lensing And Supernova survey with $Hubble$ (CLASH; \citealt{postman12}) and also has a strong-lensing model (\citealt{zitrin2015}). We use the CLASH model best-fit mass for \macsB\, as input and calculate the expected magnification curve ($\mu(\theta)$ vs. $\theta$) for a source at $z=2.8$ (our median sample redshift) following the algorithm described above. We obtain an analogous curve for the \citet{zitrin2015} model by calculating the azimuthally-averaged magnification factor at each projected radius; in Figure \ref{fig:magmap} we plot both results. We find that, for equal virial mass, the magnifications predicted by our analytical algorithm are underestimated by a factor up to $\sim40$ within the inner arcminute compared to those derived from the detailed strong lensing analysis, but at radii beyond $\sim1.2$\arcmin\ the discrepancies are reduced to $\sim 10\%$. The reported discrepancies in the inner region may be explained by the differences in the assumed density profile (elliptical vs. spherical NFW), and also by the fact that the CLASH model includes the dark matter contributions of individual galaxies in the cluster, such that the total density profile steepens towards the center, thus boosting the lensing magnification \citep{zitrin2015} . This is a more realistic representation of the cluster's mass distribution, so we can conclude that for background sources detected at projected radii under $\sim1.2$\arcmin, estimation of magnification factors based solely on a cluster's virial mass is insufficient, and lens modeling based on high-resolution imaging is required to derive intrinsic luminosities. Since such data are not available at present for most clusters in our sample, for the following number counts analysis we exclude sources located at $\theta \leq 1.2$\arcmin. As seen in Table \ref{tab:laboca_dets}, this criterion affects 6 sources in our catalog (ACTJ0215$-$3, ACTJ0235$-$2, ACTJ0245$-$2, ACTJ0546$-$2, ACTJ0546$-$8, and ACTJ0616$-$2). 

%FIGURE 5

\subsection{ Number counts}\label{sec:counts}

With the final catalog of de-magnified sources, we can construct the integral number counts for LASCAR. In total, we detect 49 SMGs, but we exclude from this analysis 6 sources (listed above) in the fields of clusters ACTJ0235, ACTJ0245, ACTJ0546 and ACTJ0616 with cluster-centric radii under 1.2\arcmin, for which we are currently unable to derive accurate intrinsic flux densities. We also identify ACTJ0235$-$3 as a foreground galaxy, so the reported number counts are based on a total of 42 sources detected over 10 fields. We bin the sources by intrinsic flux density, and use the completeness and magnification estimates obtained in Sections \ref{sec:extraction} and \ref{sec:lensing} to account for undetected sources and to calculate the total effective area surveyed by our LABOCA maps. Bin centers are defined to match the minimum and maximum de-magnified flux densities in our source catalogs, and we use a constant logarithmic bin width $\Delta \rm{log}_{10}(S_{870})=0.1$.

Completeness corrections for each cluster field are based on the curves presented in Figure \ref{fig:completeness}, which indicate the fractions of successful detections relative to the total number of sources at different flux densities, which we refer to as $C$. We assume that for each detection at observed flux density $S_{\rm obs}$, there are $N_{\rm undet}=C^{-1}-1$ undetected sources, randomly located across the LABOCA map. To estimate the distribution of \emph{intrinsic} flux densities for these missing objects, we add $N_{\rm{undet}}$ point sources with $S_{\rm obs}$ and uniform spatial distribution to each cluster map, obtain their expected magnifications using the lensing curves in Figure \ref{fig:lensing}, and calculate their intrinsic flux densities. The resulting catalogs of simulated sources are then binned in the same way as the detected SMGs; the process is repeated 1000 times to obtain average number counts for the undetected point source population. Finally, both sets of number counts (detected and undetected) are added to obtain the final completeness-corrected number counts over all fields.

\begin{figure}[tb]
\centering 
\includegraphics[width=8.5cm]{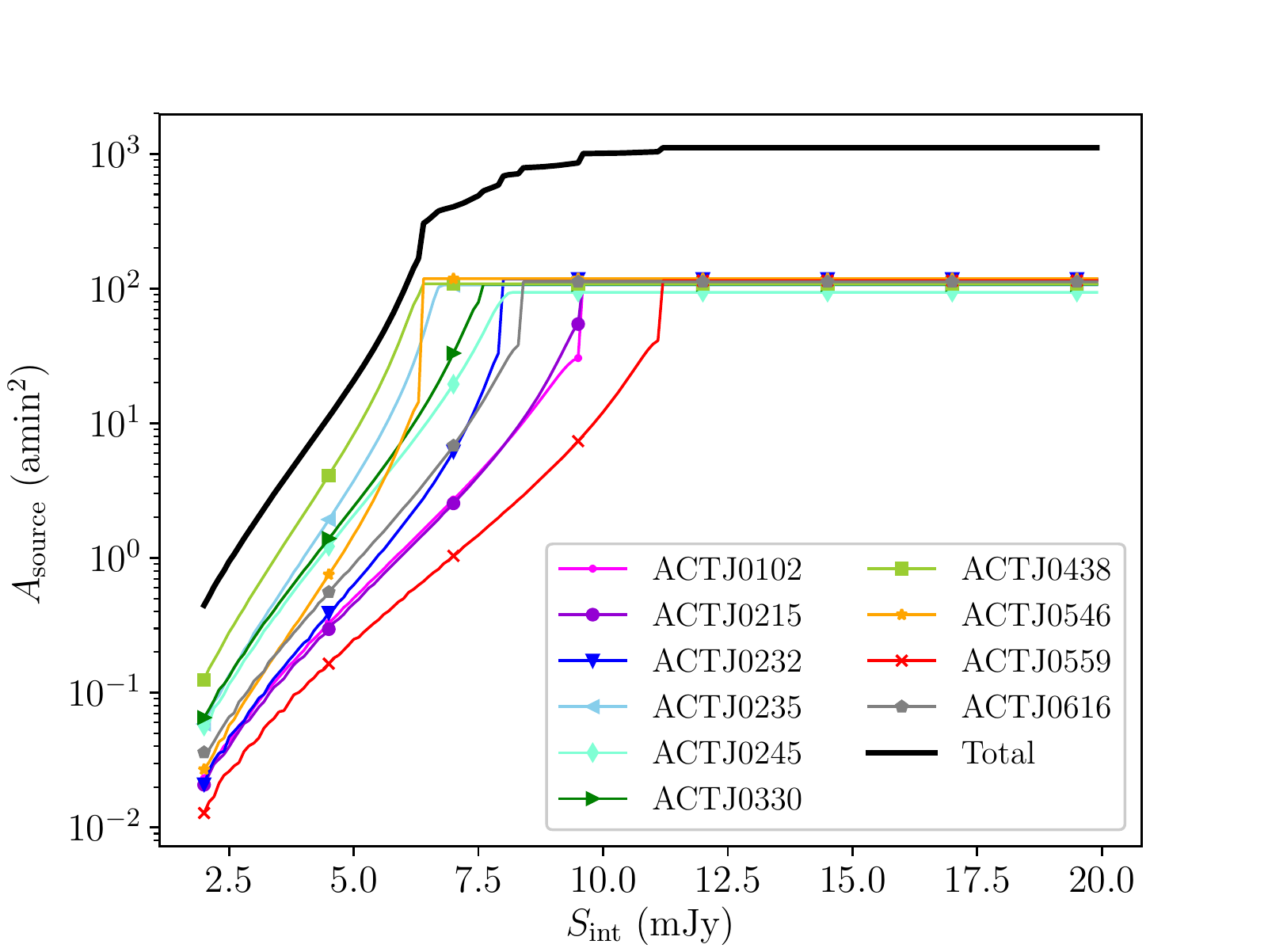}
\caption{Effective source plane area A$_{\rm source}$ where sources of intrinsic flux density $S_{\rm int}$ can be detected, considering magnification effects. Curves for individual clusters are plotted as indicated in the figure legend, and the solid black curve represents the total summed effective area for the LASCAR survey. Individual cluster curves converge to an effective area equal to the total image area for $S_{\rm int}=4\sigma$, when $\mu_{\rm min}=1$. }
 \label{fig:mag_areas}
\end{figure}

The total effective area surveyed by our submillimeter maps depends on the target  intrinsic flux density of point sources, which are magnified by the foreground clusters. In the lensing formalism, the image area is calculated as $A_{\rm image}=\mu A_{\rm source}$, where $A_{\rm source}$ is the area in the source plane and $\mu$ the magnification factor. For each intrinsic flux density $S_{\rm int}$ we can calculate the minimum magnification $\mu_{\rm min}$ required so that the observed flux density $S_{\rm obs}$ is above the $4\sigma$ detection threshold, $\mu_{\rm min}=4\sigma/S_{\rm int}$. Hence, the effective area where sources of flux density $S_{\rm int}$ can be detected corresponds to the set of pixels where $\mu \geq \mu_{\rm min}$; we use the magnification maps in Figure \ref{fig:lensing} to determine this region for each cluster assuming a median source redshift $z=2.8$, and calculate the effective source plane area as

\begin{equation}
A_{\rm source}(S_{\rm int})=\sum\limits_{\mu_i \geq \mu_{\rm min}} A_{\rm pix}/\mu_i
\end{equation}

\noindent where $i$ runs over all pixels in the detection map minus the inner 1.2\arcmin\ where magnification estimates are uncertain (except for ACTJ0102), and $A_{\rm pix}=12.96$ arcsec$^2$ is the image pixel area. We then add all clusters' effective areas to obtain the total de-magnified area for the complete survey. The completeness-corrected binned number counts are then divided by this total area. In Figure \ref{fig:mag_areas}, we show the resulting $S_{\rm int}$ vs. $A_{\rm source}$ curves for each cluster, and the total summed curve for the LASCAR survey.

%FIGURE 6

\begin{figure*}[tb]
\centering 
\includegraphics[width=13cm]{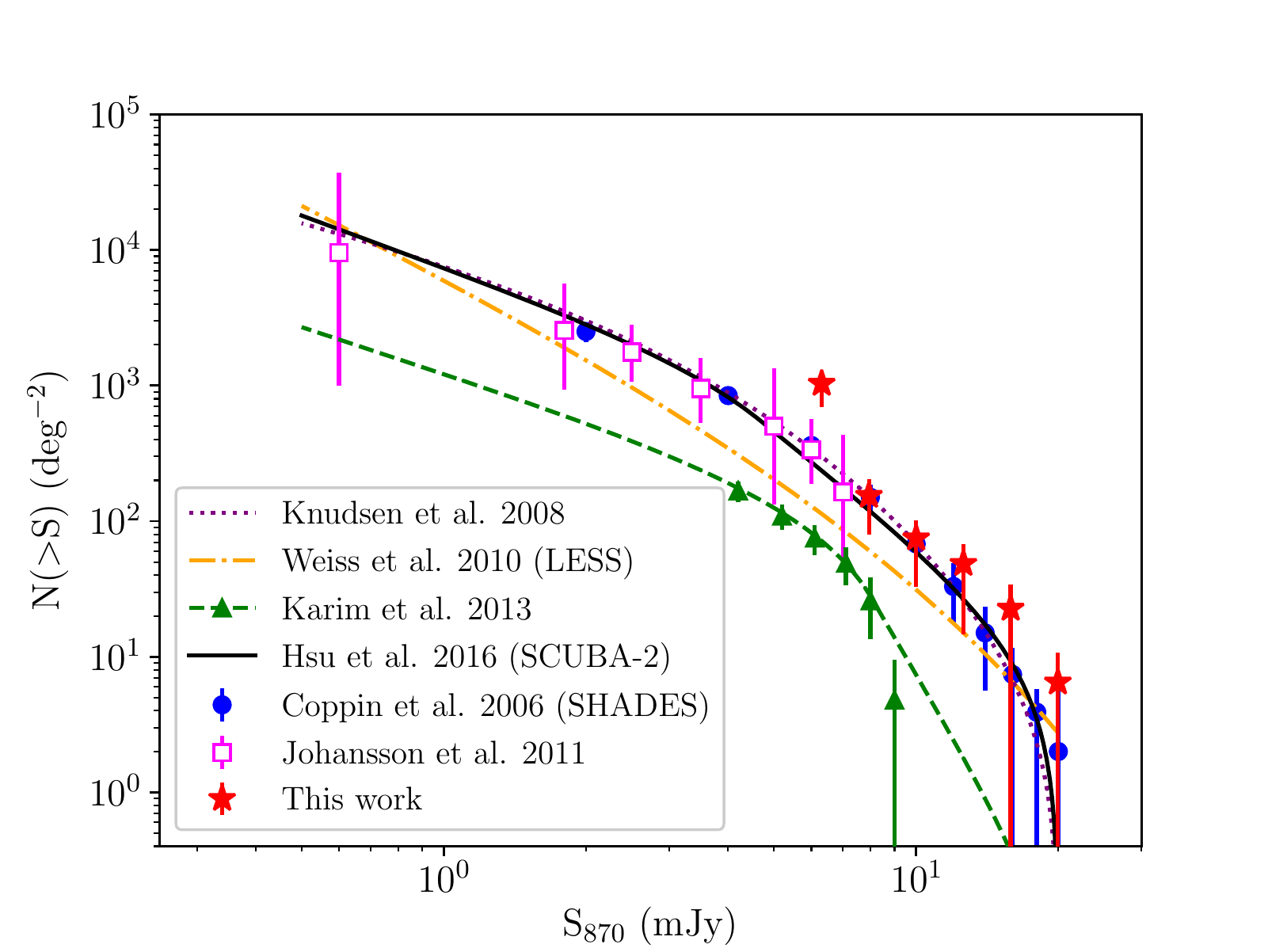}
\caption{Integral number counts for submillimeter sources detected behind 13 galaxy clusters. The plotted results consider in total 42 background sources detected at cluster-centric radii larger than 1.2\arcmin. Our counts have been corrected for completeness and gravitational magnification, using the results from Sections \ref{sec:extraction} and \ref{sec:lensing} respectively. For comparison, we also show previous results from \citet{hsu2016, karim13, johansson11,knudsen08,weiss09b}, and \citet{coppin06}. Line plots correspond to the best-fitting Schechter functions for each dataset. }
 \label{fig:number_counts}
\end{figure*}

 The uncertainties in number counts were derived from Poisson statistics, which apply  when event rates are calculated from small numbers of observed events (\citealt{gehrels86}). Our results are given in Table \ref{tab:counts} and plotted in Figure \ref{fig:number_counts}. For comparison, we also show the integral number counts from the lensing cluster surveys of \citet{knudsen08} and \citet{johansson11}, from the SCUBA Half-Degree Extragalactic Survey (SHADES; \citealt{coppin06}), from LESS (\citealt{weiss09b}), from SCUBA-2 \citep{hsu2016}, and from the high-resolution ALMA follow-up of LESS \citep{karim13}. We find that, for the intrinsic flux density range covered by our survey, results are consistent within uncertainties with previous single-dish surveys conducted in blank fields (e.g., \citealt{coppin06}), towards lensing clusters \citep{knudsen08,johansson11}, and combining both cluster and blank fields \citep{hsu2016}. The exception is the datapoint at $S_{\rm int}=6.3$ mJy, fainter than the $4\sigma$ detection threshold for all cluster, which therefore comprises  sources that are necessarily magnified. The discrepancy is possibly due to the uncertainties in our analytical lens models, which may generally underestimate magnification factors relative to those derived from strong lensing models, as suggested by the comparison presented in Figure \ref{fig:magmap}. If magnification factors are minimally increased across the field, intrinsic flux densities and binned number counts do not vary significantly, but the detectable area $A_{\rm source}$ where $\mu>4\sigma/S_{\rm int}$ is increased, thus affecting the number counts per unit area.  To test this hypothesis, we repeated our calculations but slightly scaled our analytical magnification maps by a factor $\sim1$. We find that a satisfactory match between our resulting number counts at $S_{\rm int}=6.3$ mJy  and previous surveys can be reached if all magnification estimates are varied by only $\sim5\%$, which is within the uncertainties obtained for $\mu$ and listed in Table \ref{tab:laboca_dets}.
 
Compared to number counts obtained by \citet{weiss09b} for LESS, our results are larger by a factor $\sim2-9$ depending on the intrinsic flux density, but it has been reported in the literature that bright SMGs and other rest-frame optical populations are underabundant in the ECDFS compared to other deep fields (see \citet{weiss09b} and references therein). \citet{karim13} reported 870 $\mu$m number counts derived from high resolution ($\sim1.5$\arcsec) ALMA continuum imaging of SMGs detected by LESS, which has revealed that bright SMGs with $S_{870} \gtrsim12$ mJy are actually resolved into multiple, fainter sources with individual flux densities $\lesssim9$ mJy  \citep{hodge13,karim13}. As a result, ALMA integral number counts of \citet{karim13} are lower than those of LASCAR and other single-dish surveys but comparable to the predictions of \citet{weiss09b} for $S_{870}\lesssim 9$ mJy, declining steeply at higher flux densities. Hence, like other single-dish surveys, LASCAR presumably overestimates the number of intrinsically bright sources behind the sample of ACT galaxy clusters. 

%FIGURE 7

%TABLE 6 
\begin{center}
\begin{deluxetable*}{cc}
\tablecolumns{2}
\tablewidth{7.5cm}
\tabletypesize{\scriptsize}
\tablecaption{Integral number counts at 870 \micron. \label{tab:counts}}
\tablehead{	
\colhead{$S_{870}$ (mJy)} & \colhead{$N(>S_{870})$ (deg$^{-2}$) }  }
\startdata
\vspace{0.1cm}\, 6.3 & $1035^{+339}_{-252}$ \\
\vspace{0.1cm}\, 7.9 & $154^{+75}_{-49}$ \\
\vspace{0.1cm}\, 10.0 & $75^{+42}_{-27}$ \\
\vspace{0.1cm}12.6 & $49^{+34}_{-20}$ \\
\vspace{0.1cm}15.8 & $23^{+26}_{-12}$ \\
\vspace{0.1cm}20.0 & $7^{+19}_{-4}$ \\
\enddata
\tablecomments{Counts are based on 42 foreground SMGs detected at cluster-centric radii $\geq1.2$\arcmin. Calculations include corrections for gravitational magnification and completeness.}
\end{deluxetable*}
\end{center}

\section {Conclusions} \label{sec:conc}

The LABOCA/ACT Survey of Clusters at All Redshifts (LASCAR) has obtained 870 \micron\, LABOCA and 2.1 GHz ATCA mapping for a set of ten massive SZE-selected galaxy clusters from the ACT southern survey, and \emph{Herschel} PACS and SPIRE data for sample subsets (5 and 6 clusters, respectively), with the aim of studying the properties of the clusters' SZE signals and of the background SMG population from the same dataset. In \citet{lindner14} we estimated the levels of radio source and SMG contamination of the SZE signal and constraints on the cluster peculiar velocities using the kinetic SZE effect; in this work we present a study of the submillimeter point sources detected in the fields of the LASCAR targets.

The 870 \micron\, LABOCA maps were reduced using a multi-scale iterative pipeline that successfully extracts the extended SZE increment signal and yields point-source sensitivities of $\sim2$ mJy beam$^{-1}$. We applied an enhanced matched-filter extraction algorithm to recover 49 sources at the $4\sigma$ level, and we used our radio and $Herschel$ observations plus existing near-IR and optical data to conduct a detailed analysis of their multi-wavelength properties. First, we used our combined dataset to identify the likely counterparts of the detected SMGs. We find that one corresponds to a foreground galaxy (ACTJ0235$-$3), four have double radio counterparts, 28 have single radio counterparts, and 16 have no radio counterparts. In the case of SMGs with double radio matches, which could be interpreted as merging or interacting subcomponents of a single SMG (or as double lobes of a radio-loud SMG), we can identify individual counterparts in the near- and far-IR and in the optical imaging, but we cannot disentangle their separate contributions to the emission at 870, 500, 350, and 250 \micron, where they are blended together due to coarser spatial resolutions. For SMGs that have single detections at 2.1 GHz, we can generally determine their correspondence to specific PACS, IRAC, and optical sources, but there are three cases (ACTJ0235$-$1, ACTJ0235$-$5, and ACTJ0330$-$1) in which the radio beam encompasses two or more IRAC/optical sources, so we can only identify a tentative counterpart at our shortest wavelengths. For the remaining SMGs that are not detected at the $4\sigma$ level in our 2.1 GHz maps, we analyze all $Herschel$, IRAC, and optical sources located within the LABOCA beam and try to select candidate counterparts based on PACS detections, and on comparison of near-IR colors to previous SMG observations. We succeed in identifying tentative counterparts for 9 of these systems.

For 34 sources with secure (25) and tentative (9) matches at different wavelengths, we estimate photometric redshifts through minimum$-\chi^2$ fitting of template SEDs from the library of \citet{micha10} and obtain values from $z\sim1.1$ to $z\sim4.9$. For sources with tentative counterpart identifications, the photometric redshift estimates still need to be confirmed through high-resolution continuum mapping at millimeter or submillimeter wavelengths, which is the only unbiased method for unequivocal matching to sources detected in near-IR and optical imaging. For LASCAR SMGs with counterparts classified as ``secure'', we obtain a photometric redshift distribution whose median $z= 2.8^{+2.1}_{-1.7}$ is consistent with results in the literature. We find an excess of sources at redshifts $z>3$ relative to previous studies, although we note that with the limited number of bands currently available for SED modeling (only 4 in cases where only radio and IRAC counterparts are identified), estimated photometric errors are bound to be significant, of order $\Delta z\sim1.0$.

Since our detected SMGs lie behind galaxy clusters, they are expected to be gravitationally lensed. For sources in the field of \lascarI\, we use magnification estimates from the strong-lensing model of \citet{zitrin2013}, and for the remaining clusters we apply analytical models to estimate the magnification factor at the position of each SMG using cluster virial mass measurements in the literature. The resulting amplifications should have $10\%$ or better accuracy for sources outside $\sim1.2$\arcmin\, of the cluster's core; in order to determine accurate amplifications for all sources, it is necessary to obtain $HST$ high-resolution optical/infrared imaging that can be used for strong lensing modeling of the dark matter mass distribution.

We use the final de-magnified flux densities to construct the integral number counts for LASCAR, excluding from this analysis six sources located at cluster-centric radii under 1.2\arcmin, for which amplification factors are not well determined, and one additional object identified as a foreground galaxy. Our results are in good agreement with those of previous submillimeter surveys in the overlapping flux density ranges. Comparison at fainter flux densities is hindered however by the lack of more precise strong-lensing  modeling of the clusters' central regions, where magnifications are expected to be highest and detected SMGs may prove to be intrinsically very faint. Towards the bright end, LASCAR number counts are consistent with those of previous single-dish surveys, but likely overestimate the number of sources with $S_{870}\gtrsim9$ mJy compared to high-resolution ALMA observations, which indicate that $\sim30-50\%$ of sources with single-dish flux density measurements above this limit may resolve into multiple, fainter components. High-resolution ALMA continuum mapping of SMGs detected in LASCAR is required to determine their single or multiple component nature, confirm and refine identification of near-IR and optical counterparts, and thus improve photometric redshift estimates and number counts; such data have been obtained by our team for a selection of LASCAR sources and will be reported in the near future.

\acknowledgments
We thank the anonymous referee for useful comments that have helped improved this paper. We also thank John Wu for his valuable comments on the paper; and Mauricio Carrasco and Adi Zitrin for providing lensing magnification maps for \lascarI.
P.A. acknowledges support from CONICYT through grant FONDECYT Iniciaci\'on 11130590, and A.J.B. acknowledges support from NSF grant AST-0955810. R.D.\ acknowledges support from CONICYT through grants FONDECYT 1141113, Anillo ACT-1417 and BASAL PFB-06 CATA.
J.P.H.\ acknowledges the hospitality of the Flatiron Institute which is supported by the Simons Foundation. M.L. is partially supported by CNPq and FAPESP.\\
ACT operates in the Parque Astron\'omico Atacama in northern Chile under the auspices of the Programa de Astronom\'ia de la Comisi\'on Nacional de Investigaci\'on Cient\'ifica y Tecnol\'ogica de Chile (CONICYT). This work was supported by the U.S. National Science Foundation through awards AST-0408698 and AST-0965625 for the ACT project, and PHY-0855887, PHY-1214379, AST-0707731, and PIRE-0507768 (award No. OISE-0530095). Funding was also provided by Princeton University, the University of Pennsylvania, and a Canada Foundation for Innovation (CFI) award to UBC. Computations were performed on the GPC super-computer at the SciNet HPC Consortium. SciNet is funded by the CFI under the auspices of Compute Canada, the Government of Ontario, the Ontario Research Fun--Research Excellence, and the University of Toronto. APEX is operated by the Max-Planck-Institut f\"ur Radioastronomie, the European Southern Observatory, and the Onsala Space Observatory.

\appendix
\section{Notes on individual sources}
 
\begin{figure}
\centering
\includegraphics[width=10cm]{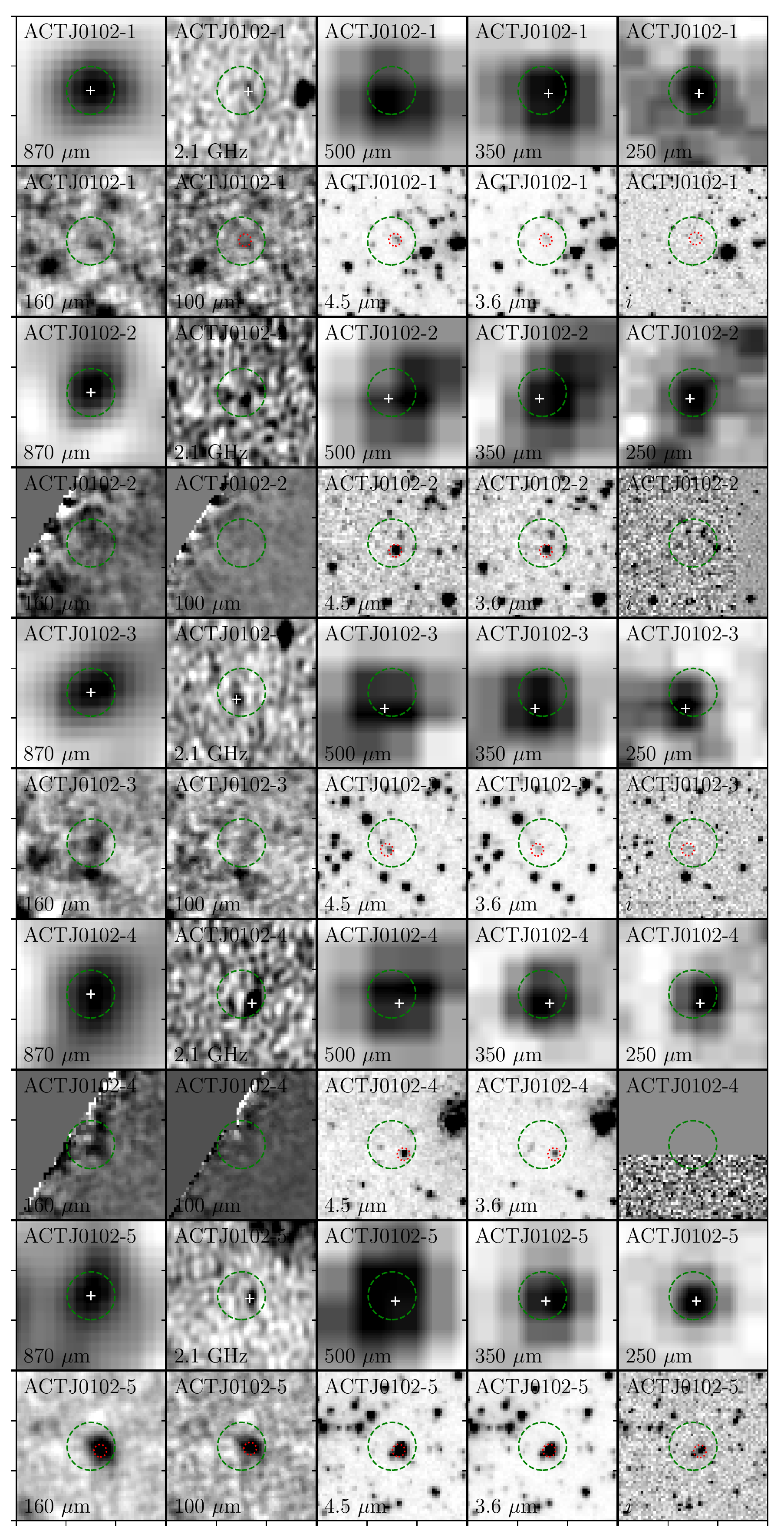}
\caption{Counterpart identification for SMGs detected in the field of cluster ACTJ0102. For each SMG, we present multi wavelength 60$ ^{\prime\prime}$$\times$60\arcsec\ postage stamps centered on the position of the LABOCA source given in Table \ref{tab:laboca_dets}. Ticks are located every 10\arcsec. Dashed circles represent the 19.6\arcsec\, FWHM LABOCA beam. The centroids of the LABOCA source and of radio (2.1 GHz) and SPIRE (500/350/250 \micron) counterparts are indicated by white crosses. In PACS (160/100 \micron), IRAC (4.5/3.6 \micron), and optical ($i$) stamps, the selected counterpart or counterpart candidate (see text in Appendix A) is indicated by a red dotted circle. Spatial resolutions for each band are: 19.2\arcsec (870 \micron), 4\arcsec (2.1 GHz), 35.0\arcsec (500 \micron), 23.9\arcsec (350 \micron), 17.6\arcsec (250 \micron), 13.0\arcsec (160 \micron), 6\arcsec (100 \micron), 2.5\arcsec (4.5 and 3.6 \micron), and 0.24\arcsec (optical). North is up and East is left.} \label{fig:count0102}
\end{figure}

\begin{figure}
\centering
\includegraphics[width=10cm]{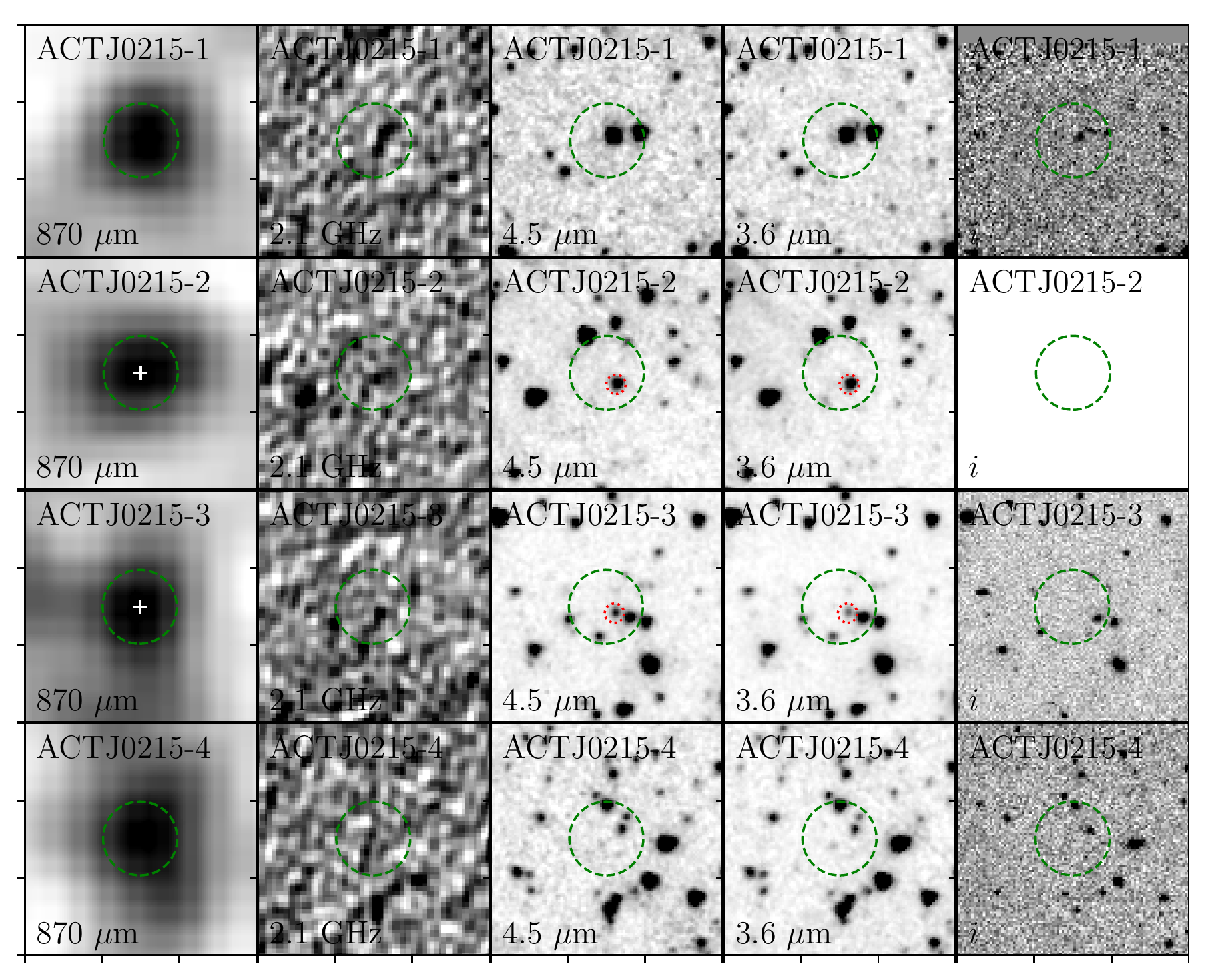}
\caption{Counterpart identification for SMGs detected in the field of cluster ACTJ0215.} \label{fig:count0215}
\end{figure}

\begin{figure}
\centering
\includegraphics[width=10cm]{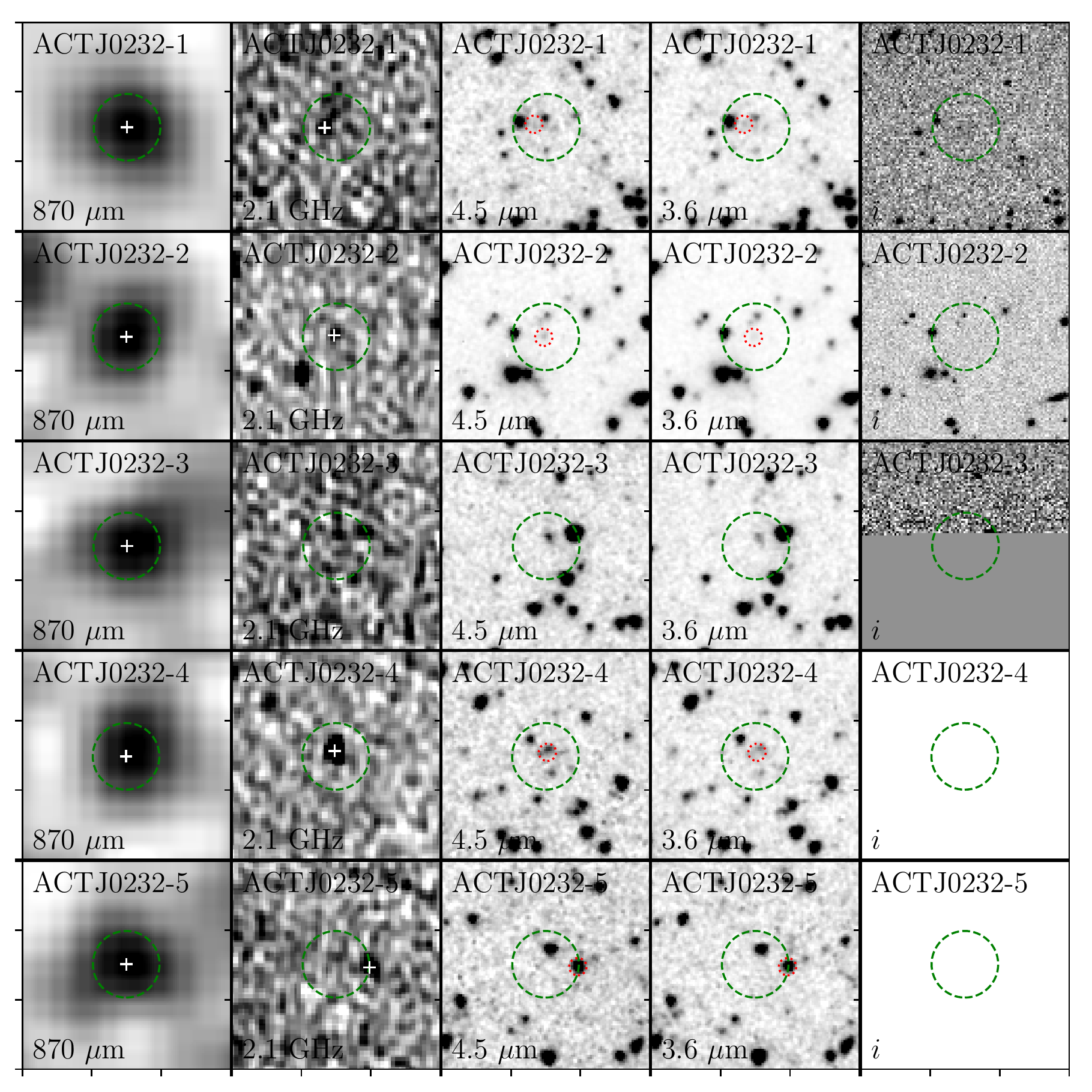}
\caption{Counterpart identification for SMGs detected in the field of cluster ACTJ0232.} \label{fig:count0232}
\end{figure}

\begin{figure}
\centering
\includegraphics[width=10cm]{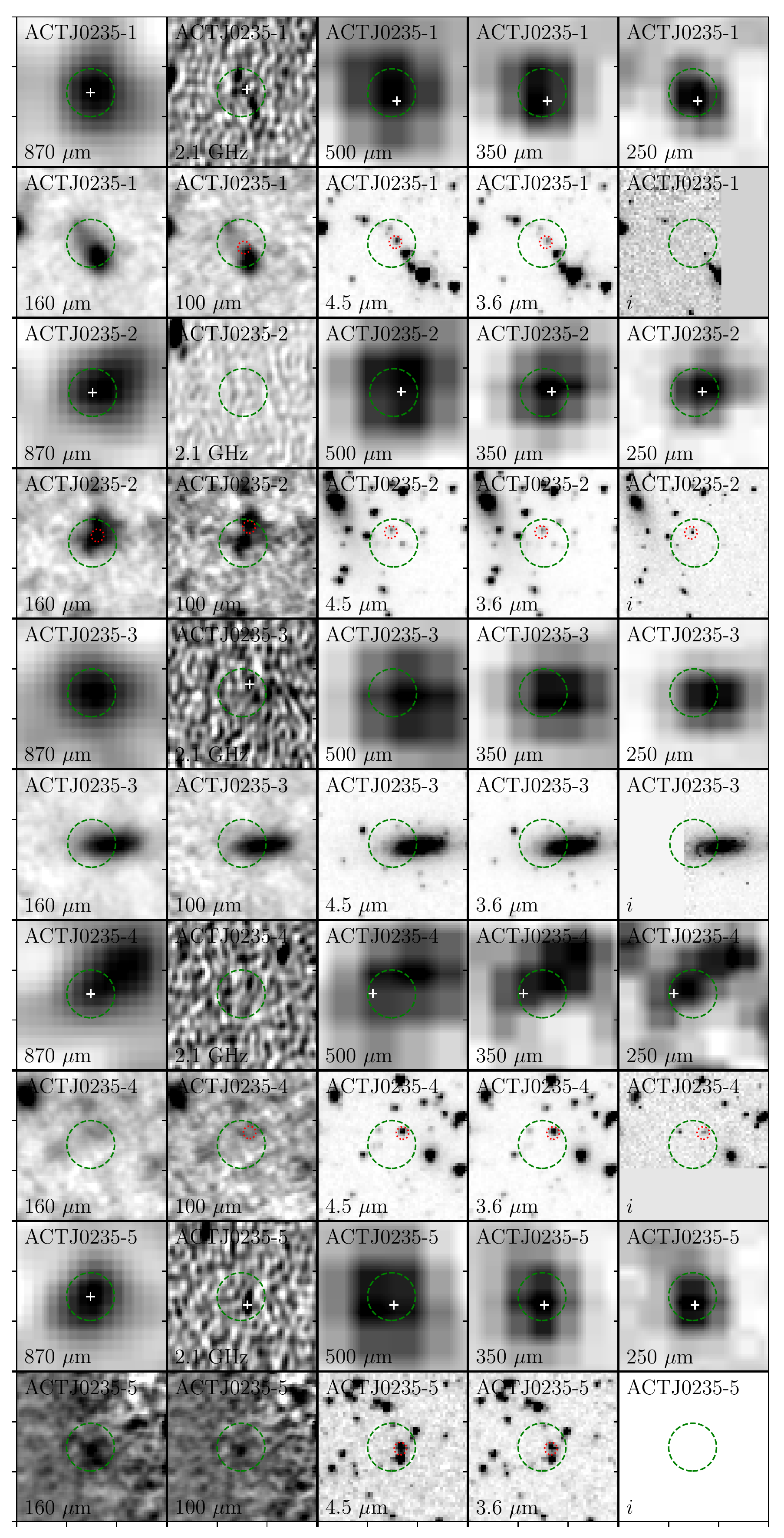}
\caption{Counterpart identification for SMGs detected in the field of cluster ACTJ0235.} \label{fig:count0235}
\end{figure}

\begin{figure}
\centering
\includegraphics[width=10cm]{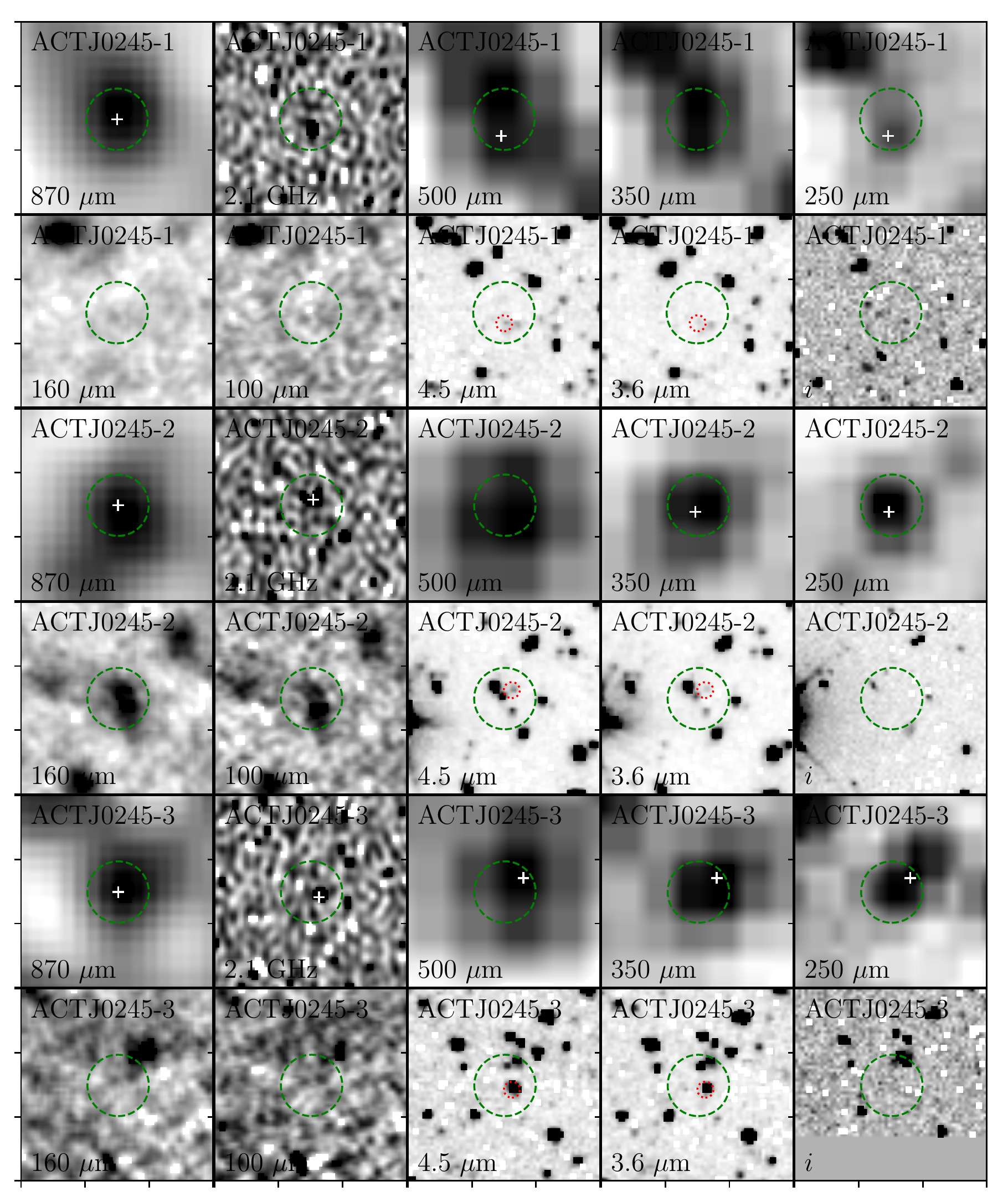}
\caption{Counterpart identification for SMGs detected in the field of cluster ACTJ0245.} \label{fig:count0245}
\end{figure}

\begin{figure}
\centering
\includegraphics[width=10cm]{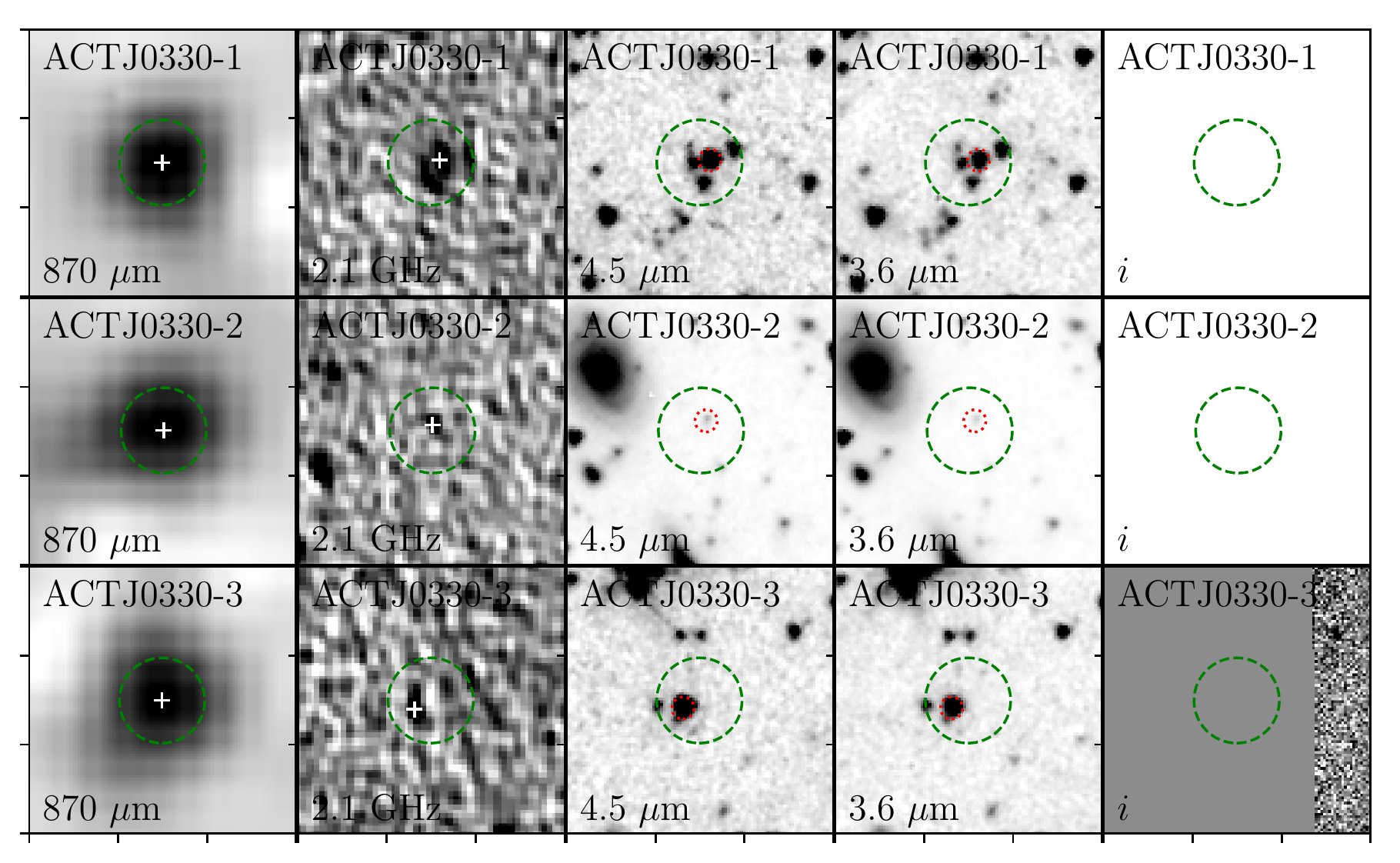}
\caption{Counterpart identification for SMGs detected in the field of cluster ACTJ0330.} \label{fig:count0330}
\end{figure}

\begin{figure}
\centering
\includegraphics[width=10cm]{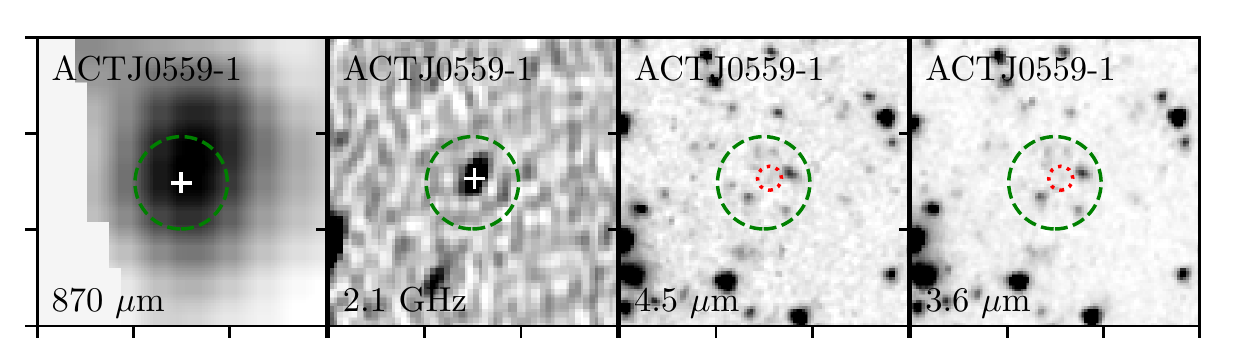}
\caption{Counterpart identification for SMGs detected in the field of cluster ACTJ0559.} \label{fig:count0559}
\end{figure}

\begin{figure}
\centering
\includegraphics[width=10cm]{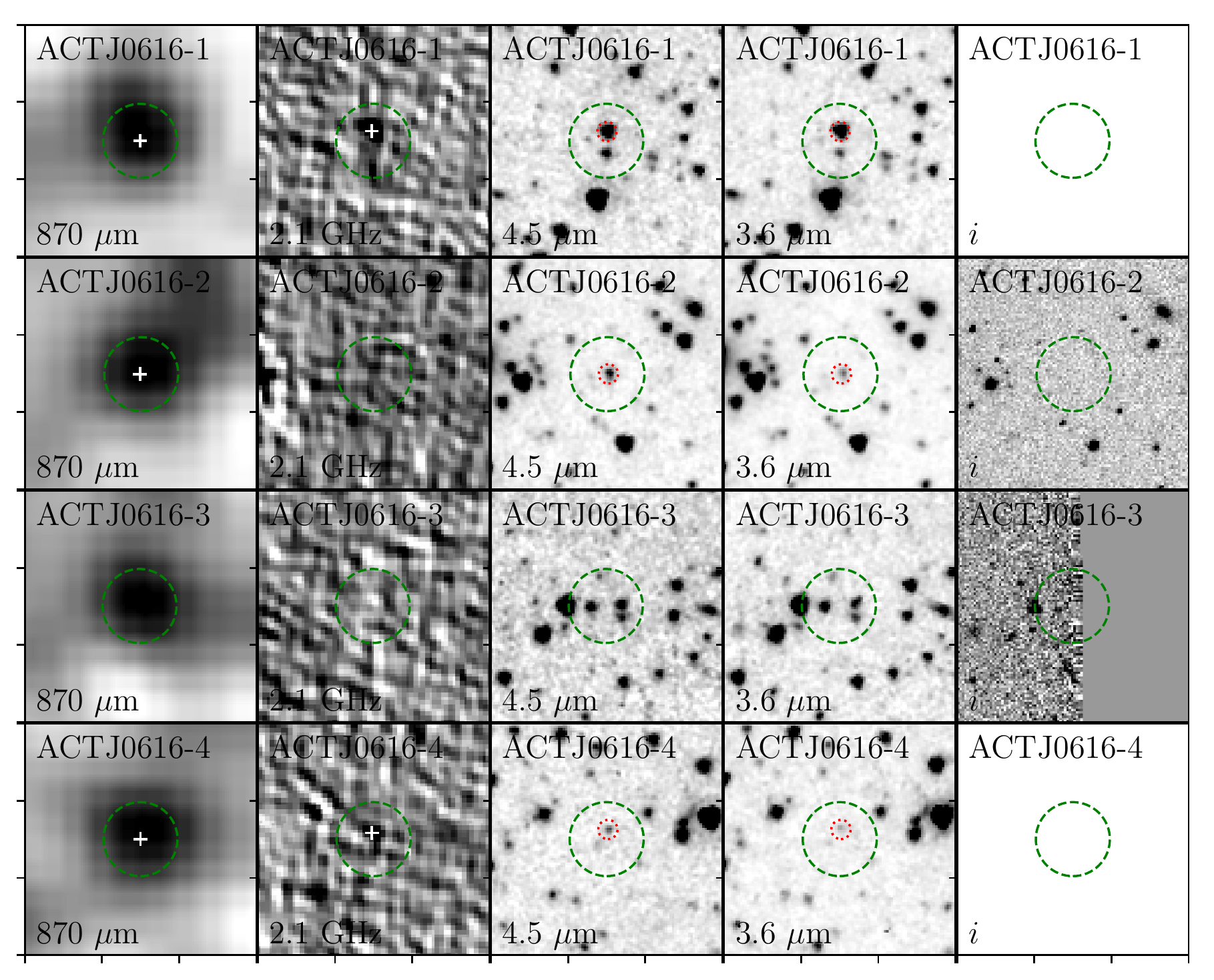}
\caption{Counterpart identification for SMGs detected in the field of cluster ACTJ0616.} \label{fig:count0616}
\end{figure}

In the following paragraphs, we report some relevant details on counterpart identification, SED modeling and estimation of photometric redshifts for each SMG in the LASCAR catalog. In Figures \ref{fig:count0102} to \ref{fig:count0616} we present 30$ ^{\prime\prime}$$\times$30\arcsec\, multi-wavelength postage stamps at the locations of the 870 \micron\, sources, and in Table \ref{tab:counterparts} we list the corresponding multi-wavelength flux densities. Finally, in Figure \ref{fig:seds1} we show SED template fits for sources with robust and tentative counterpart identifications.\\

\noindent {\bf ACTJ0102$-$1:}  This source stands out as the brightest SMG in our catalog, with $S(870)=33.9\pm1.4$ mJy. It lies at a projected cluster-centric radius of 34.82\arcsec, so it is likely magnified by the cluster's gravitational potential. Within the LABOCA beam, we detect a single radio counterpart that has matching emission at 100/4.5/3.6 \micron\, and in the optical. The best-fit SED model suggests $z_{\rm{phot}}\sim4.5$.\\

\noindent {\bf ACTJ0102$-$2:} There are no $4\sigma$ radio counterparts within the LABOCA beam, but the source is detected in SPIRE bands and we identify a tentative IRAC counterpart. The high noise of PACS imaging at the location of this source hinders detection of a 100/160 \micron\, counterpart. Assuming the IRAC match is correct, we estimate $z_{\rm{phot}}\sim4.5$, but deeper radio imaging or high-resolution submillimeter continuum mapping is required to confirm it.\\

\noindent {\bf ACTJ0102$-$3:} This source has a radio/IRAC/optical counterpart and is detected as well in SPIRE bands, but not at 100/160 \micron. The observed SED and upper flux density limits at 100/160 \micron\, are consistent with  $z_{\rm{phot}}\sim4.2$.  \\

\noindent {\bf ACTJ0102$-$4:} We identify a single radio/IRAC counterpart, which is detected in all SPIRE bands. The source is located close to the border of the cluster's PACS and optical imaging, so we are not able to determine the existence of a counterpart at these wavelengths. SED modeling points to $z_{\rm{phot}}\sim2.0$.  \\

\noindent {\bf ACTJ0102$-$5:} This source is detected as a single object at all wavelengths for which data are available; we obtain a good fit from the radio to optical bands for a low $z_{\rm{phot}}\sim1.1$. \\

\noindent {\bf ACTJ0215$-$1:} This source has no radio detection; peaks visible in the 2.1 GHz stamp in Figure \ref{fig:count0215} are consistent with noise. Two IRAC sources are found within the LABOCA search radius, but both have $S_{4.5}/S_{3.6}$ colors inconsistent with SMG templates. We are therefore unable to select a tentative counterpart.\\

\noindent {\bf ACTJ0215$-$2:} There is no radio detection at the $4\sigma$ level, but we identify a single IRAC counterpart candidate within the LABOCA beam. The source lies outside the cluster's optical mapping. With only two tentative photometric points besides the 870 \micron\, measurement, we are unable to estimate a photometric redshift.\\

\noindent {\bf ACTJ0215$-$3:} As for ACTJ0215$-$2, we have no $4\sigma$ radio detection and can only identify a tentative IRAC counterpart that lies close to the 870 \micron\, centroid and has a flux density ratio $S_{4.5}/S_{3.6}=1.1$, consistent with previous SMG near-IR observations.\\

\noindent {\bf ACTJ0215$-$4:} We do not detect a significant radio counterpart, and furthermore we are not able to select a single counterpart candidate among several IRAC/optical sources within the LABOCA beam. Our data are thus insufficient for SED modeling and estimation of a photometric redshift.\\

\noindent {\bf ACTJ0232$-$1:} A radio counterpart matches a very faint IRAC 3.6/4.5 \micron\, source, but  is undetected in the optical. These measurements are consistent with  $z_{\rm{phot}}\sim4.5$.\\

\noindent {\bf ACTJ0232$-$2:} Like ACTJ0232$-$1, this source is detected in radio imaging and faintly in the IRAC bands, but not in optical bands. We obtain a best-fit  $z_{\rm{phot}}\sim2.8$.\\

\noindent {\bf ACTJ0232$-$3:} We see no significant radio detection but identify two IRAC sources located within the LABOCA beam that may constitute a double-component counterpart to the submillimeter detection. We cannot determine the components' individual contributions to the submillimeter counterpart or model their SEDs. \\

\noindent {\bf ACTJ0232$-$4:} This source has a strong radio counterpart that matches a single IRAC 4.5/3.6 \micron\, source. There is no optical coverage at the location of this SMG, but we estimate  $z_{\rm{phot}}\sim1.5$. \\

\noindent {\bf ACTJ0232$-$5:} We detect a single radio source exactly on the edge of the LABOCA beam with $P_C= 0.03$, which is also detected in IRAC bands and has $S_{4.5}/S_{3.6}= 1.33$ consistent with an SMG. We estimate $z_{\rm{phot}}\sim1.7$.\\

\noindent {\bf ACTJ0235$-$1:} Multi-wavelength imaging reveals that this source may correspond to a complex multi-component system. We detect one radio counterpart in the 2.1 GHz $4\sigma$ catalog, but visual inspection of ATCA and IRAC imaging suggests that the submillimeter emission may actually be resolved into three sources aligned in a north-east direction. The northernmost component is on the edge of the LABOCA beam and is seen faintly in radio and IRAC imaging, but it is beneath the detection thresholds of our catalogs. The central component is a significant radio/IRAC detection, while the south-east component is also barely within the LABOCA beam, but is detected at high S/N in PACS and IRAC imaging. We also detect SPIRE 500/350/250 \micron\, emission that appears to be centered closer to the south-east component. At the depth and resolution of our imaging, we are unable to disentangle the submillimeter, radio, and far-IR emission of each individual component, so we cannot estimate the photometric redshift of the SMG. High-resolution submillimeter continuum mapping is required to correctly identify the source of the 870 \micron\, emission and correlate it with sources detected at shorter wavelengths and in radio mapping.\\

\noindent {\bf ACTJ0235$-$2:} There are no radio detections within the LABOCA beam, but the source is detected in all $Herschel$ bands. At 100 \micron\, we detect two sources, which are blended together in the broader beam at 160 \micron. The north 100 \micron\, source matches an IRAC/optical counterpart that has properties consistent with the SMG population and is selected as a tentative counterpart. Since we are unable at this point to disentangle the contribution of the counterpart candidate to the emission measured at 500, 350, 250 and 160 $\mu$m, only the 100 $\mu$m, IRAC and optical flux densities were used in SED modeling, which yields a tentative $z_{\rm{phot}}\sim2.9$. \\

\noindent {\bf ACTJ0235$-$3:} This source coincides with the position of the foreground galaxy ESO 198-G021, which has a redshift of $z=0.018$ as reported in the 6dF Galaxy Survey Database \citep{jones09}; it is bright in the optical and near-IR, but much fainter at 2.1 GHz. The emission in SPIRE, PACS, IRAC, and optical bands is dominated by this galaxy, but we also detect a radio source that lies within the LABOCA beam offset from the centroid of ESO 198$-$G021.\\

\noindent {\bf ACTJ0235$-$4:} No radio counterpart is seen in our ATCA mapping, but we detect emission in all SPIRE bands, at 100 \micron, and in IRAC/optical imaging. We consider this source a plausible counterpart to the SMG detected at 870 \micron; we perform SED modeling and estimate  $z_{\rm{phot}}\sim4.0$.\\

\noindent {\bf ACTJ0235$-$5:} This source has a clear radio counterpart that overlaps two IRAC sources and also matches closely the location of a SPIRE 250/350/500 \micron\, source. Although we also detect a PACS 100/160 \micron\, source within the LABOCA beam, its centroid lies outside the ATCA beam, so it is likely not associated with the 870 \micron/SPIRE/IRAC emission. Both IRAC sources have near-IR colors $S_{4.5}/S_{3.6}\sim1.4$ comparable to previous observations for SMGs \citep{hainline09}, so we are unable to select a single one as the SMG counterpart. If we choose the IRAC galaxy closest to the 2.1 GHz centroid, the estimated redshift is $z_{\rm{phot}}\sim3.7$, but we note that we may be looking as well at a double system. There is no optical coverage at the location of this source.\\

\noindent {\bf ACTJ0245$-$1:} We do not detect a radio or PACS counterpart. However, the source is seen in the SPIRE bands, and we identify an IRAC counterpart candidate that is not seen in the optical imaging. For the tentative IRAC counterpart, we estimate $z_{\rm{phot}}\sim4.0$, which corresponds to an SED fit that is consistent with PACS/radio upper limits.\\

\noindent {\bf ACTJ0245$-$2:} This source is detected in the SPIRE bands and has a radio counterpart that matches an IRAC source, but is undetected in optical. The IRAC counterpart is one of four sources found inside the LABOCA beam, all of which are blended together at the resolution of PACS 100/160 $\mu$m imaging. Hence, although the SMG has detectable 100/160 $\mu$m emission, we are not able to determine its individual flux density due to source blending. We do not include these bands in SED modeling, and obtain a good fit at $z_{\rm{phot}}\sim2.1$ to the 870 $\mu$m, SPIRE, radio, and IRAC photometry.\\

\noindent {\bf ACTJ0245$-$3:} This source has a single radio/SPIRE/IRAC counterpart but is undetected in PACS and optical imaging. Our SED modeling implies redshift $z_{\rm{phot}}\sim3.8$ and is consistent with PACS 100/160 $\mu$m upper  limits derived from non-detections at the $3\sigma$ level.\\

\noindent {\bf ACTJ0330$-$1:} In our ATCA 2.1 GHz catalog we identify a radio counterpart that matches the location of an IRAC source. Assuming these are the radio/near-IR counterparts for our SMG, we estimate $z_{\rm{phot}}\sim3.0$. However, visual inspection of the radio and IRAC imaging shows that the selected counterpart is accompanied by two fainter IRAC sources that may be blended together in our radio mapping, forming an elongated feature that is fainter than the detection limit of our radio catalog. Hence, the detected SMG may correspond to a multiple system; confirmation can only be achieved with high-resolution submillimeter mapping.\\

\noindent {\bf ACTJ0330$-$2:} A single radio counterpart matches the location of the only IRAC source within the LABOCA beam, which we thus identify as the likely near-IR counterpart. SED fits to these four data points give $z_{\rm{phot}}\sim3.2$.\\

\noindent {\bf ACTJ0330$-$3:} As for ACTJ0330$-$2, we identify a single radio/IRAC counterpart and estimate a photometric redshift $z_{\rm{phot}}\sim1.7$.\\

\noindent {\bf ACTJ0438$-$1:} There are no radio or PACS counterparts detected within the LABOCA beam, but we do detect the source in the SPIRE bands. However, the centroid of the SPIRE emission is located close the edge of the beam, and it is unclear whether it is associated with the 870 \micron\, source, or with a PACS 100/160 \micron\, source located just outside the beam. We identify a tentative IRAC counterpart, which is undetected in the optical. Adopting the SPIRE and IRAC counterparts, we obtain a tentative $z_{\rm{phot}}\sim3.4$. \\

\noindent {\bf ACTJ0438$-$2:} This SMG is detected in SPIRE and PACS 100 \micron\, imaging, but not at 2.1 GHz or 160 \micron. The PACS 100 \micron\, detection matches an IRAC counterpart. Our SED fit indicates $z_{\rm{phot}}\sim3.2$ and is consistent with a radio upper flux density limit corresponding to our $4\sigma$ detection threshold.
\\

\noindent {\bf ACTJ0438$-$3:} The source's radio and far-IR emission is lower than the thresholds or our ATCA and $Herschel$ catalogs, so we cannot pinpoint the location of the 870 \micron\, source. We detect three IRAC counterpart candidates within the LABOCA beam but are unable to select one of them as the SMG counterpart. \\

\noindent {\bf ACTJ0438$-$4:} Two radio sources lie close to the search radius; we denote them ACT0438$-$4A (north) and ACT0438$-$4B (south). ACT0438$-$4A is detected in PACS, IRAC, and optical imaging, while ACT0438$-$4B is only detected at 4.5 and 3.6 \micron. There is no SPIRE counterpart to the submillimeter emission.\\

\noindent {\bf ACTJ0438$-$5:} The source is detected in radio imaging, in all SPIRE bands, and at 100 \micron, and based on the precise radio positioning we are able to identify as well an IRAC counterpart. SED modeling of these data result in a $z_{\rm{phot}}\sim3.5$. \\

\noindent {\bf ACTJ0438$-$6:} This source is detected in radio mapping as a double system, and we are able to identify a PACS/IRAC counterpart for each component. We identify the northern radio counterpart as source ACTJ0438$-$6A, which is also detected at 100/4.5/3.6 \micron\, and in the optical. The southern radio source is denoted ACTJ0438$-$6B and is detected in all PACS and IRAC bands. We also detect a SPIRE 500/350/250 \micron\, source that is centered closer to the location of ACTJ0438$-$6B.\\

\noindent {\bf ACTJ0438$-$7:} There are no radio, SPIRE, or PACS counterparts above the defined detection thresholds. Out of three IRAC sources located inside the LABOCA beam, only one has an $S_{4.5}/S_{3.6}$ ratio close to that expected for SMGs, so we single it out as a counterpart candidate. However, accurate positioning at radio or submillimeter wavelengths is required to confirm this possible association.\\

\noindent {\bf ACTJ0438$-$8:} For this SMG we detect two radio counterparts within the LABOCA beam, denoted ACTJ0438$-$8A (west) and ACTJ0438$-$8B (east), which respectively match IRAC sources. We also detect SPIRE and PACS sources, but in these cases the emission from both counterparts is blended together due to the lower spatial resolution compared to radio and near-IR imaging. Only ACTJ0438$-$8A is detected in the optical. Since we are unable to disentangle the contribution of each source to the submillimeter and far-IR emission, we cannot model their SEDs individually.\\

\noindent {\bf ACTJ0546$-$1:} A radio counterpart matches a SPIRE and IRAC source, but is undetected at PACS wavelengths and has no optical imaging coverage. SED fitting based on radio, SPIRE, and IRAC measurements  suggest a high $z_{\rm{phot}}\sim4.9$.\\

\noindent {\bf ACTJ0546$-$2:} This source is detected at all wavelengths except the optical, but at the higher resolutions of the IRAC and optical imaging we find that the radio counterpart may actually encompass at least two of three neighboring galaxies that are blended in the PACS and ATCA beams. Based on the location of the radio source and the optical non-detection, we propose that the source located at the center of this system is the correct SMG counterpart, and in fact we find a good SED fit with $z_{\rm{phot}}\sim1.9$ considering only the submillimeter, SPIRE, and IRAC photometry for the source indicated in Figure \ref{fig:count0546}. However, if we try to fit as well flux densities measured for the PACS 100(160) \micron\, source detected within the LABOCA beam, we are unable to reproduce the observed radio flux density, which further supports the hypothesis that the PACS detections actually correspond to two or more near-IR/optical sources blended by the 7.2\arcsec(12\arcsec) beams.\\

\noindent {\bf ACTJ0546$-$3:} The submillimeter source has a single radio/SPIRE/PACS/IRAC counterpart, but is undetected at optical wavelengths. We infer $z_{\rm{phot}}\sim2.2$.\\

\noindent {\bf ACTJ0546$-$4:} We identify a single radio counterpart that is detected in all $Herschel$ bands except 160 \micron, and in IRAC imaging, but not in the optical. SED best-fit templates are consistent with the 160 \micron\, upper density limit and result in $z_{\rm{phot}}\sim4.5$.\\

\noindent {\bf ACTJ0546$-$5:} A single radio counterpart matches PACS and IRAC emission, but is too faint for significant detection in the optical and does not have a SPIRE counterpart. The best-fit redshift is $z_{\rm{phot}}\sim2.7$.\\

\noindent {\bf ACTJ0546$-$6:} The submillimeter emission has a double radio counterpart aligned in the north-south direction; we denote as ACTJ0546$-$6A the radio source located on the north edge of the LABOCA beam and ACTJ0546$-$6B the southern  source. Both are detected in PACS and IRAC bands, but not in the optical. We also identify a 500/350/250 \micron\, source that likely corresponds to the combined emission from both sources blended together in the SPIRE beams, although its centroid is located closer to ACTJ0546$-$6A.\\

\noindent {\bf ACTJ0546$-$7:} The location of this source falls outside PACS and optical imaging, but we identify a single radio/SPIRE/IRAC counterpart. We derive $z_{\rm{phot}}\sim3.2$. \\

\noindent {\bf ACTJ0546$-$8:}  The source has no radio counterpart but is detected in SPIRE bands and at 100 \micron. However, the PACS counterpart does not match the location of any of the IRAC/optical sources found within the LABOCA beam, so we cannot identify a candidate counterpart in these bands. Given the limited number of photometric data points, we can venture an estimate $z_{\rm{phot}}\sim2.3$, but the SED fit is very poorly constrained towards the blue extreme. \\

\noindent {\bf ACTJ0546$-$9:} We detect a single radio counterpart that matches 100 \micron\, and IRAC detections but is not recovered in our 160 \micron\, or optical catalogs. We estimate its photometric redshift is $z_{\rm{phot}}\sim2.1$. \\

\noindent {\bf ACTJ0546$-$10:} We identify a radio counterpart close to the edge of the LABOCA beam that matches an IRAC source. Although we detect some 100/160 \micron\, emission, this may be be associated to a SPIRE source located just outside the LABOCA beam rather than to the radio source. The observed photometry matches the SED of an SMG at $z_{\rm{phot}}\sim3.2$.\\

\noindent {\bf ACTJ0546$-$11:}  We detect a single radio counterpart that matches an IRAC source, but there are no significant SPIRE detections within the LABOCA beam. This SMG lies close to the edge of the PACS image for the cluster ACTJ0546, and we are thus unable to determine the existence of a 100/160 \micron\, match due to increased noise. Optical imaging does not cover the source's location either. Observed photometry is consistent with SED templates of SMGs at $z_{\rm{phot}}\sim1.9$.\\

\noindent {\bf ACTJ0559$-$1:} We identify a strong radio source that matches the location of a single faint IRAC source. There is no optical coverage at the source's location. Based on this limited photometry, this SMG is expected to lie at $z_{\rm{phot}}\sim2.9$. \\

\noindent {\bf ACTJ0616$-$1:} This source has a radio counterpart that accurately matches the position of an IRAC source, but the source's location is not covered by our optical imaging. We estimate a photometric redshift $z_{\rm{phot}}\sim1.1$.\\

\noindent {\bf ACTJ0616$-$2:} There is no $4\sigma$ radio detection, but we identify a tentative IRAC counterpart with $S_{4.5}/S_{3.6} \sim1.46$, consistent with the SMG population. These data are insufficient for SED modeling.\\

\noindent {\bf ACTJ0616$-$3:} The source is undetected at 2.1 GHz; we detect several IRAC/optical sources within the LABOCA beam, but are unable to select a probable counterpart.\\

\noindent {\bf ACTJ0616$-$4:} This source has a single radio/IRAC counterpart, and SED fitting to the available data points allows us to estimate $z_{\rm{phot}}\sim3.9$.\\

 \clearpage

%FIGURE 8

 \clearpage
 
 %FIGURE 9
\clearpage

 %FIGURE 10

 \clearpage
 
 %FIGURE 11

 \clearpage
 
 %FIGURE 12

 \clearpage
 
 %FIGURE 13

 \clearpage
 
 %FIGURE 14A
 \begin{figure}
\centering
\includegraphics[width=10cm]{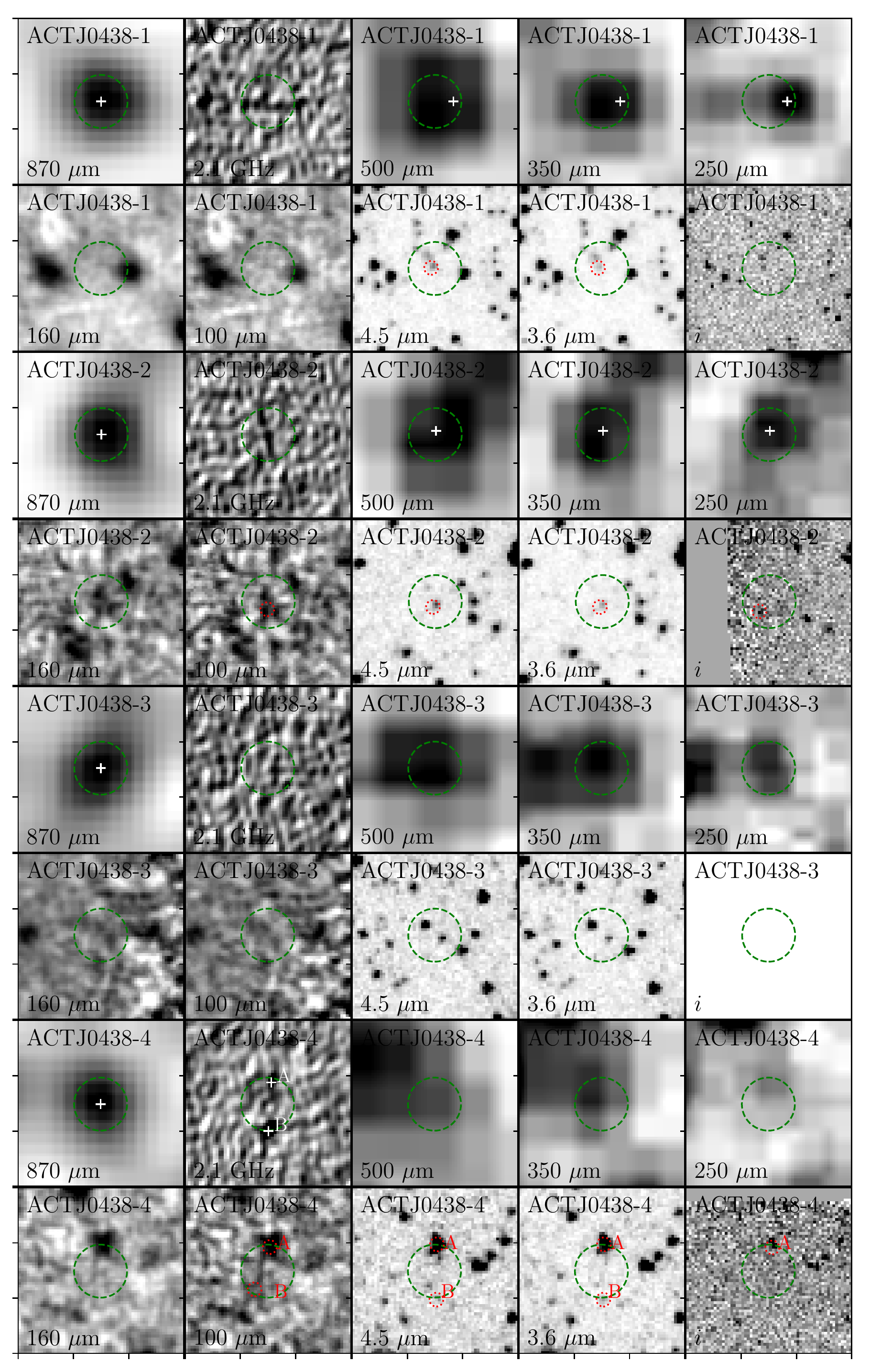}
\caption{Counterpart identification for SMGs detected in the field of cluster ACTJ0438.} \label{fig:count0438}
\end{figure}

 \clearpage
 
 %FIGURE 14B
 \begin{figure}
 \figurenum{14}
\centering
\includegraphics[width=10cm]{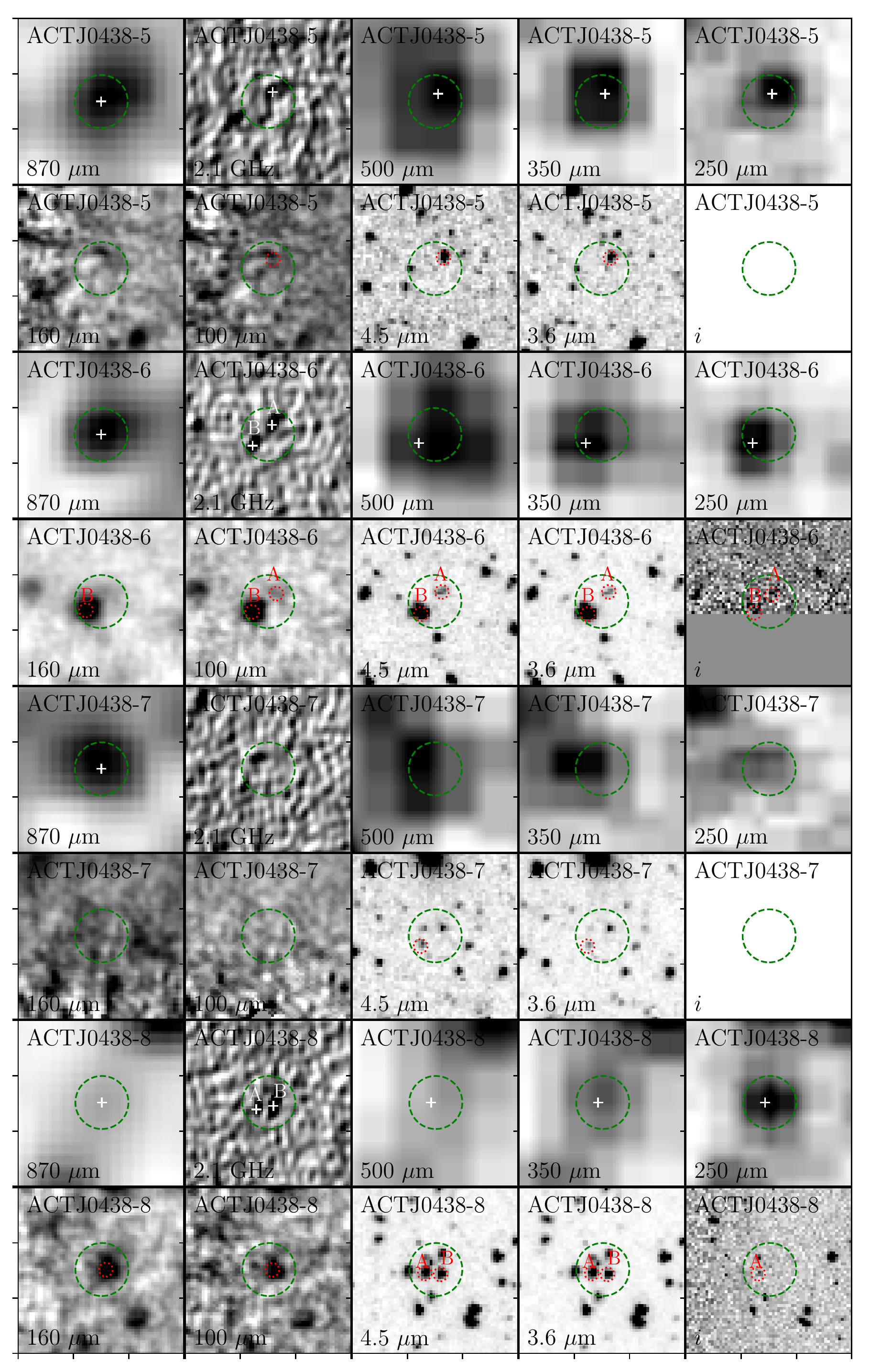}
\caption{(Continued). Counterpart identification for SMGs detected in the field of cluster ACTJ0438.} \label{fig:count0438}
\end{figure}

 \clearpage

%FIGURE 15A
\begin{figure}
\centering
\includegraphics[width=10cm]{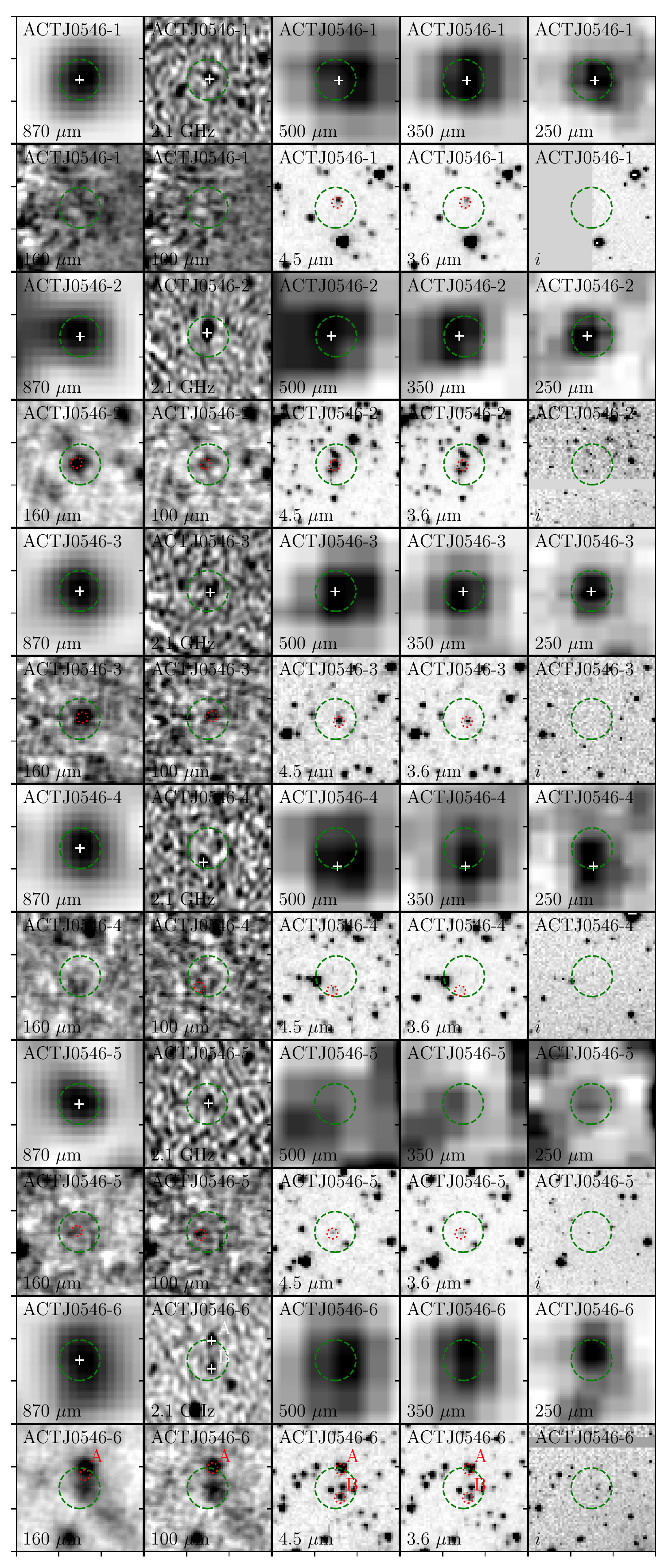}
\caption{Counterpart identification for SMGs detected in the field of cluster ACTJ0546.} \label{fig:count0546}
\end{figure}
\clearpage

%FIGURE 15B
\begin{figure}
\figurenum{15}
\centering
\includegraphics[width=10cm]{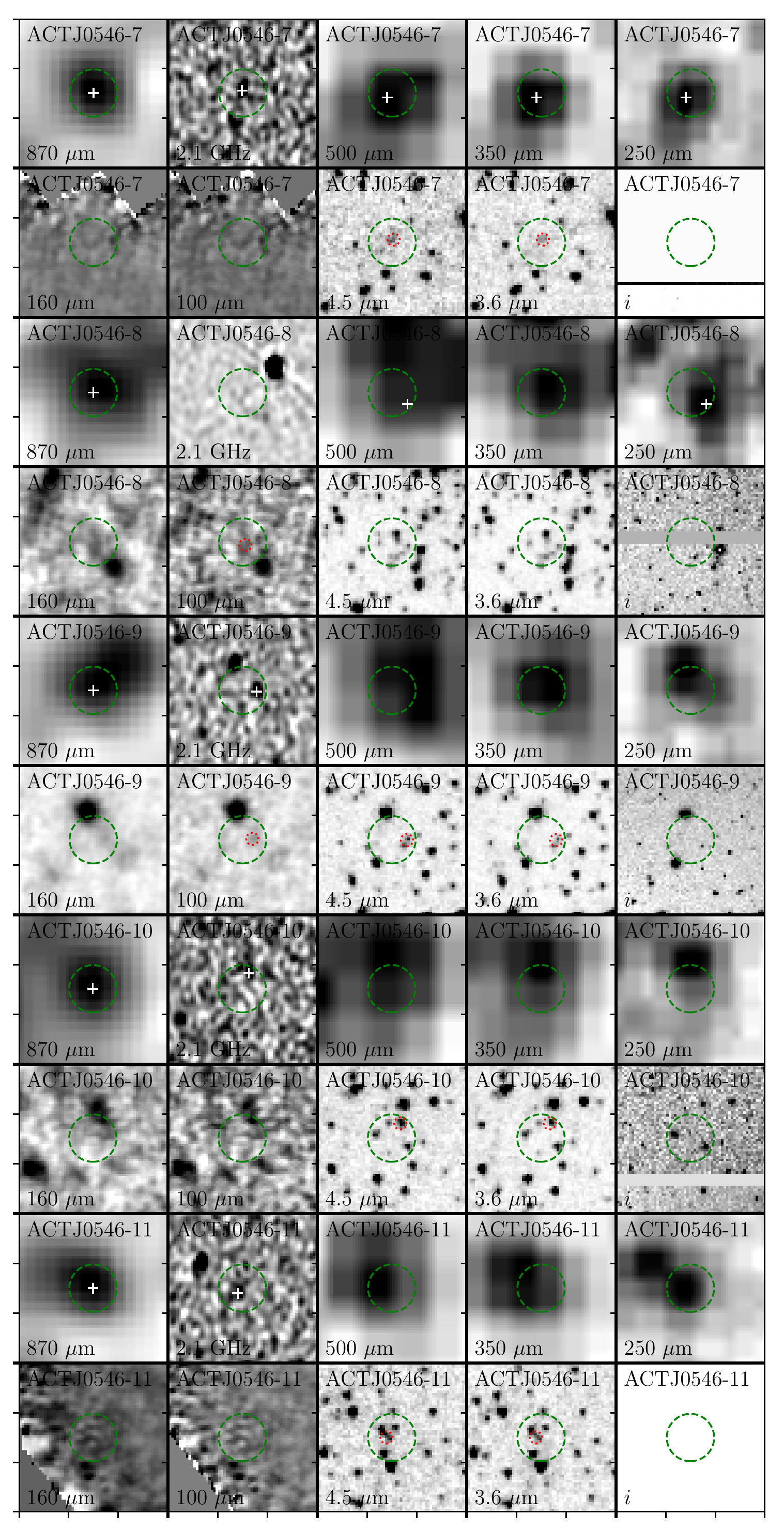}
\caption{(Continued) Counterpart identification for SMGs detected in the field of cluster ACTJ0546.} \label{fig:count0546}
\end{figure}

 \clearpage
 
 %FIGURE 16 UPDATED

 %FIGURE 17  UPDATED

 \clearpage
 
 %FIGURE 18A
 \begin{figure}
\centering
\includegraphics[width=13cm]{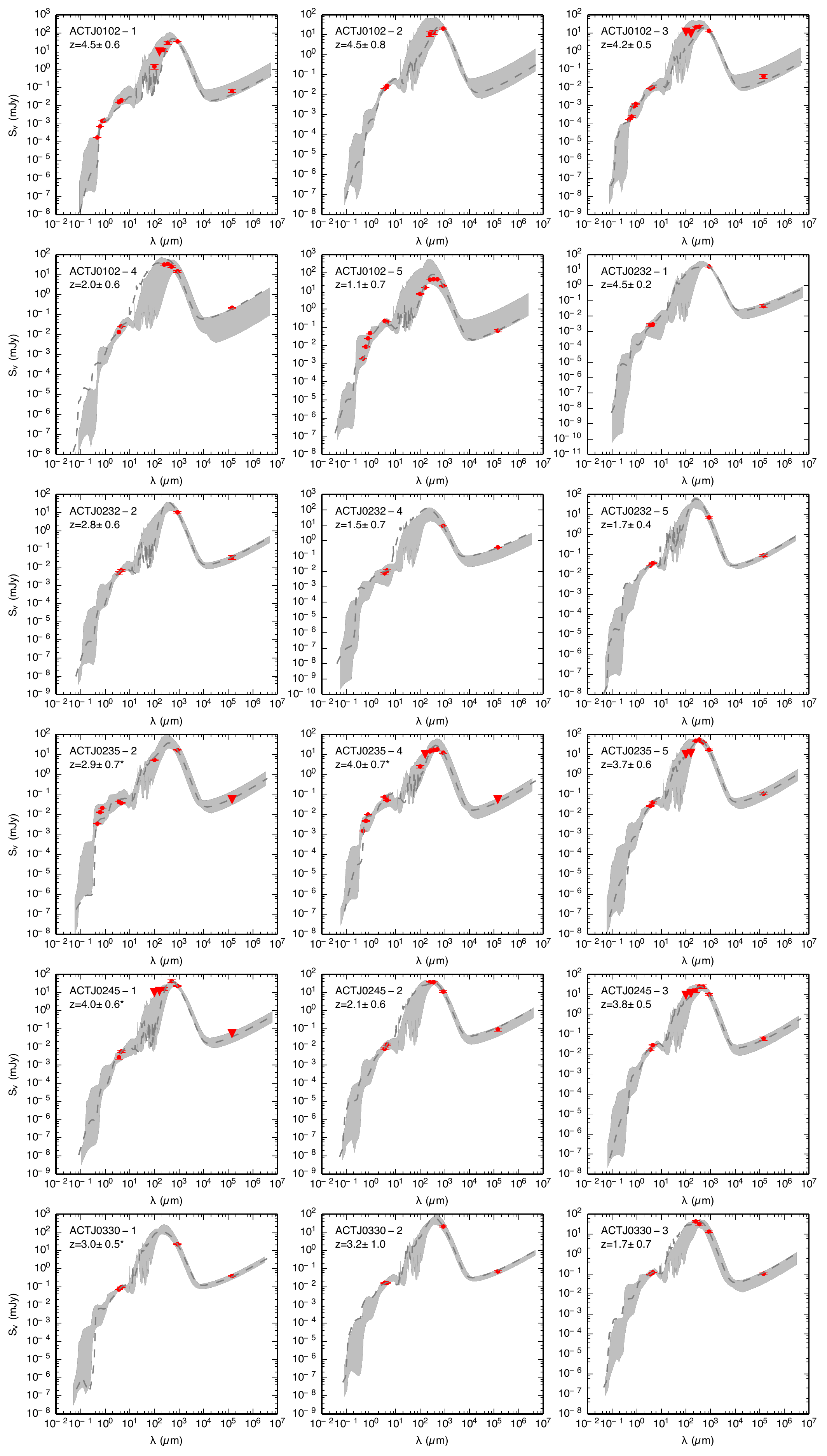}

\caption{SED fits for LASCAR SMGs with secure or tentative counterpart identifications. For each source we plot the observed flux densities (red dots and error bars), flux density upper limits when appropriate (red downward triangles), the absolute best-fit SED, and the range of redshifted templates (shaded grey) that provide similarly good fits to the observational data (i.e., models for which the resulting $\chi^2$ values are within $10\%$ of their lowest
values). The annotations in each panel indicated the estimated photometric redshift and error; values marked with (*) indicate tentative counterpart matches and SED fits.} \label{fig:seds1}
\end{figure}

 \clearpage
 
 %FIGURE 18B
 \begin{figure}
\figurenum{18}
\centering
\includegraphics[width=13cm]{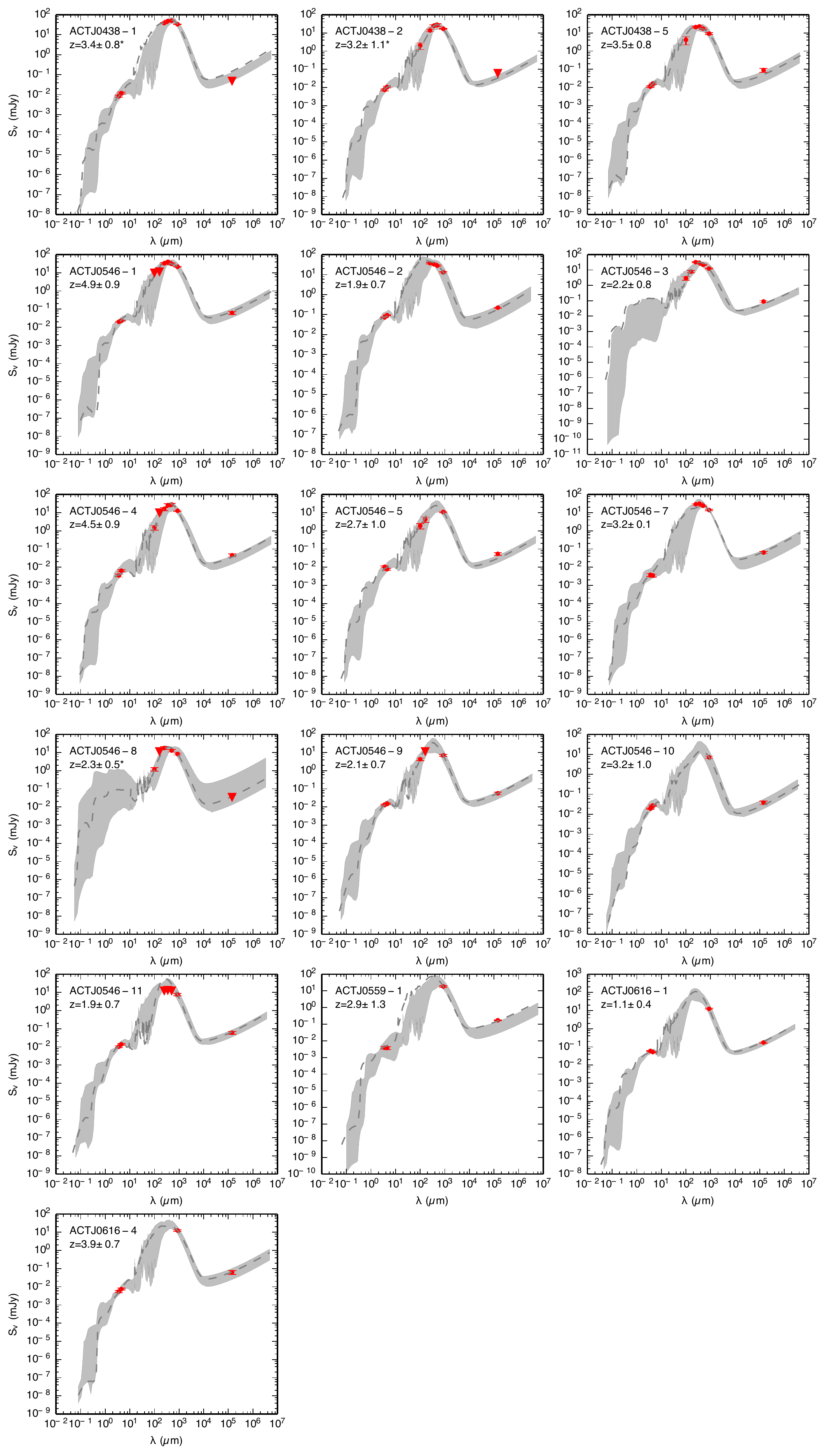}
\caption{(Continued) SED fits for LASCAR SMGs with secure or tentative counterpart identifications. }
 \end{figure}

 \clearpage
 
 %%TABLE 7
\begin{longrotatetable}
\begin{deluxetable}{lcccccccccccccccccc}
\raggedleft
%\tabletypesize{8pt}

\tablecolumns{19}
\tablewidth{700pt}
\tablecaption{ Radio, FIR, near-IR, and optical counterparts for SMGs detected in LASCAR clusters. } 
\tablehead{
\colhead{(1)}  &  \colhead{(2)}  &  \colhead{(3)} & \colhead{(4)} & \colhead{(5)} & \colhead{(6)} & \colhead{(7)} &  \colhead{(8)} &  \colhead{(9)} & \colhead{(10)} & \colhead{(11)} &\colhead{(12)} &\colhead{(13)} & \colhead{(14)} & \colhead{(15)}  &\colhead{(16)} & \colhead{(17)} & \colhead{(18)} & \colhead{(19)} \\ \\
\colhead{Source}  &  \colhead{$S_{2.1}$}  &  \colhead{$\Delta S_{2.1}$} & \colhead{$S_{500}$} & \colhead{$\Delta S_{500}$} & \colhead{$S_{350}$} & \colhead{$\Delta S_{350}$} &  \colhead{$S_{250}$} &  \colhead{$\Delta S_{250}$} &\colhead{$S_{160}$} & \colhead{$\Delta S_{160}$} &\colhead{$S_{100}$} &\colhead{$\Delta S_{100}$} & \colhead{$S_{4.5}$} & \colhead{$\Delta S_{4.5}$}  &\colhead{$S_{3.6}$} & \colhead{$\Delta S_{3.6}$} & \colhead{$i_{\rm{AB}}$} & \colhead{Comment}
\label{tab:counterparts}}
 
 % &  &  \colhead{($\mu$Jy)} & \colhead{(mJy)} & \colhead{(mJy)} & \colhead{(mJy)} & \colhead{(mJy)} & \colhead{(mJy)} & \colhead{(mJy)} & \colhead{(mJy)} & \colhead{(mJy)} 

\startdata
ACTJ0102$-$1 	&	63.0	&	10.0	&	 \nodata	&	 \nodata	&	28.6	&	6.0	&	11.4	&	2.0	&	 \nodata	&	 \nodata	&	1.4	&	0.4	&	19.57	&	0.02	&	15.46	&	0.01	&	25.78	&	S	\\
ACTJ0102$-$2 	&	 \nodata	&	 \nodata	&	13.5	&	3.0	&	13.2	&	3.0	&	11.0	&	3.0	&	 \nodata	&	 \nodata	&	 \nodata	&	 \nodata	&	27.62	&	0.02	&	20.10	&	0.02	&	 \nodata	&	T	\\
ACTJ0102$-$3 	&	41.0	&	10.0	&	19.1	&	2.0	&	21.7	&	3.0	&	20.2	&	2.0	&	 \nodata	&	 \nodata	&	 \nodata	&	 \nodata	&	10.2	&	0.02	&	8.72	&	0.01	&	25.85	&	S	\\
ACTJ0102$-$4 	&	219.0	&	11.0	&	25	&	2.0	&	34	&	3.0	&	31.7	&	2.0	&	 \nodata	&	 \nodata	&	 \nodata	&	 \nodata	&	25.7	&	0.02	&	13.30	&	0.01	&	 \nodata	&	S	\\
ACTJ0102$-$5 	&	65.0	&	10.0	&	43.9	&	2.0	&	44.5	&	2.0	&	42.0	&	2.0	&	15.4	&	0.6	&	6.9	&	0.4	&	198.1	&	0.05	&	225.90	&	0.05	&	23.2	&	S	\\
ACTJ0215$-$1 	&	 \nodata	&	 \nodata	&	 \nodata	&	 \nodata	&	 \nodata	&	 \nodata	&	 \nodata	&	 \nodata	&	 \nodata	&	 \nodata	&	 \nodata	&	 \nodata	&	 \nodata	&	 \nodata	&	 \nodata	&	 \nodata	&	 \nodata	&	N	\\
ACTJ0215$-$2 	&	\nodata	&	 \nodata	&	 \nodata	&	 \nodata	&	 \nodata	&	 \nodata	&	 \nodata	&	 \nodata	&	 \nodata	&	 \nodata	&	 \nodata	&	 \nodata	&	38.69	&	0.02	&	30.30	&	0.02	&	 \nodata	&	T	\\
ACTJ0215$-$3 	&	\nodata	&	 \nodata	&	 \nodata	&	 \nodata	&	 \nodata	&	 \nodata	&	 \nodata	&	 \nodata	&	 \nodata	&	 \nodata	&	 \nodata	&	 \nodata	&	22.1	&	0.02	&	16.76	&	0.01	&	 \nodata	&	T	\\
ACTJ0232$-$1 	&	45.0	&	11.0	&	 \nodata	&	 \nodata	&	 \nodata	&	 \nodata	&	 \nodata	&	 \nodata	&	 \nodata	&	 \nodata	&	 \nodata	&	 \nodata	&	2.89	&	0.01	&	2.61	&	0.01	&	 \nodata	&	S	\\
ACTJ0232$-$2 	&	36.0	&	9.0	&	 \nodata	&	 \nodata	&	 \nodata	&	 \nodata	&	 \nodata	&	 \nodata	&	 \nodata	&	 \nodata	&	 \nodata	&	 \nodata	&	6.63	&	0.01	&	4.87	&	0.01	&	 \nodata	&	S	\\
ACTJ0232$-$4 	&	366.0	&	11.0	&	 \nodata	&	 \nodata	&	 \nodata	&	 \nodata	&	 \nodata	&	 \nodata	&	 \nodata	&	 \nodata	&	 \nodata	&	 \nodata	&	11.44	&	0.01	&	7.52	&	0.01	&	 \nodata	&	S	\\
ACTJ0232$-$5 	&	89.0	&	11.0	&	 \nodata	&	 \nodata	&	 \nodata	&	 \nodata	&	 \nodata	&	 \nodata	&	 \nodata	&	 \nodata	&	 \nodata	&	 \nodata	&	36.75	&	0.02	&	27.77	&	0.02	&	 \nodata	&	S	\\
ACTJ0235$-$1 	&	135.0	&	21.0	&	43.2	&	2.0	&	61.5	&	3.0	&	68.7	&	3.0	&	 \nodata	&	 \nodata	&	2.3	&	0.1	&	36.44	&	0.02	&	26.18	&	0.02	&	 \nodata	&	T	\\
ACTJ0235$-$2 	&\nodata	&	 \nodata	&	\nodata	&	\nodata	&	\nodata	&	\nodata	&	\nodata&	\nodata	&	\nodata	&	\nodata	&	5.3	&	0.4	&	36.17	&	0.02	&	44.63	&	0.02	&	22.57	&	T	\\
ACTJ0235$-$3 	&49.0	&	12.0	&	 \nodata	&	 \nodata	&	46.1	&	1.0	&	126.8	&	3.0	&	95.8	&	1.7	&	24.1	&	0.7	&	262.3	&	0.06	&	403.60	&	0.07	&	15.83	&	F	\\
ACTJ0235$-$4 	&\nodata	&	 \nodata	&	17.7	&	3.0	&	16.9	&	3.0	&	14.1	&	2.0	&	 \nodata	&	 \nodata	&	2.5	&	0.4	&	50.63	&	0.02	&	73.65	&	0.03	&	23.47	&	T	\\
ACTJ0235$-$5 	&107.0	&	14.0	&	39.6	&	2.0	&	54.8	&	3.0	&	47.7	&	3.0	&	 \nodata	&	 \nodata	&	 \nodata	&	 \nodata	&	39.48	&	0.02	&	26.01	&	0.02	&	 \nodata	&	T	\\
ACTJ0245$-$1 	&\nodata	&	\nodata	&	41.8	&	7.0	&	 \nodata	&	 \nodata	&	15.5	&	3.0	&	 \nodata	&	 \nodata	&	 \nodata	&	 \nodata	&	5.67	&	0.01	&	2.67	&	0.01	&	 \nodata	&	T	\\
ACTJ0245$-$2 	&92.0	&	18.0	&	 \nodata	&	 \nodata	&	36.3	&	3.0	&	36.7	&	3.0	&	 \nodata	&	 \nodata	&	 \nodata	&	 \nodata	&	13.44	&	0.01	&	7.76	&	0.01	&	 \nodata	&	S	\\
ACTJ0245$-$3 	&61.0	&	12.0	&	24.2	&	5.0	&	24.7	&	5.0	&	15.6	&	3.0	&	 \nodata	&	 \nodata	&	 \nodata	&	 \nodata	&	28.16	&	0.02	&	17.08	&	0.01	&	 \nodata	&	S	\\
ACTJ0330$-$1 	&189.0	&	19.0	&	 \nodata	&	 \nodata	&	 \nodata	&	 \nodata	&	 \nodata	&	 \nodata	&	 \nodata	&	 \nodata	&	 \nodata	&	 \nodata	&	95.32	&	0.03	&	70.73	&	0.03	&	 \nodata	&	T	\\
ACTJ0330$-$2 	&70.0	&	13.0	&	 \nodata	&	 \nodata	&	 \nodata	&	 \nodata	&	 \nodata	&	 \nodata	&	 \nodata	&	 \nodata	&	 \nodata	&	 \nodata	&	16.22	&	0.01	&	16.60	&	0.01	&	 \nodata	&	S	\\
ACTJ0330$-$3 	&104.0	&	12.0	&	 \nodata	&	 \nodata	&	 \nodata	&	 \nodata	&	 \nodata	&	 \nodata	&	 \nodata	&	 \nodata	&	 \nodata	&	 \nodata	&	122	&	0.04	&	98.45	&	0.03	&	 \nodata	&	S	\\
ACTJ0438$-$1 	&\nodata	&	 \nodata	&	50.3	&	4.0	&	48.5	&	4.0	&	38.2	&	3.0	&	 \nodata	&	 \nodata	&	 \nodata	&	 \nodata	&	12.06	&	0.01	&	8.37	&	0.01	&	 \nodata	&	T	\\
ACTJ0438$-$2 	&\nodata	&	 \nodata	&	27.3	&	5.0	&	24	&	4.0	&	13.7	&	2.0	&	 \nodata	&	 \nodata	&	2	&	0.8	&	10.62	&	0.01	&	7.37	&	0.01	&	25.54	&	T	\\
ACTJ0438$-$4A &	57.0	&	14.0	&	 \nodata	&	 \nodata	&	 \nodata	&	 \nodata	&	 \nodata	&	 \nodata	&	 \nodata	&	 \nodata	&	4.3	&	0.8	&	57.02	&	0.03	&	58.24	&	0.03	&	21.89	&	D	\\
ACTJ0438$-$4B &	171.0	&	14.0	&	 \nodata	&	 \nodata	&	 \nodata	&	 \nodata	&	 \nodata	&	 \nodata	&	 \nodata	&	 \nodata	&	 \nodata	&	 \nodata	&	2.69	&	0.01	&	3.25	&	0.01	&	 \nodata	&	D	\\
ACTJ0438$-$5 	&90.0	&	22.0	&	17.2	&	2.0	&	22.7	&	3.0	&	20.9	&	2.0	&	 \nodata	&	 \nodata	&	4.2	&	2.0	&	16.11	&	0.00	&	11.05	&	0.01	&	 \nodata	&	S	\\
ACTJ0438$-$6A &	119.0	&	16.0	&	 \nodata	&	 \nodata	&	 \nodata	&	 \nodata	&	 \nodata	&	 \nodata	&	 \nodata	&	 \nodata	&	 \nodata	&	 \nodata	&	11.2	&	0.11	&	10.17	&	0.11	&	25.18	&	D	\\
ACTJ0438$-$6B &	110.0	&	13.0	&	6.7	&	0.7	&	29.4	&	1.0	&	49.7	&	2.0	&	30.4	&	2.8	&	16.4	&	1.0	&	105.4	&	0.04	&	154.70	&	0.04	&	23.17	&	D	\\
ACTJ0438$-$7 	&\nodata	&	 \nodata	&	 \nodata	&	 \nodata	&	 \nodata	&	 \nodata	&	 \nodata	&	 \nodata	&	 \nodata	&	 \nodata	&	 \nodata	&	 \nodata	&	8.34	&	0.01	&	6.63	&	0.01	&	 \nodata	&	T	\\
ACTJ0438$-$8A&	49.0	&	12.0	&	16.3	&	2.0	&	21.4	&	3.0	&	22.4	&	3.0	&	8.4	&	2.8	&	4.9	&	1.0	&	61.77	&	2.72	&	69.76	&	2.82	&	22.9	&	D	\\
ACTJ0438$-$8B&	158.0	&	14.0	&	 \nodata	&	 \nodata	&	 \nodata	&	 \nodata	&	 \nodata	&	 \nodata	&	 \nodata	&	 \nodata	&	 \nodata	&	 \nodata	&	77.84	&	0.03	&	74.20	&	0.03	&	 \nodata	&	D	\\
ACTJ0546$-$1 	&62.0	&	11.0	&	30.8	&	3.0	&	40.1	&	3.0	&	32.7	&	3.0	&	 \nodata	&	 \nodata	&	 \nodata	&	 \nodata	&	23.1	&	0.02	&	20.10	&	0.02	&	 \nodata	&	S	\\
ACTJ0546$-$2 	&231.0	&	10.0	&	27.7	&	2.0	&	33.8	&	3.0	&	37.3	&	3.0	&	8.4	&	0.9	&	2.7	&	0.5	&	93.15	&	0.03	&	71.85	&	0.03	&	 \nodata	&	S	\\
ACTJ0546$-$3 	&89.0	&	13.0	&	21.4	&	2.0	&	29.7	&	2.0	&	32.1	&	3.0	&	7.8	&	1.3	&	2.9	&	0.9	&	22.43	&	0.02	&	14.82	&	0.01	&	 \nodata	&	S	\\
ACTJ0546$-$4 	&45.0	&	8.0	&	28.0	&	5.0	&	25.9	&	5.0	&	15.9	&	3.0	&	 \nodata	&	 \nodata	&	1.5	&	0.4	&	6.39	&	0.01	&	3.51	&	0.01	&	 \nodata	&	S	\\
ACTJ0546$-$5 	&54.0	&	9.0	&	 \nodata	&	 \nodata	&	 \nodata	&	 \nodata	&	 \nodata	&	 \nodata	&	4.3	&	1.3	&	1.9	&	0.6	&	7.73	&	0.01	&	10.43	&	0.01	&	 \nodata	&	S	\\
ACTJ0546$-$6A &	91.0	&	10.0	&	 \nodata	&	 \nodata	&	 \nodata	&	 \nodata	&	 \nodata	&	 \nodata	&	22.3	&	1.0	&	6	&	0.4	&	84.64	&	3.18	&	68.61	&	2.79	&	 \nodata	&	D	\\
ACTJ0546$-$6B &	81.0	&	9.0	&	 \nodata	&	 \nodata	&	 \nodata	&	 \nodata	&	 \nodata	&	 \nodata	&	 \nodata	&	 \nodata	&	 \nodata	&	 \nodata	&	37.26	&	0.02	&	32.93	&	0.02	&	 \nodata	&	D	\\
ACTJ0546$-$7 	&64.0	&	9.0	&	24	&	2.0	&	31.1	&	3.0	&	28.9	&	3.0	&	 \nodata	&	 \nodata	&	 \nodata	&	 \nodata	&	3.43	&	0.01	&	3.82	&	0.01	&	 \nodata	&	S	\\
ACTJ0546$-$8 	&\nodata	&	 \nodata	&	12.8	&	2.0	&	 \nodata	&	 \nodata	&	17.7	&	3.0	&	 \nodata	&	 \nodata	&	1.2	&	0.3	&	 \nodata	&	 \nodata	&	 \nodata	&	 \nodata	&	 \nodata	&	T	\\
ACTJ0546$-$9 	&57.0	&	9.0	&	 \nodata	&	 \nodata	&	 \nodata	&	 \nodata	&	 \nodata	&	 \nodata	&	 \nodata	&	 \nodata	&	4.3	&	0.7	&	15.47	&	0.01	&	13.12	&	0.01	&	 \nodata	&	S	\\
ACTJ0546$-$10 &38.0	&	8.0	&	 \nodata	&	 \nodata	&	 \nodata	&	 \nodata	&	 \nodata	&	 \nodata	&	 \nodata	&	 \nodata	&	 \nodata	&	 \nodata	&	27.57	&	0.02	&	19.27	&	0.01	&	 \nodata	&	S	\\
ACTJ0546$-$11 &60.0	&	11.0	&	 \nodata	&	 \nodata	&	 \nodata	&	 \nodata	&	 \nodata	&	 \nodata	&	 \nodata	&	 \nodata	&	 \nodata	&	 \nodata	&	13.85	&	0.01	&	10.26	&	0.01	&	 \nodata	&	S	\\
ACTJ0559$-$1 	&172.0	&	10.0	&	 \nodata	&	 \nodata	&	 \nodata	&	 \nodata	&	 \nodata	&	 \nodata	&	 \nodata	&	 \nodata	&	 \nodata	&	 \nodata	&	3.71	&	0.01	&	3.84	&	0.01	&	 \nodata	&	S	\\
ACTJ0616$-$1 	&175.0	&	14.0	&	 \nodata	&	 \nodata	&	 \nodata	&	 \nodata	&	 \nodata	&	 \nodata	&	 \nodata	&	 \nodata	&	 \nodata	&	 \nodata	&	51.9	&	0.02	&	59.81	&	0.03	&	 \nodata	&	S	\\
ACTJ0616$-$2 	&\nodata	&	 \nodata	&	 \nodata	&	 \nodata	&	 \nodata	&	 \nodata	&	 \nodata	&	 \nodata	&	 \nodata	&	 \nodata	&	 \nodata	&	 \nodata	&	21.03	&	0.21	&	14.41	&	0.14	&	 \nodata	&	T	\\
ACTJ0616$-$4 	&63.0	&	15.0	&	 \nodata	&	 \nodata	&	 \nodata	&	 \nodata	&	 \nodata	&	 \nodata	&	 \nodata	&	 \nodata	&	 \nodata	&	 \nodata	&	7.5	&	0.01	&	5.53	&	0.01	&	 \nodata	&	S	\\
\enddata
\tablecomments{Columns are: \\
(1) Source name. 
(2),(3) 2.1 GHz flux density  and flux density uncertainty in $\mu$Jy. 
(4) to (9) SPIRE 500, 350 and 250 \micron\, flux densities and uncertainties  in mJy. 
(10) to (13) PACS 160, 100 \micron\, flux densities  and uncertainties  in mJy. 
(14) to (17) IRAC 4.5 \micron, 3.6 \micron\, flux densities  and uncertainties  in mJy. 
(18) $i$ AB magnitude of the optical counterpart. 
(19) Comment on counterpart identification:``S'' indicates a secure counterpart, ``D'' a double counterpart identification, ``T'' a tentative identification, and ``F'' a foreground source.\\
For radio to near-IR bands, $S_{\lambda}$ indicates the flux density and $\Delta S_{\lambda}$ is the associated uncertainty, with $\lambda$ the observing frequency in GHz (for radio measurements) or wavelength in microns (for all other measurements).}
\end{deluxetable}
\end{longrotatetable}

 %%TABLE 8%
%\begin{longrotatetable}
\begin{center}
\begin{deluxetable*}{|c|cc|cc|cc|cc|cc|}
%\raggedleft
%\tabletypesize{8pt}
\tablecolumns{11}
\tablecaption{ Positions of radio, SPIRE, PACS, IRAC,  and optical counterparts for SMGs detected in LASCAR clusters. } 
\tablehead{\colhead{}  &\multicolumn{2}{c}{2.1 GHz}& \multicolumn{2}{c}{SPIRE} &\multicolumn{2}{c}{PACS}  & \multicolumn{2}{c}{IRAC}  &  \multicolumn{2}{c}{Optical} \\
\colhead{Source}  &  \colhead{R.A.}  &  \colhead{Dec.} &  \colhead{R.A.}  &  \colhead{Dec.}& \colhead{R.A.}  &  \colhead{Dec.} &   \colhead{R.A.}  &  \colhead{Dec.}&  \colhead{R.A.}  &  \colhead{Dec.}}

\label{tab:counterparts_coords}
 \startdata
ACTJ0102$-$1 & 01:02:55.6 & -49:15:9.4 & 01:02:55.6 & -49:15:10.3 & 01:02:55.7 & -49:15:08.8 & 01:02:55.7 & -49:15:8.6 & 01:02:55.7 & -49:15:8.0 \\
ACTJ0102$-$2 & \nodata & \nodata & 01:03:14.4 & -49:13:32.7 & \nodata & \nodata & 01:03:14.1 & -49:13:33.4 & \nodata & \nodata \\
ACTJ0102$-$3 & 01:03:5.4 & -49:17:11.6 & 01:03:5.5 & -49:17:15.3 & \nodata & \nodata & 01:03:5.4 & -49:17:11.6 & 01:03:5.4 & -49:17:11.5 \\
ACTJ0102$-$4 & 01:03:8.3 & -49:11:47.8 & 01:03:8.4 & -49:11:47.8 & \nodata & \nodata & 01:03:8.2 & -49:11:48.0 & \nodata & \nodata \\
ACTJ0102$-$5 & 01:02:49.3 & -49:15:5.5 & 01:02:49.5 & -49:15:6.5 & 01:02:49.2 & -49:15:6.2 & 01:02:49.4 & -49:15:6.1 & 01:02:49.3 & -49:15:6.4 \\
ACTJ0215$-$2 & \nodata & \nodata & \nodata & \nodata & \nodata & \nodata & 02:15:19.2 & -52:17:9.9 & \nodata & \nodata \\
ACTJ0215$-$3 & \nodata & \nodata & \nodata & \nodata & \nodata & \nodata & 02:15:11.1 & -52:11:18.4 & \nodata & \nodata \\
ACTJ0232$-$1 & 02:32:50.0 & -52:56:15.9 & \nodata & \nodata & \nodata & \nodata & 02:32:50.0 & -52:56:14.9 & \nodata & \nodata \\
ACTJ0232$-$2 & 02:32:57.5 & -52:56:34.4 & \nodata & \nodata & \nodata & \nodata & 02:32:57.5 & -52:56:35.0 & \nodata & \nodata \\
ACTJ0232$-$4 & 02:33:0.1 & -53:02:31.5 & \nodata & \nodata & \nodata & \nodata & 02:32:60.0 & -53:02:31.9 & \nodata & \nodata \\
ACTJ0232$-$5 & 02:33:9.7 & -52:57:34.0 & \nodata & \nodata & \nodata & \nodata & 02:33:9.8 & -52:57:33.8 & \nodata & \nodata \\
ACTJ0235$-$1 & 02:35:38.8 & -51:19:3.1 & 02:35:38.9 & -51:19:7.9 & 02:35:39.0 & -51:19:06.3 & 02:35:38.9 & -51:19:4.1 & \nodata & \nodata \\
ACTJ0235$-$2 & \nodata & \nodata & \nodata & \nodata & 02:35:42.5 & -51:21:17.9 & 02:35:42.9 & -51:21:16.6 & 02:35:42.9 & -51:21:16.8 \\
ACTJ0235$-$4 & \nodata & \nodata & 02:35:43.8 & -51:22:44.5 & 02:35:42.7&  -51:22:40.0 & 02:35:42.6 & -51:22:40.1 & 02:35:42.6 & -51:22:40.1 \\
ACTJ0235$-$5 & 02:35:27.4 & -51:18:49.5 & 02:35:27.6 & -51:18:49.5 & \nodata & \nodata & 02:35:27.3 & -51:18:46.7 & \nodata & \nodata \\
ACTJ0245$-$1 & \nodata & \nodata & 02:45:30.7 & -53:04:13.8 & \nodata & \nodata & 02:45:30.6 & -53:04:11.9 & \nodata & \nodata \\
ACTJ0245$-$2 & 02:45:42.1 & -53:02:0.6 & 02:45:42.2 & -53:02:4.5 & \nodata & \nodata & 02:45:41.9 & -53:01:59.7 & \nodata & \nodata \\
ACTJ0245$-$3 & 02:45:35.6 & -53:05:3.0 & 02:45:35.3 & -53:04:57.0 & \nodata & \nodata & 02:45:35.7 & -53:05:2.7 & \nodata & \nodata \\
ACTJ0330$-$1 & 03:30:30.1 & -52:27:23.5 & \nodata & \nodata & \nodata & \nodata & 03:30:30.0 & -52:27:23.6 & \nodata & \nodata \\
ACTJ0330$-$2 & 03:30:54.0 & -52:24:33.7 & \nodata & \nodata & \nodata & \nodata & 03:30:53.9 & -52:24:32.8 & \nodata & \nodata \\
ACTJ0330$-$3 & 03:31:14.0 & -52:28:35.4 & \nodata & \nodata & \nodata & \nodata & 03:31:14.0 & -52:28:35.1 & \nodata & \nodata \\
ACTJ0438$-$1 & \nodata & \nodata & 04:38:29.8 & -54:18:32.3 & \nodata & \nodata & 04:38:30.8 & -54:18:32.1 & \nodata & \nodata \\
ACTJ0438$-$2 & \nodata & \nodata & 04:38:34.9 & -54:19:40.6 & \nodata & \nodata & 04:38:35.0 & -54:19:43.9 & 04:38:35.4 & -54:19:45.5 \\
ACTJ0438$-$4A & 04:38:24.4 & -54:17:18.8 & \nodata & \nodata & 04:38:24.5 & -54:17:17.9 & 04:38:24.5 & -54:17:16.9 & \nodata &\nodata \\
ACTJ0438$-$4B & 04:38:24.5 & -54:17:36.3 & \nodata & \nodata & 04:38:25.1 & -54:17:33.2 & 04:38:24.5 & -54:17:36.9 & 04:38:24.4 & -54:17:18.1 \\
ACTJ0438$-$5 & 04:38:42.3 & -54:21:9.8 & 04:38:42.4 & -54:21:10.4 & \nodata & \nodata & 04:38:42.2 & -54:21:9.4 & \nodata & \nodata \\
ACTJ0438$-$6A & 04:38:24.5 & -54:21:17.6 & 04:38:25.3 & -54:21:24.1 & 04:38:24.3 &-54:21:18.2  & 04:38:24.4 & -54:21:17.6 & 04:38:24.5 & -54:21:18.5 \\
ACTJ0438$-$6B &04:38:25.3 & -54:21:25.1 & 04:38:25.3 & -54:21:24.1 &04:38:25.3 -54:21:24.2 & 04:38:25.3 &-54:21:25.6 & 04:38:25.3 & -54:21:25.1 \\
ACTJ0438$-$7 & \nodata & \nodata & \nodata & \nodata & \nodata & \nodata & 04:38:19.5 & -54:23:10.9 & \nodata & \nodata \\
ACTJ0438$-$8A & 04:38:33.6 & -54:19:10.5 & 04:38:33.3 & -54:19:7.9 & 04:38:32.9 & -54:19:8.0 & 04:38:33.6 & -54:19:09.4 & 04:38:33.6 & -54:19:9.5 \\
ACTJ0438$-$8B & 04:38:32.9 & -54:19:09.2 & 04:38:33.3 & -54:19:7.9 & 04:38:32.9 & -54:19:8.0 & 04:38:32.9 & -54:19:9.7 & 04:38:33.4 & -54:19:6.2 \\
ACTJ0546$-$1 & 05:47:1.3 & -53:45:24.2 & 05:47:1.3 & -53:45:24.7 & \nodata & \nodata & 05:47:1.4 & -53:45:22.1 & \nodata & \nodata \\
ACTJ0546$-$2 & 05:46:34.5 & -53:45:50.2 & 05:46:34.7 & -53:45:51.7 & 05:46:34.7 & -53:45:51.1 & 05:46:34.5 & -53:45:52.4 & \nodata & \nodata \\
ACTJ0546$-$3 & 05:46:53.9 & -53:44:12.9 & 05:46:54.0 & -53:44:12.5 & 05:46:53.8 & -53:44:11.5 & 05:46:53.8 & -53:44:13.6 & \nodata & \nodata \\
ACTJ0546$-$4 & 05:46:37.9 & -53:43:14.4 & 05:46:37.5 & -53:43:16.3 & \nodata & \nodata & 05:46:37.8 & -53:43:14.7 & \nodata & \nodata \\
ACTJ0546$-$5 & 05:46:55.0 & -53:46:48.4 & \nodata & \nodata & 05:46:55.2 & -53:46:48.3 & 05:46:55.2 & -53:46:49.5 & \nodata & \nodata \\
ACTJ0546$-$6A & 05:46:49.7 & -53:46:13.7 & \nodata & \nodata & 05:46:49.7 & -53:46:13.4 & 05:46:49.7 & -53:46:13.9 & \nodata & \nodata \\
ACTJ0546$-$6B & 05:46:49.7 -& 53:46:27.0 & \nodata & \nodata &\nodata & \nodata &05:46:49.7 & -53:46:27.4 & \nodata & \nodata \\
ACTJ0546$-$7 & 05:46:30.1 & -53:41:39.7 & 05:46:30.3 & -53:41:42.4 & \nodata & \nodata & 05:46:30.0 & -53:41:39.4 & \nodata & \nodata \\
ACTJ0546$-$8 & \nodata & \nodata & 05:46:39.2 & -53:46:7.6 & \nodata & \nodata & \nodata & \nodata & \nodata & \nodata \\
ACTJ0546$-$9 & 05:46:53.1 & -53:47:47.0 & \nodata & \nodata & \nodata & \nodata & 05:46:53.1 & -53:47:46.6 & \nodata & \nodata \\
ACTJ0546$-$10 & 05:46:28.1 & -53:45:38.1 & \nodata & \nodata & \nodata & \nodata & 05:46:28.0 & -53:45:38.4 & \nodata & \nodata \\
ACTJ0546$-$11 & 05:47:6.2 & -53:47:57.5 & \nodata & \nodata & \nodata & \nodata & 05:47:6.2 & -53:47:55.4 & \nodata & \nodata \\
ACTJ0559$-$1 & 06:00:18.4 & -52:49:56.9 & \nodata & \nodata & \nodata & \nodata & 06:00:18.3 & -52:49:56.8 & \nodata & \nodata \\
ACTJ0616$-$1 & 06:16:14.0 & -52:27:21.0 & \nodata & \nodata & \nodata & \nodata & 06:16:13.9 & -52:27:21.1 & \nodata & \nodata \\
ACTJ0616$-$2 & \nodata & \nodata & \nodata & \nodata & \nodata & \nodata & 06:16:30.2 & -52:27:10.5 & \nodata & \nodata \\
ACTJ0616$-$4 & 06:16:39.4 & -52:22:39.3 & \nodata & \nodata & \nodata & \nodata & 06:16:39.3 & -52:22:38.4 & \nodata & \nodata \\
\enddata
\end{deluxetable*}
\end{center}
%\end{longrotatetable}
\clearpage
 
%%%%%%%%%%%%%%%%%%%%%%%%
\bibliographystyle{aasjournal}

\begin{thebibliography}{}
\expandafter\ifx\csname natexlab\endcsname\relax\def\natexlab#1{#1}\fi
\providecommand{\url}[1]{\href{#1}{#1}}

\bibitem[{{Abell} {et~al.}(1989){Abell}, {Corwin}, \& {Olowin}}]{abell1989}
{Abell}, G.~O., {Corwin}, Jr., H.~G., \& {Olowin}, R.~P. 1989, \apjs, 70, 1

\bibitem[{{Barger} {et~al.}(1999){Barger}, {Cowie}, \& {Sanders}}]{barger99a}
{Barger}, A.~J., {Cowie}, L.~L., \& {Sanders}, D.~B. 1999, \apjl, 518, L5

\bibitem[{{Baugh} {et~al.}(2005){Baugh}, {Lacey}, {Frenk}, {Granato}, {Silva},
  {Bressan}, {Benson}, \& {Cole}}]{baug05}
{Baugh}, C.~M., {Lacey}, C.~G., {Frenk}, C.~S., {et~al.} 2005, \mnras, 356,
  1191

\bibitem[{{Berta} {et~al.}(2010){Berta}, {Magnelli}, {Lutz}, {Altieri},
  {Aussel}, {Andreani}, {Bauer}, {Bongiovanni}, {Cava}, {Cepa}, {Cimatti},
  {Daddi}, {Dominguez}, {Elbaz}, {Feuchtgruber}, {F{\"o}rster Schreiber},
  {Genzel}, {Gruppioni}, {Katterloher}, {Magdis}, {Maiolino}, {Nordon},
  {P{\'e}rez Garc{\'{\i}}a}, {Poglitsch}, {Popesso}, {Pozzi}, {Riguccini},
  {Rodighiero}, {Saintonge}, {Santini}, {Sanchez-Portal}, {Shao}, {Sturm},
  {Tacconi}, {Valtchanov}, {Wetzstein}, \& {Wieprecht}}]{berta10}
{Berta}, S., {Magnelli}, B., {Lutz}, D., {et~al.} 2010, \aap, 518, L30

\bibitem[{{Blain} {et~al.}(2002){Blain}, {Smail}, {Ivison}, {Kneib}, \&
  {Frayer}}]{blain02}
{Blain}, A.~W., {Smail}, I., {Ivison}, R.~J., {Kneib}, J.-P., \& {Frayer},
  D.~T. 2002, \physrep, 369, 111

\bibitem[{{Bleem} {et~al.}(2015){Bleem}, {Stalder}, {de Haan}, {Aird}, {Allen},
  {Applegate}, {Ashby}, {Bautz}, {Bayliss}, {Benson}, {Bocquet}, {Brodwin},
  {Carlstrom}, {Chang}, {Chiu}, {Cho}, {Clocchiatti}, {Crawford}, {Crites},
  {Desai}, {Dietrich}, {Dobbs}, {Foley}, {Forman}, {George}, {Gladders},
  {Gonzalez}, {Halverson}, {Hennig}, {Hoekstra}, {Holder}, {Holzapfel},
  {Hrubes}, {Jones}, {Keisler}, {Knox}, {Lee}, {Leitch}, {Liu}, {Lueker},
  {Luong-Van}, {Mantz}, {Marrone}, {McDonald}, {McMahon}, {Meyer}, {Mocanu},
  {Mohr}, {Murray}, {Padin}, {Pryke}, {Reichardt}, {Rest}, {Ruel}, {Ruhl},
  {Saliwanchik}, {Saro}, {Sayre}, {Schaffer}, {Schrabback}, {Shirokoff},
  {Song}, {Spieler}, {Stanford}, {Staniszewski}, {Stark}, {Story}, {Stubbs},
  {Vanderlinde}, {Vieira}, {Vikhlinin}, {Williamson}, {Zahn}, \&
  {Zenteno}}]{bleem2015}
{Bleem}, L.~E., {Stalder}, B., {de Haan}, T., {et~al.} 2015, \apjs, 216, 27

\bibitem[{{Bolzonella} {et~al.}(2000){Bolzonella}, {Miralles}, \&
  {Pell{\'o}}}]{bolzonella2000}
{Bolzonella}, M., {Miralles}, J.-M., \& {Pell{\'o}}, R. 2000, \aap, 363, 476

\bibitem[{{Bondi} {et~al.}(2008){Bondi}, {Ciliegi}, {Schinnerer}, {Smol{\v
  c}i{\'c}}, {Jahnke}, {Carilli}, \& {Zamorani}}]{bondi08}
{Bondi}, M., {Ciliegi}, P., {Schinnerer}, E., {et~al.} 2008, \aa, 681, 1129

\bibitem[{{Brisbin} {et~al.}(2017){Brisbin}, {Miettinen}, {Aravena}, {Smol{\v
  c}i{\'c}}, {Delvecchio}, {Jiang}, \& {Magnelli}}]{brisbin2017}
{Brisbin}, D., {Miettinen}, O., {Aravena}, M., {et~al.} 2017, \apj, 608, 15

\bibitem[{{Bruzual} \& {Charlot}(2003)}]{bruz03}
{Bruzual}, G., \& {Charlot}, S. 2003, \mnras, 344, 1000

\bibitem[{{Capak} {et~al.}(2008){Capak}, {Carilli}, {Lee}, {Aldcroft},
  {Aussel}, {Schinnerer}, {Wilson}, {Yun}, {Blain}, {Giavalisco}, {Ilbert},
  {Kartaltepe}, {Lee}, {McCracken}, {Mobasher}, {Salvato}, {Sasaki}, {Scott},
  {Sheth}, {Shioya}, {Thompson}, {Elvis}, {Sanders}, {Scoville}, \&
  {Tanaguchi}}]{capak08}
{Capak}, P., {Carilli}, C.~L., {Lee}, N., {et~al.} 2008, \apjl, 681, L53

\bibitem[{{Carilli} \& {Walter}(2013)}]{carilli2013}
{Carilli}, C.~L., \& {Walter}, F. 2013, \araa, 51, 105

\bibitem[{{Carlstrom} {et~al.}(2011){Carlstrom}, {Ade}, {Aird}, {Benson},
  {Bleem}, {Busetti}, {Chang}, {Chauvin}, {Cho}, {Crawford}, {Crites}, {Dobbs},
  {Halverson}, {Heimsath}, {Holzapfel}, {Hrubes}, {Joy}, {Keisler}, {Lanting},
  {Lee}, {Leitch}, {Leong}, {Lu}, {Lueker}, {Luong-Van}, {McMahon}, {Mehl},
  {Meyer}, {Mohr}, {Montroy}, {Padin}, {Plagge}, {Pryke}, {Ruhl}, {Schaffer},
  {Schwan}, {Shirokoff}, {Spieler}, {Staniszewski}, {Stark}, {Tucker},
  {Vanderlinde}, {Vieira}, \& {Williamson}}]{carlstrom11}
{Carlstrom}, J.~E., {Ade}, P.~A.~R., {Aird}, K.~A., {et~al.} 2011, \pasp, 123,
  568

\bibitem[{{Casey} {et~al.}(2014){Casey}, {Narayanan}, \& {Cooray}}]{casey2014}
{Casey}, C.~M., {Narayanan}, D., \& {Cooray}, A. 2014, \physrep, 541, 45

\bibitem[{{Chapman} {et~al.}(2005){Chapman}, {Blain}, {Smail}, \&
  {Ivison}}]{chapman2005}
{Chapman}, S.~C., {Blain}, A.~W., {Smail}, I., \& {Ivison}, R.~J. 2005, \apj,
  622, 772

\bibitem[{{Chapman} {et~al.}(2002){Chapman}, {Scott}, {Borys}, \&
  {Fahlman}}]{chapman2002b}
{Chapman}, S.~C., {Scott}, D., {Borys}, C., \& {Fahlman}, G.~G. 2002, \mnras,
  330, 92

\bibitem[{{Chen} {et~al.}(2013){Chen}, {Cowie}, {Barger}, {Casey}, {Lee},
  {Sanders}, {Wang}, \& {Williams}}]{chen2013}
{Chen}, C.-C., {Cowie}, L.~L., {Barger}, A.~J., {et~al.} 2013, \apj, 776, 131

\bibitem[{{Condon}(1992)}]{condon92}
{Condon}, J.~J. 1992, \araa, 30, 575

\bibitem[{{Coppin} {et~al.}(2006){Coppin}, {Chapin}, {Mortier}, {Scott},
  {Borys}, {Dunlop}, {Halpern}, {Hughes}, {Pope}, {Scott}, {Serjeant}, {Wagg},
  {Alexander}, {Almaini}, {Aretxaga}, {Babbedge}, {Best}, {Blain}, {Chapman},
  {Clements}, {Crawford}, {Dunne}, {Eales}, {Edge}, {Farrah}, {Gazta{\~n}aga},
  {Gear}, {Granato}, {Greve}, {Fox}, {Ivison}, {Jarvis}, {Jenness}, {Lacey},
  {Lepage}, {Mann}, {Marsden}, {Martinez-Sansigre}, {Oliver}, {Page},
  {Peacock}, {Pearson}, {Percival}, {Priddey}, {Rawlings}, {Rowan-Robinson},
  {Savage}, {Seigar}, {Sekiguchi}, {Silva}, {Simpson}, {Smail}, {Stevens},
  {Takagi}, {Vaccari}, {van Kampen}, \& {Willott}}]{coppin06}
{Coppin}, K., {Chapin}, E.~L., {Mortier}, A.~M.~J., {et~al.} 2006, \mnras, 372,
  1621

\bibitem[{{Cowie} {et~al.}(2002){Cowie}, {Barger}, \& {Kneib}}]{cowie2002}
{Cowie}, L.~L., {Barger}, A.~J., \& {Kneib}, J.-P. 2002, \aj, 123, 2197


\bibitem[{{Danielson} {et~al.}(2017){Danielson}, {Swinbank}, \& {Smail}}]{danielson2017}
{Danielson}, A.~L.~R., {Swinbank}, A.~M., \& {Smail}, I. 2017, \apj, 840, 78


\bibitem[{{Dannerbauer} {et~al.}(2010){Dannerbauer}, {Daddi}, {Morrison},
  {Altieri}, {Andreani}, {Aussel}, {Berta}, {Bongiovanni}, {Cava}, {Cepa},
  {Cimatti}, {Dominguez}, {Elbaz}, {F{\"o}rster Schreiber}, {Genzel},
  {Gruppioni}, {Horeau}, {Hwang}, {Le Floc'h}, {Le Pennec}, {Lutz}, {Magdis},
  {Magnelli}, {Maiolino}, {Nordon}, {P{\'e}rez Garc{\'{\i}}a}, {Poglitsch},
  {Popesso}, {Pozzi}, {Riguccini}, {Rodighiero}, {Saintonge}, {Santini},
  {Sanchez-Portal}, {Shao}, {Sturm}, {Tacconi}, \& {Valtchanov}}]{danner10}
{Dannerbauer}, H., {Daddi}, E., {Morrison}, G.~E., {et~al.} 2010, \apjl, 720,
  L144

\bibitem[{{Downes} {et~al.}(1986){Downes}, {Peacock}, {Savage}, \&
  {Carrie}}]{downes86}
{Downes}, A.~J.~B., {Peacock}, J.~A., {Savage}, A., \& {Carrie}, D.~R. 1986,
  \mnras, 218, 31

\bibitem[{{Edge} {et~al.}(1994){Edge}, {Boehringer}, {Guzzo}, {Collins},
  {Neumann}, {Chincarini}, {de Grandi}, {Duemmler}, {Ebeling}, {Schindler},
  {Seitter}, {Vettolani}, {Briel}, {Cruddace}, {Gruber}, {Gursky}, {Hartner},
  {MacGillivray}, {Schuecker}, {Shaver}, {Voges}, {Wallin}, {Wolter}, \&
  {Zamorani}}]{edge1994}
{Edge}, A.~C., {Boehringer}, H., {Guzzo}, L., {et~al.} 1994, \aap, 289, L34

\bibitem[{{Egami} {et~al.}(2010){Egami}, {Rex}, {Rawle},
  {P{\'e}rez-Gonz{\'a}lez}, {Richard}, {Kneib}, {Schaerer}, {Altieri},
  {Valtchanov}, {Blain}, {Fadda}, {Zemcov}, {Bock}, {Boone}, {Bridge},
  {Clement}, {Combes}, {Dessauges-Zavadsky}, {Dowell}, {Ilbert}, {Ivison},
  {Jauzac}, {Lutz}, {Metcalfe}, {Omont}, {Pell{\'o}}, {Pereira}, {Rieke},
  {Rodighiero}, {Smail}, {Smith}, {Tramoy}, {Walth}, {van der Werf}, \&
  {Werner}}]{egami10}
{Egami}, E., {Rex}, M., {Rawle}, T.~D., {et~al.} 2010, \aap, 518, L12

\bibitem[{{Fazio} {et~al.}(2004){Fazio}, {Hora}, {Allen}, {Ashby}, {Barmby},
  {Deutsch}, {Huang}, {Kleiner}, {Marengo}, {Megeath}, {Melnick}, {Pahre},
  {Patten}, {Polizotti}, {Smith}, {Taylor}, {Wang}, {Willner}, {Hoffmann},
  {Pipher}, {Forrest}, {McMurty}, {McCreight}, {McKelvey}, {McMurray}, {Koch},
  {Moseley}, {Arendt}, {Mentzell}, {Marx}, {Losch}, {Mayman}, {Eichhorn},
  {Krebs}, {Jhabvala}, {Gezari}, {Fixsen}, {Flores}, {Shakoorzadeh}, {Jungo},
  {Hakun}, {Workman}, {Karpati}, {Kichak}, {Whitley}, {Mann}, {Tollestrup},
  {Eisenhardt}, {Stern}, {Gorjian}, {Bhattacharya}, {Carey}, {Nelson},
  {Glaccum}, {Lacy}, {Lowrance}, {Laine}, {Reach}, {Stauffer}, {Surace},
  {Wilson}, {Wright}, {Hoffman}, {Domingo}, \& {Cohen}}]{fazio04}
{Fazio}, G.~G., {Hora}, J.~L., {Allen}, L.~E., {et~al.} 2004, \apjs, 154, 10

\bibitem[{{Geach} {et~al.}(2017){Geach}, {Dunlop}, {Halpern}, {Smail}, {van der
  Werf}, {Alexander}, {Almaini}, {Aretxaga}, {Arumugam}, {Asboth}, {Banerji},
  {Beanlands}, {Best}, {Blain}, {Birkinshaw}, {Chapin}, {Chapman}, {Chen},
  {Chrysostomou}, {Clarke}, {Clements}, {Conselice}, {Coppin}, {Cowley},
  {Danielson}, {Eales}, {Edge}, {Farrah}, {Gibb}, {Harrison}, {Hine}, {Hughes},
  {Ivison}, {Jarvis}, {Jenness}, {Jones}, {Karim}, {Koprowski}, {Knudsen},
  {Lacey}, {Mackenzie}, {Marsden}, {McAlpine}, {McMahon}, {Meijerink},
  {Micha{\l}owski}, {Oliver}, {Page}, {Peacock}, {Rigopoulou}, {Robson},
  {Roseboom}, {Rotermund}, {Scott}, {Serjeant}, {Simpson}, {Simpson}, {Smith},
  {Spaans}, {Stanley}, {Stevens}, {Swinbank}, {Targett}, {Thomson}, {Valiante},
  {Wake}, {Webb}, {Willott}, {Zavala}, \& {Zemcov}}]{geach2017}
{Geach}, J.~E., {Dunlop}, J.~S., {Halpern}, M., {et~al.} 2017, \mnras, 465,
  1789

\bibitem[{{Gehrels}(1986)}]{gehrels86}
{Gehrels}, N. 1986, \apj, 303, 336

\bibitem[{{Griffin} {et~al.}(2010){Griffin}, {Abergel}, {Abreu}, {Ade},
  {Andr{\'e}}, {Augueres}, {Babbedge}, {Bae}, {Baillie}, {Baluteau}, {Barlow},
  {Bendo}, {Benielli}, {Bock}, {Bonhomme}, {Brisbin}, {Brockley-Blatt},
  {Caldwell}, {Cara}, {Castro-Rodriguez}, {Cerulli}, {Chanial}, {Chen},
  {Clark}, {Clements}, {Clerc}, {Coker}, {Communal}, {Conversi}, {Cox},
  {Crumb}, {Cunningham}, {Daly}, {Davis}, {de Antoni}, {Delderfield}, {Devin},
  {di Giorgio}, {Didschuns}, {Dohlen}, {Donati}, {Dowell}, {Dowell}, {Duband},
  {Dumaye}, {Emery}, {Ferlet}, {Ferrand}, {Fontignie}, {Fox}, {Franceschini},
  {Frerking}, {Fulton}, {Garcia}, {Gastaud}, {Gear}, {Glenn}, {Goizel},
  {Griffin}, {Grundy}, {Guest}, {Guillemet}, {Hargrave}, {Harwit}, {Hastings},
  {Hatziminaoglou}, {Herman}, {Hinde}, {Hristov}, {Huang}, {Imhof}, {Isaak},
  {Israelsson}, {Ivison}, {Jennings}, {Kiernan}, {King}, {Lange}, {Latter},
  {Laurent}, {Laurent}, {Leeks}, {Lellouch}, {Levenson}, {Li}, {Li},
  {Lilienthal}, {Lim}, {Liu}, {Lu}, {Madden}, {Mainetti}, {Marliani}, {McKay},
  {Mercier}, {Molinari}, {Morris}, {Moseley}, {Mulder}, {Mur}, {Naylor},
  {Nguyen}, {O'Halloran}, {Oliver}, {Olofsson}, {Olofsson}, {Orfei}, {Page},
  {Pain}, {Panuzzo}, {Papageorgiou}, {Parks}, {Parr-Burman}, {Pearce},
  {Pearson}, {P{\'e}rez-Fournon}, {Pinsard}, {Pisano}, {Podosek}, {Pohlen},
  {Polehampton}, {Pouliquen}, {Rigopoulou}, {Rizzo}, {Roseboom}, {Roussel},
  {Rowan-Robinson}, {Rownd}, {Saraceno}, {Sauvage}, {Savage}, {Savini},
  {Sawyer}, {Scharmberg}, {Schmitt}, {Schneider}, {Schulz}, {Schwartz},
  {Shafer}, {Shupe}, {Sibthorpe}, {Sidher}, {Smith}, {Smith}, {Smith},
  {Spencer}, {Stobie}, {Sudiwala}, {Sukhatme}, {Surace}, {Stevens}, {Swinyard},
  {Trichas}, {Tourette}, {Triou}, {Tseng}, {Tucker}, {Turner}, {Vaccari},
  {Valtchanov}, {Vigroux}, {Virique}, {Voellmer}, {Walker}, {Ward}, {Waskett},
  {Weilert}, {Wesson}, {White}, {Whitehouse}, {Wilson}, {Winter}, {Woodcraft},
  {Wright}, {Xu}, {Zavagno}, {Zemcov}, {Zhang}, \& {Zonca}}]{griffin10}
{Griffin}, M.~J., {Abergel}, A., {Abreu}, A., {et~al.} 2010, \aap, 518, L3

\bibitem[{{G{\"u}sten} {et~al.}(2006){G{\"u}sten}, {Nyman}, {Schilke},
  {Menten}, {Cesarsky}, \& {Booth}}]{gusten2006}
{G{\"u}sten}, R., {Nyman}, L.~{\AA}., {Schilke}, P., {et~al.} 2006, \aap, 454,
  L13

\bibitem[{{Hainline} {et~al.}(2009){Hainline}, {Blain}, {Smail}, {Frayer},
  {Chapman}, {Ivison}, \& {Alexander}}]{hainline09}
{Hainline}, L.~J., {Blain}, A.~W., {Smail}, I., {et~al.} 2009, \apj, 699, 1610

\bibitem[{{Hasselfield} {et~al.}(2013){Hasselfield}, {Hilton}, {Marriage},
  {Addison}, {Barrientos}, {Battaglia}, {Battistelli}, {Bond}, {Crichton},
  {Das}, {Devlin}, {Dicker}, {Dunkley}, {D{\"u}nner}, {Fowler}, {Gralla},
  {Hajian}, {Halpern}, {Hincks}, {Hlozek}, {Hughes}, {Infante}, {Irwin},
  {Kosowsky}, {Marsden}, {Menanteau}, {Moodley}, {Niemack}, {Nolta}, {Page},
  {Partridge}, {Reese}, {Schmitt}, {Sehgal}, {Sherwin}, {Sievers}, {Sif{\'o}n},
  {Spergel}, {Staggs}, {Swetz}, {Switzer}, {Thornton}, {Trac}, \&
  {Wollack}}]{hasselfield2013}
{Hasselfield}, M., {Hilton}, M., {Marriage}, T.~A., {et~al.} 2013, \jcap, 7,
  008

\bibitem[{{Hauser} \& {Dwek}(2001)}]{hauser2001}
{Hauser}, M.~G., \& {Dwek}, E. 2001, \araa, 39, 249

\bibitem[{{Hilton} {et~al.}(2013){Hilton}, {Hasselfield}, {Sif{\'o}n}, {Baker},
  {Barrientos}, {Battaglia}, {Bond}, {Crichton}, {Das}, {Devlin}, {Gralla},
  {Hajian}, {Hincks}, {Hughes}, {Infante}, {Irwin}, {Kosowsky}, {Lin},
  {Marriage}, {Marsden}, {Menanteau}, {Moodley}, {Niemack}, {Nolta}, {Page},
  {Reese}, {Sievers}, {Spergel}, \& {Wollack}}]{hilton13}
{Hilton}, M., {Hasselfield}, M., {Sif{\'o}n}, C., {et~al.} 2013, \mnras, 435,
  3469

\bibitem[{{Hodge} {et~al.}(2013){Hodge}, {Karim}, {Smail}, {Swinbank},
  {Walter}, {Biggs}, {Ivison}, {Weiss}, {Alexander}, {Bertoldi}, {Brandt},
  {Chapman}, {Coppin}, {Cox}, {Danielson}, {Dannerbauer}, {De Breuck},
  {Decarli}, {Edge}, {Greve}, {Knudsen}, {Menten}, {Rix}, {Schinnerer},
  {Simpson}, {Wardlow}, \& {van der Werf}}]{hodge13}
{Hodge}, J.~A., {Karim}, A., {Smail}, I., {et~al.} 2013, \apj, 768, 91

\bibitem[{{Holland} {et~al.}(1999){Holland}, {Robson}, {Gear}, {Cunningham},
  {Lightfoot}, {Jenness}, {Ivison}, {Stevens}, {Ade}, {Griffin}, {Duncan},
  {Murphy}, \& {Naylor}}]{holland99}
{Holland}, W.~S., {Robson}, E.~I., {Gear}, W.~K., {et~al.} 1999, \mnras, 303,
  659

\bibitem[{{Holland} {et~al.}(2013){Holland}, {Bintley}, {Chapin},
  {Chrysostomou}, {Davis}, {Dempsey}, {Duncan}, {Fich}, {Friberg}, {Halpern},
  {Irwin}, {Jenness}, {Kelly}, {MacIntosh}, {Robson}, {Scott}, {Ade},
  {Atad-Ettedgui}, {Berry}, {Craig}, {Gao}, {Gibb}, {Hilton}, {Hollister},
  {Kycia}, {Lunney}, {McGregor}, {Montgomery}, {Parkes}, {Tilanus}, {Ullom},
  {Walther}, {Walton}, {Woodcraft}, {Amiri}, {Atkinson}, {Burger}, {Chuter},
  {Coulson}, {Doriese}, {Dunare}, {Economou}, {Niemack}, {Parsons},
  {Reintsema}, {Sibthorpe}, {Smail}, {Sudiwala}, \& {Thomas}}]{holland2013}
{Holland}, W.~S., {Bintley}, D., {Chapin}, E.~L., {et~al.} 2013, \mnras, 430,
  2513

\bibitem[{{Hsu} {et~al.}(2016){Hsu}, {Cowie}, {Chen}, {Barger}, \&
  {Wang}}]{hsu2016}
{Hsu}, L.-Y., {Cowie}, L.~L., {Chen}, C.-C., {Barger}, A.~J., \& {Wang}, W.-H.
  2016, \apj, 829, 25

\bibitem[{{Hughes} {et~al.}(1998){Hughes}, {Serjeant}, {Dunlop},
  {Rowan-Robinson}, {Blain}, {Mann}, {Ivison}, {Peacock}, {Efstathiou}, {Gear},
  {Oliver}, {Lawrence}, {Longair}, {Goldschmidt}, \& {Jenness}}]{hughes98}
{Hughes}, D.~H., {Serjeant}, S., {Dunlop}, J., {et~al.} 1998, \nat, 394, 241

\bibitem[{{Ibar} {et~al.}(2010){Ibar}, {Ivison}, {Best}, {Coppin}, {Pope},
  {Smail}, \& {Dunlop}}]{ibar10}
{Ibar}, E., {Ivison}, R.~J., {Best}, P.~N., {et~al.} 2010, \mnras, 401, L53

\bibitem[{{Iglesias-P{\'a}ramo} {et~al.}(2007){Iglesias-P{\'a}ramo}, {Buat},
  {Hern{\'a}ndez-Fern{\'a}ndez}, {Xu}, {Burgarella}, {Takeuchi}, {Boselli},
  {Shupe}, {Rowan-Robinson}, {Babbedge}, {Conrow}, {Fang}, {Farrah},
  {Gonz{\'a}lez-Solares}, {Lonsdale}, {Smith}, {Surace}, {Barlow}, {Forster},
  {Friedman}, {Martin}, {Morrissey}, {Neff}, {Schiminovich}, {Seibert},
  {Small}, {Wyder}, {Bianchi}, {Donas}, {Heckman}, {Lee}, {Madore}, {Milliard},
  {Rich}, {Szalay}, {Welsh}, \& {Yi}}]{iglesias07}
{Iglesias-P{\'a}ramo}, J., {Buat}, V., {Hern{\'a}ndez-Fern{\'a}ndez}, J.,
  {et~al.} 2007, \apj, 670, 279

\bibitem[{{Ivison} {et~al.}(2000){Ivison}, {Smail}, {Barger}, {Kneib}, {Blain},
  {Owen}, {Kerr}, \& {Cowie}}]{ivison2000}
{Ivison}, R.~J., {Smail}, I., {Barger}, A.~J., {et~al.} 2000, \mnras, 315, 209

\bibitem[{{Ivison} {et~al.}(1998){Ivison}, {Smail}, {Le Borgne}, {Blain},
  {Kneib}, {Bezecourt}, {Kerr}, \& {Davies}}]{ivison1998}
{Ivison}, R.~J., {Smail}, I., {Le Borgne}, J.-F., {et~al.} 1998, \mnras, 298,
  583

\bibitem[{{Ivison} {et~al.}(2002){Ivison}, {Greve}, {Smail}, {Dunlop}, {Roche},
  {Scott}, {Page}, {Stevens}, {Almaini}, {Blain}, {Willott}, {Fox}, {Gilbank},
  {Serjeant}, \& {Hughes}}]{ivison2002}
{Ivison}, R.~J., {Greve}, T.~R., {Smail}, I., {et~al.} 2002, \mnras, 337, 1

\bibitem[{{Johansson} {et~al.}(2011){Johansson}, {Sigurdarson}, \&
  {Horellou}}]{johansson11}
{Johansson}, D., {Sigurdarson}, H., \& {Horellou}, C. 2011, \aap, 527, A117+

\bibitem[{{Johansson} {et~al.}(2010){Johansson}, {Horellou}, {Sommer}, {Basu},
  {Bertoldi}, {Birkinshaw}, {Lancaster}, {Lopez-Cruz}, \&
  {Quintana}}]{johansson10}
{Johansson}, D., {Horellou}, C., {Sommer}, M.~W., {et~al.} 2010, \aap, 514,
  A77+

\bibitem[{{Jones} {et~al.}(2009){Jones}, {Read}, {Saunders}, {Colless},
  {Jarrett}, {Parker}, {Fairall}, {Mauch}, {Sadler}, {Watson}, {Burton},
  {Campbell}, {Cass}, {Croom}, {Dawe}, {Fiegert}, {Frankcombe}, {Hartley},
  {Huchra}, {James}, {Kirby}, {Lahav}, {Lucey}, {Mamon}, {Moore}, {Peterson},
  {Prior}, {Proust}, {Russell}, {Safouris}, {Wakamatsu}, {Westra}, \&
  {Williams}}]{jones09}
{Jones}, D.~H., {Read}, M.~A., {Saunders}, W., {et~al.} 2009, \mnras, 399, 683

\bibitem[{{Jullo} {et~al.}(2007){Jullo}, {Kneib}, {Limousin},
  {El{\'{\i}}asd{\'o}ttir}, {Marshall}, \& {Verdugo}}]{jullo07}
{Jullo}, E., {Kneib}, J.-P., {Limousin}, M., {et~al.} 2007, New Journal of
  Physics, 9, 447

\bibitem[{{Karim} {et~al.}(2013){Karim}, {Swinbank}, {Hodge}, {Smail},
  {Walter}, {Biggs}, {Simpson}, {Danielson}, {Alexander}, {Bertoldi}, {de
  Breuck}, {Chapman}, {Coppin}, {Dannerbauer}, {Edge}, {Greve}, {Ivison},
  {Knudsen}, {Menten}, {Schinnerer}, {Wardlow}, {Wei{\ss}}, \& {van der
  Werf}}]{karim13}
{Karim}, A., {Swinbank}, A.~M., {Hodge}, J.~A., {et~al.} 2013, \mnras, 432, 2

\bibitem[{{Knudsen} {et~al.}(2008){Knudsen}, {van der Werf}, \&
  {Kneib}}]{knudsen08}
{Knudsen}, K.~K., {van der Werf}, P.~P., \& {Kneib}, J.-P. 2008, \mnras, 384,
  1611

\bibitem[{{Komatsu} {et~al.}(2011){Komatsu}, {Smith}, {Dunkley}, {Bennett},
  {Gold}, {Hinshaw}, {Jarosik}, {Larson}, {Nolta}, {Page}, {Spergel},
  {Halpern}, {Hill}, {Kogut}, {Limon}, {Meyer}, {Odegard}, {Tucker}, {Weiland},
  {Wollack}, \& {Wright}}]{komatsu11}
{Komatsu}, E., {Smith}, K.~M., {Dunkley}, J., {et~al.} 2011, \apjs, 192, 18

\bibitem[{{Lima} {et~al.}(2010{\natexlab{a}}){Lima}, {Jain}, \&
  {Devlin}}]{lima10}
{Lima}, M., {Jain}, B., \& {Devlin}, M. 2010{\natexlab{a}}, \mnras, 406, 2352

\bibitem[{{Lima} {et~al.}(2010{\natexlab{b}}){Lima}, {Jain}, {Devlin}, \&
  {Aguirre}}]{lima10a}
{Lima}, M., {Jain}, B., {Devlin}, M., \& {Aguirre}, J. 2010{\natexlab{b}},
  \apjl, 717, L31

\bibitem[{{Limousin} {et~al.}(2007){Limousin}, {Richard}, {Jullo}, {Kneib},
  {Fort}, {Soucail}, {El{\'{\i}}asd{\'o}ttir}, {Natarajan}, {Ellis}, {Smail},
  {Czoske}, {Smith}, {Hudelot}, {Bardeau}, {Ebeling}, {Egami}, \&
  {Knudsen}}]{limousin07}
{Limousin}, M., {Richard}, J., {Jullo}, E., {et~al.} 2007, \apj, 668, 643

\bibitem[{{Lindner} {et~al.}(2015){Lindner}, {Aguirre}, {Baker}, {Bond},
  {Crichton}, {Devlin}, {Essinger-Hileman}, {Gallardo}, {Gralla}, {Hilton},
  {Hincks}, {Huffenberger}, {Hughes}, {Infante}, {Lima}, {Marriage},
  {Menanteau}, {Niemack}, {Page}, {Schmitt}, {Sehgal}, {Sievers}, {Sif{\'o}n},
  {Staggs}, {Swetz}, {Wei{\ss}}, \& {Wollack}}]{lindner14}
{Lindner}, R.~R., {Aguirre}, P., {Baker}, A.~J., {et~al.} 2015, \apj, 803, 79

\bibitem[{{Magnelli} {et~al.}(2010){Magnelli}, {Lutz}, {Berta}, {Altieri},
  {Andreani}, {Aussel}, {Casta{\~n}eda}, {Cava}, {Cepa}, {Cimatti}, {Daddi},
  {Dannerbauer}, {Dominguez}, {Elbaz}, {F{\"o}rster Schreiber}, {Genzel},
  {Grazian}, {Gruppioni}, {Magdis}, {Maiolino}, {Nordon}, {P{\'e}rez Fournon},
  {P{\'e}rez Garc{\'{\i}}a}, {Poglitsch}, {Popesso}, {Pozzi}, {Riguccini},
  {Rodighiero}, {Saintonge}, {Santini}, {Sanchez-Portal}, {Shao}, {Sturm},
  {Tacconi}, {Valtchanov}, {Wieprecht}, \& {Wiezorrek}}]{magnelli10}
{Magnelli}, B., {Lutz}, D., {Berta}, S., {et~al.} 2010, \aap, 518, L28

\bibitem[{{Markevitch} {et~al.}(2002){Markevitch}, {Gonzalez}, {David},
  {Vikhlinin}, {Murray}, {Forman}, {Jones}, \& {Tucker}}]{marke02}
{Markevitch}, M., {Gonzalez}, A.~H., {David}, L., {et~al.} 2002, \apjl, 567,
  L27

\bibitem[{{Marriage} {et~al.}(2011{\natexlab{a}}){Marriage}, {Baptiste Juin},
  {Lin}, {Marsden}, {Nolta}, {Partridge}, {Ade}, {Aguirre}, {Amiri}, {Appel},
  {Barrientos}, {Battistelli}, {Bond}, {Brown}, {Burger}, {Chervenak}, {Das},
  {Devlin}, {Dicker}, {Bertrand Doriese}, {Dunkley}, {D{\"u}nner},
  {Essinger-Hileman}, {Fisher}, {Fowler}, {Hajian}, {Halpern}, {Hasselfield},
  {Hern{\'a}ndez-Monteagudo}, {Hilton}, {Hilton}, {Hincks}, {Hlozek},
  {Huffenberger}, {Handel Hughes}, {Hughes}, {Infante}, {Irwin}, {Kaul},
  {Klein}, {Kosowsky}, {Lau}, {Limon}, {Lupton}, {Martocci}, {Mauskopf},
  {Menanteau}, {Moodley}, {Moseley}, {Netterfield}, {Niemack}, {Page},
  {Parker}, {Quintana}, {Reid}, {Sehgal}, {Sherwin}, {Sievers}, {Spergel},
  {Staggs}, {Swetz}, {Switzer}, {Thornton}, {Trac}, {Tucker}, {Warne},
  {Wilson}, {Wollack}, \& {Zhao}}]{marriage11a}
{Marriage}, T.~A., {Baptiste Juin}, J., {Lin}, Y.-T., {et~al.}
  2011{\natexlab{a}}, \apj, 731, 100

\bibitem[{{Marriage} {et~al.}(2011{\natexlab{b}}){Marriage}, {Acquaviva},
  {Ade}, {Aguirre}, {Amiri}, {Appel}, {Barrientos}, {Battistelli}, {Bond},
  {Brown}, {Burger}, {Chervenak}, {Das}, {Devlin}, {Dicker}, {Bertrand
  Doriese}, {Dunkley}, {D{\"u}nner}, {Essinger-Hileman}, {Fisher}, {Fowler},
  {Hajian}, {Halpern}, {Hasselfield}, {Hern{\'a}ndez-Monteagudo}, {Hilton},
  {Hilton}, {Hincks}, {Hlozek}, {Huffenberger}, {Handel Hughes}, {Hughes},
  {Infante}, {Irwin}, {Baptiste Juin}, {Kaul}, {Klein}, {Kosowsky}, {Lau},
  {Limon}, {Lin}, {Lupton}, {Marsden}, {Martocci}, {Mauskopf}, {Menanteau},
  {Moodley}, {Moseley}, {Netterfield}, {Niemack}, {Nolta}, {Page}, {Parker},
  {Partridge}, {Quintana}, {Reese}, {Reid}, {Sehgal}, {Sherwin}, {Sievers},
  {Spergel}, {Staggs}, {Swetz}, {Switzer}, {Thornton}, {Trac}, {Tucker},
  {Warne}, {Wilson}, {Wollack}, \& {Zhao}}]{marriage11b}
{Marriage}, T.~A., {Acquaviva}, V., {Ade}, P.~A.~R., {et~al.}
  2011{\natexlab{b}}, \apj, 737, 61

\bibitem[{{Marsden} {et~al.}(2014){Marsden}, {Gralla}, {Marriage}, {Switzer},
  {Partridge}, {Massardi}, {Morales}, {Addison}, {Bond}, {Crichton}, {Das},
  {Devlin}, {D{\"u}nner}, {Hajian}, {Hilton}, {Hincks}, {Hughes}, {Irwin},
  {Kosowsky}, {Menanteau}, {Moodley}, {Niemack}, {Page}, {Reese}, {Schmitt},
  {Sehgal}, {Sievers}, {Staggs}, {Swetz}, {Thornton}, \&
  {Wollack}}]{marsden2014}
{Marsden}, D., {Gralla}, M., {Marriage}, T.~A., {et~al.} 2014, \mnras, 439,
  1556

\bibitem[{{McMullin} {et~al.}(2007){McMullin}, {Waters}, {Schiebel}, {Young},
  \& {Golap}}]{mcmullin07}
{McMullin}, J.~P., {Waters}, B., {Schiebel}, D., {Young}, W., \& {Golap}, K.
  2007, in Astronomical Society of the Pacific Conference Series, Vol. 376,
  Astronomical Data Analysis Software and Systems XVI, ed. R.~A. {Shaw},
  F.~{Hill}, \& D.~J. {Bell}, 127

\bibitem[{{Menanteau} {et~al.}(2009){Menanteau}, {Hughes}, {Jimenez},
  {Hernandez-Monteagudo}, {Verde}, {Kosowsky}, {Moodley}, {Infante}, \&
  {Roche}}]{menan09a}
{Menanteau}, F., {Hughes}, J.~P., {Jimenez}, R., {et~al.} 2009, \apj, 698, 1221

\bibitem[{{Menanteau} {et~al.}(2010{\natexlab{a}}){Menanteau}, {Gonz{\'a}lez},
  {Juin}, {Marriage}, {Reese}, {Acquaviva}, {Aguirre}, {Appel}, {Baker},
  {Barrientos}, {Battistelli}, {Bond}, {Das}, {Deshpande}, {Devlin}, {Dicker},
  {Dunkley}, {D{\"u}nner}, {Essinger-Hileman}, {Fowler}, {Hajian}, {Halpern},
  {Hasselfield}, {Hern{\'a}ndez-Monteagudo}, {Hilton}, {Hincks}, {Hlozek},
  {Huffenberger}, {Hughes}, {Infante}, {Irwin}, {Klein}, {Kosowsky}, {Lin},
  {Marsden}, {Moodley}, {Niemack}, {Nolta}, {Page}, {Parker}, {Partridge},
  {Sehgal}, {Sievers}, {Spergel}, {Staggs}, {Swetz}, {Switzer}, {Thornton},
  {Trac}, {Warne}, \& {Wollack}}]{menan10b}
{Menanteau}, F., {Gonz{\'a}lez}, J., {Juin}, J.-B., {et~al.}
  2010{\natexlab{a}}, \apj, 723, 1523

\bibitem[{{Menanteau} {et~al.}(2010{\natexlab{b}}){Menanteau}, {Hughes},
  {Barrientos}, {Deshpande}, {Hilton}, {Infante}, {Jimenez}, {Kosowsky},
  {Moodley}, {Spergel}, \& {Verde}}]{menan10a}
{Menanteau}, F., {Hughes}, J.~P., {Barrientos}, L.~F., {et~al.}
  2010{\natexlab{b}}, \apjs, 191, 340

\bibitem[{{Menanteau} {et~al.}(2012){Menanteau}, {Hughes}, {Sif{\'o}n},
  {Hilton}, {Gonz{\'a}lez}, {Infante}, {Barrientos}, {Baker}, {Bond}, {Das},
  {Devlin}, {Dunkley}, {Hajian}, {Hincks}, {Kosowsky}, {Marsden}, {Marriage},
  {Moodley}, {Niemack}, {Nolta}, {Page}, {Reese}, {Sehgal}, {Sievers},
  {Spergel}, {Staggs}, \& {Wollack}}]{menanteau12}
{Menanteau}, F., {Hughes}, J.~P., {Sif{\'o}n}, C., {et~al.} 2012, \apj, 748, 7

\bibitem[{{Menanteau} {et~al.}(2013){Menanteau}, {Sif{\'o}n}, {Barrientos},
  {Battaglia}, {Bond}, {Crichton}, {Das}, {Devlin}, {Dicker}, {D{\"u}nner},
  {Gralla}, {Hajian}, {Hasselfield}, {Hilton}, {Hincks}, {Hughes}, {Infante},
  {Kosowsky}, {Marriage}, {Marsden}, {Moodley}, {Niemack}, {Nolta}, {Page},
  {Partridge}, {Reese}, {Schmitt}, {Sievers}, {Spergel}, {Staggs}, {Switzer},
  \& {Wollack}}]{menan13}
{Menanteau}, F., {Sif{\'o}n}, C., {Barrientos}, L.~F., {et~al.} 2013, \apj,
  765, 67

\bibitem[{{Micha{\l}owski} {et~al.}(2010){Micha{\l}owski}, {Hjorth}, \&
  {Watson}}]{micha10}
{Micha{\l}owski}, M., {Hjorth}, J., \& {Watson}, D. 2010, \aap, 514, A67

\bibitem[{{Micha{\l}owski} {et~al.}(2008){Micha{\l}owski}, {Hjorth}, {Castro
  Cer{\'o}n}, \& {Watson}}]{micha08}
{Micha{\l}owski}, M.~J., {Hjorth}, J., {Castro Cer{\'o}n}, J.~M., \& {Watson},
  D. 2008, \apj, 672, 817

\bibitem[{{Mocanu} {et~al.}(2013){Mocanu}, {Crawford}, {Vieira}, {Aird},
  {Aravena}, {Austermann}, {Benson}, {B{\'e}thermin}, {Bleem}, {Bothwell},
  {Carlstrom}, {Chang}, {Chapman}, {Cho}, {Crites}, {de Haan}, {Dobbs},
  {Everett}, {George}, {Halverson}, {Harrington}, {Hezaveh}, {Holder},
  {Holzapfel}, {Hoover}, {Hrubes}, {Keisler}, {Knox}, {Lee}, {Leitch},
  {Lueker}, {Luong-Van}, {Marrone}, {McMahon}, {Mehl}, {Meyer}, {Mohr},
  {Montroy}, {Natoli}, {Padin}, {Plagge}, {Pryke}, {Rest}, {Reichardt}, {Ruhl},
  {Sayre}, {Schaffer}, {Shirokoff}, {Spieler}, {Spilker}, {Stalder},
  {Staniszewski}, {Stark}, {Story}, {Switzer}, {Vanderlinde}, \&
  {Williamson}}]{mocanu2013}
{Mocanu}, L.~M., {Crawford}, T.~M., {Vieira}, J.~D., {et~al.} 2013, \apj, 779,
  61

\bibitem[{{Navarro} {et~al.}(1997){Navarro}, {Frenk}, \& {White}}]{nfw97}
{Navarro}, J.~F., {Frenk}, C.~S., \& {White}, S.~D.~M. 1997, \apj, 490, 493

\bibitem[{{Ott}(2010)}]{ott10}
{Ott}, S. 2010, in Astronomical Society of the Pacific Conference Series, Vol.
  434, Astronomical Data Analysis Software and Systems XIX, ed. Y.~{Mizumoto},
  K.-I. {Morita}, \& M.~{Ohishi}, 139

\bibitem[{{Pilbratt} {et~al.}(2010){Pilbratt}, {Riedinger}, {Passvogel},
  {Crone}, {Doyle}, {Gageur}, {Heras}, {Jewell}, {Metcalfe}, {Ott}, \&
  {Schmidt}}]{pilbratt10}
{Pilbratt}, G.~L., {Riedinger}, J.~R., {Passvogel}, T., {et~al.} 2010, \aap,
  518, L1

\bibitem[{{Poglitsch} {et~al.}(2010){Poglitsch}, {Waelkens}, {Geis},
  {Feuchtgruber}, {Vandenbussche}, {Rodriguez}, {Krause}, {Renotte}, {van
  Hoof}, {Saraceno}, {Cepa}, {Kerschbaum}, {Agn{\`e}se}, {Ali}, {Altieri},
  {Andreani}, {Augueres}, {Balog}, {Barl}, {Bauer}, {Belbachir}, {Benedettini},
  {Billot}, {Boulade}, {Bischof}, {Blommaert}, {Callut}, {Cara}, {Cerulli},
  {Cesarsky}, {Contursi}, {Creten}, {De Meester}, {Doublier}, {Doumayrou},
  {Duband}, {Exter}, {Genzel}, {Gillis}, {Gr{\"o}zinger}, {Henning},
  {Herreros}, {Huygen}, {Inguscio}, {Jakob}, {Jamar}, {Jean}, {de Jong},
  {Katterloher}, {Kiss}, {Klaas}, {Lemke}, {Lutz}, {Madden}, {Marquet},
  {Martignac}, {Mazy}, {Merken}, {Montfort}, {Morbidelli}, {M{\"u}ller},
  {Nielbock}, {Okumura}, {Orfei}, {Ottensamer}, {Pezzuto}, {Popesso},
  {Putzeys}, {Regibo}, {Reveret}, {Royer}, {Sauvage}, {Schreiber}, {Stegmaier},
  {Schmitt}, {Schubert}, {Sturm}, {Thiel}, {Tofani}, {Vavrek}, {Wetzstein},
  {Wieprecht}, \& {Wiezorrek}}]{pogli10}
{Poglitsch}, A., {Waelkens}, C., {Geis}, N., {et~al.} 2010, \aap, 518, L2

\bibitem[{{Postman} {et~al.}(2012){Postman}, {Coe}, {Ben{\'{\i}}tez},
  {Bradley}, {Broadhurst}, {Donahue}, {Ford}, {Graur}, {Graves}, {Jouvel},
  {Koekemoer}, {Lemze}, {Medezinski}, {Molino}, {Moustakas}, {Ogaz}, {Riess},
  {Rodney}, {Rosati}, {Umetsu}, {Zheng}, {Zitrin}, {Bartelmann}, {Bouwens},
  {Czakon}, {Golwala}, {Host}, {Infante}, {Jha}, {Jimenez-Teja}, {Kelson},
  {Lahav}, {Lazkoz}, {Maoz}, {McCully}, {Melchior}, {Meneghetti}, {Merten},
  {Moustakas}, {Nonino}, {Patel}, {Reg{\"o}s}, {Sayers}, {Seitz}, \& {Van der
  Wel}}]{postman12}
{Postman}, M., {Coe}, D., {Ben{\'{\i}}tez}, N., {et~al.} 2012, \apjs, 199, 25

\bibitem[{{Reichardt} {et~al.}(2013){Reichardt}, {Stalder}, {Bleem}, {Montroy},
  {Aird}, {Andersson}, {Armstrong}, {Ashby}, {Bautz}, {Bayliss}, {Bazin},
  {Benson}, {Brodwin}, {Carlstrom}, {Chang}, {Cho}, {Clocchiatti}, {Crawford},
  {Crites}, {de Haan}, {Desai}, {Dobbs}, {Dudley}, {Foley}, {Forman}, {George},
  {Gladders}, {Gonzalez}, {Halverson}, {Harrington}, {High}, {Holder},
  {Holzapfel}, {Hoover}, {Hrubes}, {Jones}, {Joy}, {Keisler}, {Knox}, {Lee},
  {Leitch}, {Liu}, {Lueker}, {Luong-Van}, {Mantz}, {Marrone}, {McDonald},
  {McMahon}, {Mehl}, {Meyer}, {Mocanu}, {Mohr}, {Murray}, {Natoli}, {Padin},
  {Plagge}, {Pryke}, {Rest}, {Ruel}, {Ruhl}, {Saliwanchik}, {Saro}, {Sayre},
  {Schaffer}, {Shaw}, {Shirokoff}, {Song}, {Spieler}, {Staniszewski}, {Stark},
  {Story}, {Stubbs}, {{\v S}uhada}, {van Engelen}, {Vanderlinde}, {Vieira},
  {Vikhlinin}, {Williamson}, {Zahn}, \& {Zenteno}}]{reichardt2013}
{Reichardt}, C.~L., {Stalder}, B., {Bleem}, L.~E., {et~al.} 2013, \apj, 763,
  127

\bibitem[{{Reynolds}(1994)}]{reynolds94}
{Reynolds}, J. 1994, A Revised Flux Scale for the AT Compact Array, ATNF Memo
  AT/39.3/040 (\url{http://www.atnf.csiro.au/observers/memos/d96783~1.pdf})

\bibitem[{{Ruel} {et~al.}(2014){Ruel}, {Bazin}, {Bayliss}, {Brodwin}, {Foley},
  {Stalder}, {Aird}, {Armstrong}, {Ashby}, {Bautz}, {Benson}, {Bleem},
  {Bocquet}, {Carlstrom}, {Chang}, {Chapman}, {Cho}, {Clocchiatti}, {Crawford},
  {Crites}, {de Haan}, {Desai}, {Dobbs}, {Dudley}, {Forman}, {George},
  {Gladders}, {Gonzalez}, {Halverson}, {Harrington}, {High}, {Holder},
  {Holzapfel}, {Hrubes}, {Jones}, {Joy}, {Keisler}, {Knox}, {Lee}, {Leitch},
  {Liu}, {Lueker}, {Luong-Van}, {Mantz}, {Marrone}, {McDonald}, {McMahon},
  {Mehl}, {Meyer}, {Mocanu}, {Mohr}, {Montroy}, {Murray}, {Natoli},
  {Nurgaliev}, {Padin}, {Plagge}, {Pryke}, {Reichardt}, {Rest}, {Ruhl},
  {Saliwanchik}, {Saro}, {Sayre}, {Schaffer}, {Shaw}, {Shirokoff}, {Song}, {{\v
  S}uhada}, {Spieler}, {Stanford}, {Staniszewski}, {Starsk}, {Story}, {Stubbs},
  {van Engelen}, {Vanderlinde}, {Vieira}, {Vikhlinin}, {Williamson}, {Zahn}, \&
  {Zenteno}}]{ruel2014}
{Ruel}, J., {Bazin}, G., {Bayliss}, M., {et~al.} 2014, \apj, 792, 45

\bibitem[{{Sanders} \& {Mirabel}(1996)}]{sanders96}
{Sanders}, D.~B., \& {Mirabel}, I.~F. 1996, \araa, 34, 749

\bibitem[{{Sault} {et~al.}(1995){Sault}, {Teuben}, \& {Wright}}]{sault95}
{Sault}, R.~J., {Teuben}, P.~J., \& {Wright}, M.~C.~H. 1995, in Astronomical
  Society of the Pacific Conference Series, Vol.~77, Astronomical Data Analysis
  Software and Systems IV, ed. R.~A. {Shaw}, H.~E. {Payne}, \& J.~J.~E.
  {Hayes}, 433

\bibitem[{{Scott} {et~al.}(2002){Scott}, {Fox}, {Dunlop}, {Serjeant},
  {Peacock}, {Ivison}, {Oliver}, {Mann}, {Lawrence}, {Efstathiou},
  {Rowan-Robinson}, {Hughes}, {Archibald}, {Blain}, \& {Longair}}]{scott2002}
{Scott}, S.~E., {Fox}, M.~J., {Dunlop}, J.~S., {et~al.} 2002, \mnras, 331, 817

\bibitem[{{Sehgal} {et~al.}(2007){Sehgal}, {Trac}, {Huffenberger}, \&
  {Bode}}]{sehgal07}
{Sehgal}, N., {Trac}, H., {Huffenberger}, K., \& {Bode}, P. 2007, \apj, 664,
  149

\bibitem[{{Sehgal} {et~al.}(2011){Sehgal}, {Trac}, {Acquaviva}, {Ade},
  {Aguirre}, {Amiri}, {Appel}, {Barrientos}, {Battistelli}, {Bond}, {Brown},
  {Burger}, {Chervenak}, {Das}, {Devlin}, {Dicker}, {Bertrand Doriese},
  {Dunkley}, {D{\"u}nner}, {Essinger-Hileman}, {Fisher}, {Fowler}, {Hajian},
  {Halpern}, {Hasselfield}, {Hern{\'a}ndez-Monteagudo}, {Hilton}, {Hilton},
  {Hincks}, {Hlozek}, {Holtz}, {Huffenberger}, {Hughes}, {Hughes}, {Infante},
  {Irwin}, {Jones}, {Baptiste Juin}, {Klein}, {Kosowsky}, {Lau}, {Limon},
  {Lin}, {Lupton}, {Marriage}, {Marsden}, {Martocci}, {Mauskopf}, {Menanteau},
  {Moodley}, {Moseley}, {Netterfield}, {Niemack}, {Nolta}, {Page}, {Parker},
  {Partridge}, {Reid}, {Sherwin}, {Sievers}, {Spergel}, {Staggs}, {Swetz},
  {Switzer}, {Thornton}, {Tucker}, {Warne}, {Wollack}, \& {Zhao}}]{sehgal11}
{Sehgal}, N., {Trac}, H., {Acquaviva}, V., {et~al.} 2011, \apj, 732, 44

\bibitem[{{Serjeant} {et~al.}(2003){Serjeant}, {Dunlop}, {Mann},
  {Rowan-Robinson}, {Hughes}, {Efstathiou}, {Blain}, {Fox}, {Ivison},
  {Jenness}, {Lawrence}, {Longair}, {Oliver}, \& {Peacock}}]{serjeant03}
{Serjeant}, S., {Dunlop}, J.~S., {Mann}, R.~G., {et~al.} 2003, \mnras, 344, 887

\bibitem[{{Sif{\'o}n} {et~al.}(2013){Sif{\'o}n}, {Menanteau}, {Hasselfield},
  {Marriage}, {Hughes}, {Barrientos}, {Gonz{\'a}lez}, {Infante}, {Addison},
  {Baker}, {Battaglia}, {Bond}, {Crichton}, {Das}, {Devlin}, {Dunkley},
  {D{\"u}nner}, {Gralla}, {Hajian}, {Hilton}, {Hincks}, {Kosowsky}, {Marsden},
  {Moodley}, {Niemack}, {Nolta}, {Page}, {Partridge}, {Reese}, {Sehgal},
  {Sievers}, {Spergel}, {Staggs}, {Thornton}, {Trac}, \& {Wollack}}]{sifon2013}
{Sif{\'o}n}, C., {Menanteau}, F., {Hasselfield}, M., {et~al.} 2013, \apj, 772,
  25

\bibitem[{{Sif{\'o}n} {et~al.}(2016){Sif{\'o}n}, {Battaglia}, {Hasselfield},
  {Menanteau}, {Barrientos}, {Bond}, {Crichton}, {Devlin}, {D{\"u}nner},
  {Hilton}, {Hincks}, {Hlozek}, {Huffenberger}, {Hughes}, {Infante},
  {Kosowsky}, {Marsden}, {Marriage}, {Moodley}, {Niemack}, {Page}, {Spergel},
  {Staggs}, {Trac}, \& {Wollack}}]{sifon2016}
{Sif{\'o}n}, C., {Battaglia}, N., {Hasselfield}, M., {et~al.} 2016, \mnras,
  461, 248

\bibitem[{{Silva} {et~al.}(1998){Silva}, {Granato}, {Bressan}, \&
  {Danese}}]{silva98}
{Silva}, L., {Granato}, G.~L., {Bressan}, A., \& {Danese}, L. 1998, \apj, 509,
  103

\bibitem[{{Simpson} {et~al.}(2015{\natexlab{a}}){Simpson}, {Smail}, {Swinbank},
  {Almaini}, {Blain}, {Bremer}, {Chapman}, {Chen}, {Conselice}, {Coppin},
  {Danielson}, {Dunlop}, {Edge}, {Farrah}, {Geach}, {Hartley}, {Ivison},
  {Karim}, {Lani}, {Ma}, {Meijerink}, {Micha{\l}owski}, {Mortlock}, {Scott},
  {Simpson}, {Spaans}, {Thomson}, {van Kampen}, \& {van der
  Werf}}]{simpson2015a}
{Simpson}, J.~M., {Smail}, I., {Swinbank}, A.~M., {et~al.} 2015{\natexlab{a}},
  \apj, 799, 81

\bibitem[{{Simpson} {et~al.}(2015{\natexlab{b}}){Simpson}, {Smail}, {Swinbank},
  {Chapman}, {Geach}, {Ivison}, {Thomson}, {Aretxaga}, {Blain}, {Cowley},
  {Chen}, {Coppin}, {Dunlop}, {Edge}, {Farrah}, {Ibar}, {Karim}, {Knudsen},
  {Meijerink}, {Micha{\l}owski}, {Scott}, {Spaans}, \& {van der
  Werf}}]{simpson2015b}
---. 2015{\natexlab{b}}, \apj, 807, 128

\bibitem[{{Siringo} {et~al.}(2009){Siringo}, {Kreysa}, {Kov{\'a}cs},
  {Schuller}, {Wei{\ss}}, {Esch}, {Gem{\"u}nd}, {Jethava}, {Lundershausen},
  {Colin}, {G{\"u}sten}, {Menten}, {Beelen}, {Bertoldi}, {Beeman}, \&
  {Haller}}]{siringo09}
{Siringo}, G., {Kreysa}, E., {Kov{\'a}cs}, A., {et~al.} 2009, \aap, 497, 945

\bibitem[{{Smail} {et~al.}(1997{\natexlab{a}}){Smail}, {Ivison}, \&
  {Blain}}]{smail1997}
{Smail}, I., {Ivison}, R.~J., \& {Blain}, A.~W. 1997{\natexlab{a}}, \apjl, 490,
  L5

\bibitem[{{Smail} {et~al.}(1997{\natexlab{b}}){Smail}, {Ivison}, \&
  {Blain}}]{smai97}
---. 1997{\natexlab{b}}, \apjl, 490, L5+

\bibitem[{{Smail} {et~al.}(2000){Smail}, {Ivison}, {Owen}, {Blain}, \&
  {Kneib}}]{smail2000}
{Smail}, I., {Ivison}, R.~J., {Owen}, F.~N., {Blain}, A.~W., \& {Kneib}, J.-P.
  2000, \apj, 528, 612

\bibitem[{{Smol{\v c}i{\'c}} {et~al.}(2012){Smol{\v c}i{\'c}}, {Aravena},
  {Navarrete}, {Schinnerer}, {Riechers}, {Bertoldi}, {Feruglio}, {Finoguenov},
  {Salvato}, {Sargent}, {McCracken}, {Albrecht}, {Karim}, {Capak}, {Carilli},
  {Cappelluti}, {Elvis}, {Ilbert}, {Kartaltepe}, {Lilly}, {Sanders}, {Sheth},
  {Scoville}, \& {Taniguchi}}]{smolcic12}
{Smol{\v c}i{\'c}}, V., {Aravena}, M., {Navarrete}, F., {et~al.} 2012, \aap,
  548, A4

\bibitem[{{Smol{\v c}i{\'c}} {et~al.}(2015){Smol{\v c}i{\'c}}, {Karim},
  {Miettinen}, {Novak}, {Magnelli}, {Riechers}, {Schinnerer}, {Capak}, {Bondi},
  {Ciliegi}, {Aravena}, {Bertoldi}, {Bourke}, {Banfield}, {Carilli}, {Civano},
  {Ilbert}, {Intema}, {Le F{\`e}vre}, {Finoguenov}, {Hallinan}, {Kl{\"o}ckner},
  {Koekemoer}, {Laigle}, {Masters}, {McCracken}, {Mooley}, {Murphy},
  {Navarette}, {Salvato}, {Sargent}, {Sheth}, {Toft}, \&
  {Zamorani}}]{smolcic15}
{Smol{\v c}i{\'c}}, V., {Karim}, A., {Miettinen}, O., {et~al.} 2015, \aap, 576,
  A127

\bibitem[{{Strandet} {et~al.}(2016){Strandet}, {Weiss}, {Vieira}, {de Breuck},
  {Aguirre}, {Aravena}, {Ashby}, {B{\'e}thermin}, {Bradford}, {Carlstrom},
  {Chapman}, {Crawford}, {Everett}, {Fassnacht}, {Furstenau}, {Gonzalez},
  {Greve}, {Gullberg}, {Hezaveh}, {Kamenetzky}, {Litke}, {Ma}, {Malkan},
  {Marrone}, {Menten}, {Murphy}, {Nadolski}, {Rotermund}, {Spilker}, {Stark},
  \& {Welikala}}]{strandet2016}
{Strandet}, M.~L., {Weiss}, A., {Vieira}, J.~D., {et~al.} 2016, \apj, 822, 80

\bibitem[{{Su} {et~al.}(2017){Su}, {Marriage}, {Asboth}, {Baker}, {Bond},
  {Crichton}, {Devlin}, {D{\"u}nner}, {Farrah}, {Frayer}, {Gralla}, {Hall},
  {Halpern}, {Harris}, {Hilton}, {Hincks}, {Hughes}, {Niemack}, {Page},
  {Partridge}, {Rivera}, {Scott}, {Sievers}, {Thornton}, {Viero}, {Wang},
  {Wollack}, \& {Zemcov}}]{su2017}
{Su}, T., {Marriage}, T.~A., {Asboth}, V., {et~al.} 2017, \mnras, 464, 968

\bibitem[{{Sunyaev} \& {Zel'dovich}(1972)}]{sunyaev72}
{Sunyaev}, R.~A., \& {Zel'dovich}, Y.~B. 1972, Comments on Astrophysics and
  Space Physics, 4, 173

\bibitem[{{Swetz} {et~al.}(2011){Swetz}, {Ade}, {Amiri}, {Appel},
  {Battistelli}, {Burger}, {Chervenak}, {Devlin}, {Dicker}, {Doriese},
  {D{\"u}nner}, {Essinger-Hileman}, {Fisher}, {Fowler}, {Halpern},
  {Hasselfield}, {Hilton}, {Hincks}, {Irwin}, {Jarosik}, {Kaul}, {Klein},
  {Lau}, {Limon}, {Marriage}, {Marsden}, {Martocci}, {Mauskopf}, {Moseley},
  {Netterfield}, {Niemack}, {Nolta}, {Page}, {Parker}, {Staggs}, {Stryzak},
  {Switzer}, {Thornton}, {Tucker}, {Wollack}, \& {Zhao}}]{swetz11}
{Swetz}, D.~S., {Ade}, P.~A.~R., {Amiri}, M., {et~al.} 2011, \apjs, 194, 41

\bibitem[{{Vieira} {et~al.}(2010){Vieira}, {Crawford}, {Switzer}, {Ade},
  {Aird}, {Ashby}, {Benson}, {Bleem}, {Brodwin}, {Carlstrom}, {Chang}, {Cho},
  {Crites}, {de Haan}, {Dobbs}, {Everett}, {George}, {Gladders}, {Hall},
  {Halverson}, {High}, {Holder}, {Holzapfel}, {Hrubes}, {Joy}, {Keisler},
  {Knox}, {Lee}, {Leitch}, {Lueker}, {Marrone}, {McIntyre}, {McMahon}, {Mehl},
  {Meyer}, {Mohr}, {Montroy}, {Padin}, {Plagge}, {Pryke}, {Reichardt}, {Ruhl},
  {Schaffer}, {Shaw}, {Shirokoff}, {Spieler}, {Stalder}, {Staniszewski},
  {Stark}, {Vanderlinde}, {Walsh}, {Williamson}, {Yang}, {Zahn}, \&
  {Zenteno}}]{vieira2010}
{Vieira}, J.~D., {Crawford}, T.~M., {Switzer}, E.~R., {et~al.} 2010, \apj, 719,
  763

\bibitem[{{Voges} {et~al.}(1999){Voges}, {Aschenbach}, {Boller},
  {Br{\"a}uninger}, {Briel}, {Burkert}, {Dennerl}, {Englhauser}, {Gruber},
  {Haberl}, {Hartner}, {Hasinger}, {K{\"u}rster}, {Pfeffermann}, {Pietsch},
  {Predehl}, {Rosso}, {Schmitt}, {Tr{\"u}mper}, \& {Zimmermann}}]{voges1999}
{Voges}, W., {Aschenbach}, B., {Boller}, T., {et~al.} 1999, \aap, 349, 389

\bibitem[{{Wardlow} {et~al.}(2011){Wardlow}, {Smail}, {Coppin}, {Alexander},
  {Brandt}, {Danielson}, {Luo}, {Swinbank}, {Walter}, {Wei{\ss}}, {Xue},
  {Zibetti}, {Bertoldi}, {Biggs}, {Chapman}, {Dannerbauer}, {Dunlop},
  {Gawiser}, {Ivison}, {Knudsen}, {Kov{\'a}cs}, {Lacey}, {Menten}, {Padilla},
  {Rix}, \& {van der Werf}}]{wardlow11}
{Wardlow}, J.~L., {Smail}, I., {Coppin}, K.~E.~K., {et~al.} 2011, \mnras, 415,
  1479

\bibitem[{{Webb} {et~al.}(2003){Webb}, {Eales}, {Foucaud}, {Lilly},
  {McCracken}, {Adelberger}, {Steidel}, {Shapley}, {Clements}, {Dunne}, {Le
  F{\`e}vre}, {Brodwin}, \& {Gear}}]{webb03}
{Webb}, T.~M., {Eales}, S., {Foucaud}, S., {et~al.} 2003, \apj, 582, 6

\bibitem[{{Wei{\ss}} {et~al.}(2009){Wei{\ss}}, {Kov{\'a}cs}, {Coppin}, {Greve},
  {Walter}, {Smail}, {Dunlop}, {Knudsen}, {Alexander}, {Bertoldi}, {Brandt},
  {Chapman}, {Cox}, {Dannerbauer}, {De Breuck}, {Gawiser}, {Ivison}, {Lutz},
  {Menten}, {Koekemoer}, {Kreysa}, {Kurczynski}, {Rix}, {Schinnerer}, \& {van
  der Werf}}]{weiss09b}
{Wei{\ss}}, A., {Kov{\'a}cs}, A., {Coppin}, K., {et~al.} 2009, \apj, 707, 1201

\bibitem[{{Wei{\ss}} {et~al.}(2013){Wei{\ss}}, {De Breuck}, {Marrone},
  {Vieira}, {Aguirre}, {Aird}, {Aravena}, {Ashby}, {Bayliss}, {Benson},
  {B{\'e}thermin}, {Biggs}, {Bleem}, {Bock}, {Bothwell}, {Bradford}, {Brodwin},
  {Carlstrom}, {Chang}, {Chapman}, {Crawford}, {Crites}, {de Haan}, {Dobbs},
  {Downes}, {Fassnacht}, {George}, {Gladders}, {Gonzalez}, {Greve},
  {Halverson}, {Hezaveh}, {High}, {Holder}, {Holzapfel}, {Hoover}, {Hrubes},
  {Husband}, {Keisler}, {Lee}, {Leitch}, {Lueker}, {Luong-Van}, {Malkan},
  {McIntyre}, {McMahon}, {Mehl}, {Menten}, {Meyer}, {Murphy}, {Padin},
  {Plagge}, {Reichardt}, {Rest}, {Rosenman}, {Ruel}, {Ruhl}, {Schaffer},
  {Shirokoff}, {Spilker}, {Stalder}, {Staniszewski}, {Stark}, {Story},
  {Vanderlinde}, {Welikala}, \& {Williamson}}]{weiss13}
{Wei{\ss}}, A., {De Breuck}, C., {Marrone}, D.~P., {et~al.} 2013, \apj, 767, 88

\bibitem[{{Werner} {et~al.}(2004){Werner}, {Roellig}, {Low}, {Rieke}, {Rieke},
  {Hoffmann}, {Young}, {Houck}, {Brandl}, {Fazio}, {Hora}, {Gehrz}, {Helou},
  {Soifer}, {Stauffer}, {Keene}, {Eisenhardt}, {Gallagher}, {Gautier}, {Irace},
  {Lawrence}, {Simmons}, {Van Cleve}, {Jura}, {Wright}, \&
  {Cruikshank}}]{werner04}
{Werner}, M.~W., {Roellig}, T.~L., {Low}, F.~J., {et~al.} 2004, \apjs, 154, 1

\bibitem[{{Wu} \& {Fang}(1997)}]{wu97}
{Wu}, X.-P., \& {Fang}, L.-Z. 1997, \apj, 483, 62

\bibitem[{{Zitrin} {et~al.}(2013){Zitrin}, {Menanteau}, {Hughes}, {Coe},
  {Barrientos}, {Infante}, \& {Mandelbaum}}]{zitrin2013}
{Zitrin}, A., {Menanteau}, F., {Hughes}, J.~P., {et~al.} 2013, \apjl, 770, L15

\bibitem[{{Zitrin} {et~al.}(2015){Zitrin}, {Fabris}, {Merten}, {Melchior},
  {Meneghetti}, {Koekemoer}, {Coe}, {Maturi}, {Bartelmann}, {Postman},
  {Umetsu}, {Seidel}, {Sendra}, {Broadhurst}, {Balestra}, {Biviano}, {Grillo},
  {Mercurio}, {Nonino}, {Rosati}, {Bradley}, {Carrasco}, {Donahue}, {Ford},
  {Frye}, \& {Moustakas}}]{zitrin2015}
{Zitrin}, A., {Fabris}, A., {Merten}, J., {et~al.} 2015, \apj, 801, 44

\end{thebibliography}

\end{document}